\documentclass{article}
\PassOptionsToPackage{numbers, sort&compress}{natbib}

\usepackage[main, final]{neurips_2026}
\usepackage[utf8]{inputenc} % allow utf-8 input
\usepackage[T1]{fontenc}    % use 8-bit T1 fonts
\usepackage[hidelinks]{hyperref}
\usepackage{url}            % simple URL typesetting
\usepackage{booktabs}       % professional-quality tables
\usepackage{amsfonts}       % blackboard math symbols
\usepackage{nicefrac}       % compact symbols for 1/2, etc.
\usepackage{microtype}      % microtypography
\usepackage{xcolor}         % colors

\usepackage{comment}
\usepackage{algorithm}
\usepackage{algpseudocode}
\usepackage{acronym}
\usepackage{siunitx}
\usepackage[table]{xcolor}
\usepackage{subcaption}
\usepackage{amsmath,amssymb}
\usepackage[capitalize]{cleveref}
\usepackage{etoolbox}
\makeatletter
\patchcmd{\AC@verridelabel}{\ltx@label{#1}}{\ltx@label{{#1}}}{}{%
  \PackageError{sips-paper}{Acronym label compatibility patch failed}{}%
}
\makeatother
\crefname{figure}{Figure}{Figures}
\Crefname{figure}{Figure}{Figures}
\usepackage{tikz}
\usepackage{xcolor}

\usepackage{pgfplots}
\pgfplotsset{compat=1.18}

\definecolor{cbBlue}{HTML}{0072B2}
\definecolor{cbOrange}{HTML}{D55E00}
\definecolor{cbGreen}{HTML}{009E73}

\usepackage[table]{xcolor}
\definecolor{lightblue}{RGB}{235,245,255}
\usepackage{tikz}
\usetikzlibrary{backgrounds,fit,arrows.meta,positioning,calc,decorations.markings,shapes.geometric}
\usepackage{threeparttable}
\newcommand{\trimgrayrow}{\noalign{\kern-0.75bp{\color{white}\hrule height 0.75bp}}}

\acrodef{ODE}{ordinary differential equation}
\acrodef{SDE}{stochastic differential equation}
\acrodef{STFT}{short-time Fourier transform}
\acrodef{SNR}{signal-to-noise ratio}
\acrodef{GPU}{graphics processing unit}
\acrodef{ASR}{automatic speech recognition}
\acrodef{WER}{word error rate}
\acrodef{SGM}{score-based generative model}
\acrodef{DDRM}{Denoising Diffusion Restoration Models}
\acrodef{CI}{confidence interval}
\acrodef{CMOS}{comparison mean opinion score}
\acrodef{CCR}{comparison category rating}

\title{Predictive–Generative Drift Decomposition for \\Speech Enhancement and Separation}

\author{%
Julius Richter \\
MERL\\
Cambridge, MA, USA\\
\texttt{richter@merl.com} \\
\And
Yoshiki Masuyama \\
MERL \\
Cambridge, MA, USA \\
\texttt{masuyama@merl.com} \\
\And
Christoph Boeddeker \\
MERL \\
Cambridge, MA, USA \\
\texttt{boeddeker@merl.com} \\
\And
Takahiro Edo \\
MERL \\
Cambridge, MA, USA \\
\texttt{tedo@merl.com} \\
\And
Gordon Wichern \\
MERL \\
Cambridge, MA, USA \\
\texttt{wichern@merl.com} \\
\And
Jonathan Le Roux \\
MERL \\
Cambridge, MA, USA \\
\texttt{leroux@merl.com} \\
}

\begin{document}

\maketitle

\begin{abstract}
We propose Stochastic Interpolant Prior for Speech (SIPS), a plug-and-play framework for speech enhancement and separation that augments pretrained predictors with a generative clean speech prior.
Building on stochastic interpolants, we decompose the sampling dynamics into a predictor-induced drift and a score-based denoising component.
The predictor guides sampling toward a task-consistent estimate, while a score model trained exclusively on clean speech promotes perceptual naturalness.
Unlike prior hybrid approaches, which typically rely on architecture-specific conditioning and are tied to particular predictors or degradation settings, SIPS employs a degradation-agnostic prior that can be reused across predictors and additive degradation tasks without retraining.
We evaluate SIPS on speech enhancement and separation using recent predictors such as SEMamba and FlexIO.
Across both tasks, SIPS consistently improves non-intrusive speech quality metrics while maintaining competitive performance on reference-based distortion measures and word error rate.
For speech enhancement, these improvements are further supported by a subjective listening test.
\end{abstract}

\section{Introduction}

Learning-based methods for speech enhancement and speech separation have achieved remarkable progress over the past decades~\cite{araki202530+}. 
Modern neural networks trained in a supervised manner are capable of producing highly accurate reconstructions and have substantially improved both objective speech quality metrics and subjective human ratings~\cite{wang2018supervised,reddy2020interspeech}.
Existing approaches can be broadly categorized into predictive and generative models.

Predictive models dominate the field~\cite{stoller2018waveunet, luo2019conv,subakan2021attention,choi2019phase,chao2024investigation}. 
These methods directly estimate the clean speech signal from degraded observations using regression-based objectives.
With sufficient training data and architectural capacity, predictive models achieve strong performance on signal-level metrics. 
However, because they are optimized for point estimates, they can produce perceptually unpleasant artifacts in challenging scenarios, such as extremely low \acp{SNR}, heavy reverberation, or domain mismatch. 
In such cases, these models may over-suppress, leak the interfering signal, or generate unnatural distortions that degrade perceptual quality~\cite{de2023behavior,leglaive2025objective}.

Generative models offer a complementary perspective.
Instead of directly estimating clean speech through regression, they model distributions over clean speech signals.
Prior-based approaches learn a clean speech distribution $p_S$ that can be combined with an observation model to enable approximate inference from the posterior $p_{S|Y}$ given a degraded observation $Y$~\cite{lemercier2025diffusion}.
Alternatively, conditional generative approaches directly model the posterior $p_{S|Y}$ by conditioning the generative process on the observation, as in conditional diffusion or score-based methods such as SGMSE~\cite{welker2022speech,richter2023speech}.
However, this expressive power comes at a cost: generative restoration methods are prone to hallucinations, inconsistencies with the observation, and reduced faithfulness to the underlying signal, especially when the conditioning signal is weak or ambiguous~\cite{de2023behavior, chhaglani2026artifree}.
 
These observations suggest a trade-off between reconstruction fidelity and perceptual quality.
Predictive models are typically faithful to the underlying signal and excel on reference-based metrics such as SI-SDR~\cite{leroux2019sdr} and PESQ~\cite{rix2001perceptual}.
However, strong performance on these metrics does not necessarily guarantee perceptually natural speech, particularly when models are explicitly optimized for the evaluation metrics~\cite{deoliveira2024pesqetarian}.
In contrast, generative models can better capture the distribution of clean speech and produce more natural-sounding outputs~\cite{richter2023speech}, often reflected in strong performance on non-intrusive (reference-free) quality metrics such as DNSMOS~\cite{reddy2021dnsmos}.
Nevertheless, generative models may alter linguistic content or introduce hallucinations, which can be indirectly assessed through \ac{ASR} using \ac{WER} as a complementary metric~\cite{oliveira26b_interspeech}.
Ideally, a speech restoration system should combine the strengths of both paradigms, achieving high perceptual quality while preserving reconstruction fidelity and linguistic content. Bridging this gap remains an open challenge.

\begin{figure}
    \centering
    \begin{tikzpicture}[
    scale=0.8,
    font=\footnotesize,
    block/.style={
        draw=black!80,
        line width=0.8pt,
        rectangle,
        rounded corners,
        align=center,
        minimum width=2.0cm,
        minimum height=0.8cm
    },
    smallblock/.style={
        draw=black!80,
        line width=0.8pt,
        rectangle,
        rounded corners,
        align=center,
        minimum width=1.0cm,
        minimum height=0.7cm
    },
    signal/.style={},
    >=Stealth,
    arrow/.style={->, line width=0.7pt, >={Stealth[length=4pt,width=4pt]}},
]

% Main block centered at (0,0)
\node[block, minimum width=3.0cm, minimum height=1.4cm, fill=black!6] (euler) at (0,0)
    {Sampling Step\\[0.5ex]cf.\ Eq.~\eqref{eq:sde}};

% Input/output anchor points
\coordinate (in1) at ($(euler.west)+(0,0.6)$);
\coordinate (in2) at ($(euler.west)+(0,0)$);
\coordinate (in3) at ($(euler.west)+(0,-0.6)$);
\coordinate (out) at (euler.east);

\node[above] at ($(in1)+(-0.3,0)$) {$x_t$};
\node[above] at ($(in2)+(-0.3,0)$) {$\hat{z}$};
\node[above] at ($(in3)+(-0.3,0)$) {$\hat{v}$};

% Output node
\coordinate (outR) at ($(out)+(1.2,0)$);

\node[right] at (outR) {$\hat{s}$};

% D_theta block on upper-left branch
\node[smallblock, fill=cbBlue!15] (dtheta) at ($(euler.north west)+(-2.1,-0.2)$) {$D_\theta$};

% P_phi(Y)-Y block for third input
\node[smallblock, fill=cbOrange!15] (pphi) at ($(euler.west)+(-4.0,-0.6)$) {$P_\phi$};

% Input signal Y
\node[signal] (Y) at ($(pphi.west)+(-1.0,0)$) {$y$};

% Output arrow
\draw[arrow] (out) -- (outR);

% --- Rectangular feedback loop (right → up → left → down) ---
\coordinate (loopA) at ($(out)+(0.3,0)$);     % right
\coordinate (loopB) at ($(loopA)+(0,1.5)$);   % up
\coordinate (loopC) at ($(loopB)+(-5.1,0)$);  % left

\draw[arrow]
  (out)
  -- (loopA)
  -- (loopB)
  -- (loopC)
  |- (in1);

\draw[arrow] (loopC) -| (dtheta.north);

% Branch to D_theta and second input
\draw[arrow] (dtheta.south) |- (in2);

% Third input path
\draw[arrow] (Y) -- (pphi.west);
\draw[arrow] (pphi.east) -- (in3);

\begin{scope}[on background layer]
\node[
  draw=black!80,
  dash pattern=on 3pt off 2.03pt,
  rounded corners,
  fill=black!1,
  fit=(euler)(dtheta)(loopB),
  inner sep=5pt,
  label={[black,font=\small]above:Sampling with SIPS}
] (sipsbox) {};
\end{scope}

\end{tikzpicture}
    \caption{Overview of the proposed plug-and-play framework. The predictor $P_\phi$ defines a deterministic drift $\hat{v} = P_\phi(y) - y$ from the observation $y$, while the denoiser $D_\theta$ estimates the Gaussian noise component $\hat{z}$ to incorporate the learned clean speech prior. Both components are combined in each sampling step to update the current state $x_t$, ultimately yielding the final clean speech estimate $\hat{s}$.}%
    \label{fig:overview}
\end{figure}

In this work, we propose Stochastic Interpolant Prior for Speech (SIPS), a plug-and-play framework that combines the strengths of predictive and generative modeling.
Our approach augments pretrained predictive speech restoration models with a generative prior learned exclusively from clean speech via score modeling~\cite{song2021score}.
This yields a degradation-agnostic prior that can be reused across different restoration tasks.
Building on stochastic interpolants~\cite{albergo2023building,albergo2025stochastic}, we decompose the sampling dynamics into a predictor-induced drift and a score-based denoising component.
At inference time, the output of a predictive model $P_\phi$ defines a deterministic drift term $\hat{v}$ that guides the sampling process toward a task-consistent estimate (see \cref{fig:overview}).  
In parallel, a denoiser $D_\theta$ estimates the Gaussian noise component $\hat z$, implicitly modeling the score of the perturbed clean speech distribution and promoting perceptual naturalness during sampling.
This decomposition helps preserve reconstruction fidelity through the predictor, while the generative prior mitigates artifacts and enhances naturalness.

We evaluate SIPS on speech enhancement and separation using recent predictors, including SEMamba~\cite{chao2024investigation} and FlexIO~\cite{masuyama2026flexio}.
Across both tasks and different predictors, SIPS consistently improves non-intrusive speech quality metrics such as NISQA~\cite{mittag2021nisqa} and UTMOS~\cite{saeki2022utmos}, while maintaining competitive performance on reference-based distortion measures and \ac{WER}.
For speech enhancement, these improvements are further supported by a subjective listening test.
These results demonstrate the potential of combining predictive and generative modeling to improve perceptual speech quality while preserving reconstruction fidelity.
Our implementation is publicly available\footnote{\url{https://github.com/merlresearch/sips-speech}}.

\section{Related Work}

A significant breakthrough in audio restoration tasks, including speech enhancement and separation, came with the introduction of diffusion models, which have demonstrated substantial improvements in perceptual quality and robustness~\cite{lemercier2025diffusion}.
Since then, several generative formulations have been explored for speech enhancement and separation, including flow matching~\cite{lee2025flowse, scheibler2025source,welker2025real} and Schrödinger bridges~\cite{jukic2024schrodinger,richter2025investigating}.

A parallel line of research investigates hybrid approaches that combine the strengths of predictive and generative models. 
\Cref{fig:modeling_approaches} provides an overview of different modeling paradigms for speech enhancement, including predictive, generative, and hybrid approaches. 
StoRM~\cite{lemercier2023storm} introduces a stochastic regeneration model in which a diffusion model is conditioned on both the noisy observation and the output of a predictor network, with the reverse diffusion process initialized from the predictor's output perturbed by Gaussian noise. Both the predictor and the score model use the NCSN++~\cite{song2021score} architecture.
Diffiner~\cite{sawata2023diffiner} is a diffusion-based refinement framework trained solely on clean speech. 
Building on \ac{DDRM}~\cite{kawar2022denoising}, it combines pretrained speech enhancement models, including Wave-U-Net~\cite{stoller2018waveunet} and DCUnet~\cite{choi2019phase}, with a diffusion prior through posterior sampling.

\begin{figure}[t]
\centering

\subcaptionbox{Predictive\label{fig:pred_approach}}[0.195\textwidth]{
    \centering
    \begin{tikzpicture}[
    scale=0.5,
    font=\small,
    dot/.style = {circle, fill, minimum size=#1, inner sep=0pt, outer sep=0pt},
    myptr/.style={
    line width=0.8pt,
    -{Stealth[length=2.5pt,width=3pt]}
}
]

% Input
\coordinate (y) at (-0.8, -0.4);

\node[dot=2pt, fill=black] at (y) {};
\node[left] at (y) {$Y$};

% Mean    
\node [
  star,
  fill=cbOrange,
  minimum size=4pt,
  inner sep=0pt,
  star point ratio=2.6
] at (1.3, -0.1) (mean) {};

\node[above left=0pt of mean, xshift=3pt, yshift=0pt, text=cbOrange] {$\mathbb{E}[S|Y]$}; 

% Arrow (corrected)
\draw[myptr] (y) to (mean);

% Invisible  point 
\node [dot=0pt, fill=blue!0] at (1.9, -0.6) (X) {};

% Label for the distribution
\node at (2.4, 2.0) [text=cbBlue] {$p_{S|Y}$};

% Then apply the clip path (invisible)
\path[clip] plot[smooth cycle, tension=1.4] coordinates{(2.5,0) (0,-2) (1.5, 0) (0,2) (1.7, 1.7)};

% First cluster (thesisblue, semi-transparent)
\shade[inner color=cbBlue!60, outer color=blue!0, opacity=0.5, shading=radial] 
    (2.0, 0.2) circle [radius=1.2cm];

% Second cluster (thesisblue, semi-transparent)
\shade[inner color=cbBlue!30, outer color=blue!0, opacity=0.2, shading=radial] 
    (1.2, -1.4) circle [radius=0.6cm];

% Third cluster (thesisblue, semi-transparent)
\shade[inner color=cbBlue!50, outer color=blue!0, opacity=0.2, shading=radial] 
    (0.8, 1.7) circle [radius=0.9cm];

% Points (same as before)
\node [dot=2pt, fill=cbBlue] at (1.0,1.5) (point1) {};
\node [dot=2pt, fill=cbBlue] at (1.2, 1.3) (point2) {};
\node [dot=2pt, fill=cbBlue] at (1.5, 1.0) (point3) {};
\node [dot=2pt, fill=cbBlue] at (1.8, 0.8) (point5) {};
\node [dot=2pt, fill=cbBlue] at (1.9, 0.7) (point6) {};
\node [dot=2pt, fill=cbBlue] at (1.7, 0.7) (point7) {};
\node [dot=2pt, fill=cbBlue] at (1.9, 0.2) (point8) {};
\node [dot=2pt, fill=cbBlue] at (1.95, 0.1) (point9) {};
\node [dot=2pt, fill=cbBlue] at (1.95, 0.02) (point10) {};
\node [dot=2pt, fill=cbBlue] at (2.1, 0.0) (point11) {};
\node [dot=2pt, fill=cbBlue] at (2.2, 0.01) (point12) {};
\node [dot=2pt, fill=cbBlue] at (1.12,-1.4) (point13) {};
\node [dot=2pt, fill=cbBlue] at (1.23, -1.3) (point14) {};
\node [dot=2pt, fill=cbBlue] at (1.57, -1.0) (point15) {};
\node [dot=2pt, fill=cbBlue] at (1.98, -0.25) (point16) {};
\node [dot=2pt, fill=cbBlue] at (2.01, -0.1) (point17) {};
\node [dot=2pt, fill=cbBlue] at (1.82, -0.02) (point18) {};
\node [dot=2pt, fill=cbBlue] at (2.08, -0.0) (point19) {};
\node [dot=2pt, fill=cbBlue] at (2.25, -0.01) (point20) {};

% % Draw the dashed line first (visible)
% \path[draw=black!30, dash pattern=on 0.8pt off 0.8pt, line width=1.8pt] plot[smooth cycle, tension=1.4] coordinates{(2.5,0) (0,-2) (1.5, 0) (0,2) (1.7, 1.7)};

% Draw the dashed line first (visible)
\path[draw=cbBlue!60, line width=1.0pt] plot[smooth cycle, tension=1.4] coordinates{(2.5,0) (0,-2) (1.5, 0) (0,2) (1.7, 1.7)};

% \draw[myptr] (y) to [out=330,in=190] node [below] {} (X);

\end{tikzpicture}
}
\subcaptionbox{SGMSE~[\citenum{richter2023speech}]\label{fig:gen_approach}}[0.195\textwidth]{
    \centering
    \begin{tikzpicture}[
    scale=0.5,
    font=\small,
    dot/.style = {circle, fill, minimum size=#1, inner sep=0pt, outer sep=0pt},
    myptr/.style={
    line width=0.8pt,
    -{Stealth[length=2.5pt,width=3pt]}
}
]

% Input 0c5da5

\coordinate (y) at (-0.8, -0.4);

\node[dot=2pt, fill=black] at (y) {};
\node[left] at (y) {$Y$};

% Invisible  point 
\node [dot=0pt, fill=blue!0] at (1.9, -0.6) (X) {};

\coordinate (p1) at ($(y)!0.33!(X)$);
\coordinate (p2) at ($(y)!0.66!(X)$);

% \node[dot=2pt, fill=black] at (p1) (dot1) {};
% \node[dot=2pt, fill=black] at (p2) (dot2) {};s

\draw[myptr] (y) to [out=330,in=190] node [below] {} (p1);
\draw[myptr] (p1) to [out=330,in=190] node [below] {} (p2);
\draw[myptr] (p2) to [out=330,in=190] node [below] {} (X);

% Label for the distribution
\node at (2.4, 2.0) [text=cbBlue] {$p_{S|Y}$};

% Then apply the clip path (invisible)
\path[clip] plot[smooth cycle, tension=1.4] coordinates{(2.5,0) (0,-2) (1.5, 0) (0,2) (1.7, 1.7)};

% First cluster (thesisblue, semi-transparent)
\shade[inner color=cbBlue!60, outer color=blue!0, opacity=0.5, shading=radial] 
    (2.0, 0.2) circle [radius=1.2cm];

% Second cluster (thesisblue, semi-transparent)
\shade[inner color=cbBlue!30, outer color=blue!0, opacity=0.2, shading=radial] 
    (1.2, -1.4) circle [radius=0.6cm];

% Third cluster (thesisblue, semi-transparent)
\shade[inner color=cbBlue!50, outer color=blue!0, opacity=0.2, shading=radial] 
    (0.8, 1.7) circle [radius=0.9cm];

% Points (same as before)
\node [dot=2pt, fill=cbBlue] at (1.2, 1.3) (point2) {};
\node [dot=2pt, fill=cbBlue] at (1.5, 1.0) (point3) {};
\node [dot=2pt, fill=cbBlue] at (1.8, 0.8) (point5) {};
\node [dot=2pt, fill=cbBlue] at (1.9, 0.7) (point6) {};
\node [dot=2pt, fill=cbBlue] at (1.7, 0.7) (point7) {};
\node [dot=2pt, fill=cbBlue] at (1.9, 0.2) (point8) {};
\node [dot=2pt, fill=cbBlue] at (1.95, 0.1) (point9) {};
\node [dot=2pt, fill=cbBlue] at (1.95, 0.02) (point10) {};
\node [dot=2pt, fill=cbBlue] at (2.1, 0.0) (point11) {};
\node [dot=2pt, fill=cbBlue] at (2.2, 0.01) (point12) {};
\node [dot=2pt, fill=cbBlue] at (1.12,-1.4) (point13) {};
\node [dot=2pt, fill=cbBlue] at (1.23, -1.3) (point14) {};
\node [dot=2pt, fill=cbBlue] at (1.57, -1.0) (point15) {};
\node [dot=2pt, fill=cbBlue] at (1.98, -0.25) (point16) {};
\node [dot=2pt, fill=cbBlue] at (2.01, -0.1) (point17) {};
\node [dot=2pt, fill=cbBlue] at (1.82, -0.02) (point18) {};
\node [dot=2pt, fill=cbBlue] at (2.08, -0.0) (point19) {};
\node [dot=2pt, fill=cbBlue] at (2.25, -0.01) (point20) {};

% Draw the dashed line first (visible)
\path[draw=cbBlue!60, line width=1.0pt] plot[smooth cycle, tension=1.4] coordinates{(2.5,0) (0,-2) (1.5, 0) (0,2) (1.7, 1.7)};

\draw[myptr] (p2) to [out=330,in=190] node [below] {} (X);

\end{tikzpicture}
}
\subcaptionbox{StoRM~[\citenum{lemercier2023storm}]\label{fig:storm}}[0.195\textwidth]{
    \centering
    \begin{tikzpicture}[
    scale=0.5,
    font=\small,
    dot/.style = {circle, fill, minimum size=#1, inner sep=0pt, outer sep=0pt},
    myptr/.style={
    line width=0.8pt,
    -{Stealth[length=2.5pt,width=3pt]}
}
]

% Input
\coordinate (y) at (-0.8, -0.4);

\node[dot=2pt, fill=black] at (y) {};
\node[left] at (y) {$Y$};

\coordinate (starcoord) at (1.3, -0.1);

\coordinate (p1) at ($(mean)!0.50!(X)$);

\draw[myptr] (starcoord) to [out=300,in=150,looseness=1.0] node [below] {} (p1);
\draw[myptr] (p1) to [out=300,in=150,looseness=1.0] node [below] {} (X);

% Mean    
\node [
  star,
  fill=cbOrange,
  minimum size=4pt,
  inner sep=0pt,
  star point ratio=2.6
] at (1.3, -0.1) (mean) {};

\node[above left=0pt of mean, xshift=3pt, yshift=0pt, text=cbOrange] {$\mathbb{E}[S|Y]$}; 

% Arrow (corrected)
\draw[myptr] (y) to (mean);

% Invisible  point 
\node [dot=0pt, fill=blue!0] at (1.9, -0.6) (X) {};

% \node[dot=2pt, fill=black] at (p1) (dot1) {};

% Label for the distribution
\node at (2.4, 2.0) [text=cbBlue] {$p_{S|Y}$};

% Then apply the clip path (invisible)
\path[clip] plot[smooth cycle, tension=1.4] coordinates{(2.5,0) (0,-2) (1.5, 0) (0,2) (1.7, 1.7)};

% First cluster (thesisblue, semi-transparent)
\shade[inner color=cbBlue!60, outer color=blue!0, opacity=0.5, shading=radial] 
    (2.0, 0.2) circle [radius=1.2cm];

% Second cluster (thesisblue, semi-transparent)
\shade[inner color=cbBlue!30, outer color=blue!0, opacity=0.2, shading=radial] 
    (1.2, -1.4) circle [radius=0.6cm];

% Third cluster (thesisblue, semi-transparent)
\shade[inner color=cbBlue!50, outer color=blue!0, opacity=0.2, shading=radial] 
    (0.8, 1.7) circle [radius=0.9cm];

% Points (same as before)
\node [dot=2pt, fill=cbBlue] at (1.0,1.5) (point1) {};
\node [dot=2pt, fill=cbBlue] at (1.2, 1.3) (point2) {};
\node [dot=2pt, fill=cbBlue] at (1.5, 1.0) (point3) {};
\node [dot=2pt, fill=cbBlue] at (1.8, 0.8) (point5) {};
\node [dot=2pt, fill=cbBlue] at (1.9, 0.7) (point6) {};
\node [dot=2pt, fill=cbBlue] at (1.7, 0.7) (point7) {};
\node [dot=2pt, fill=cbBlue] at (1.9, 0.2) (point8) {};
\node [dot=2pt, fill=cbBlue] at (1.95, 0.1) (point9) {};
\node [dot=2pt, fill=cbBlue] at (1.95, 0.02) (point10) {};
\node [dot=2pt, fill=cbBlue] at (2.1, 0.0) (point11) {};
\node [dot=2pt, fill=cbBlue] at (2.2, 0.01) (point12) {};
\node [dot=2pt, fill=cbBlue] at (1.12,-1.4) (point13) {};
\node [dot=2pt, fill=cbBlue] at (1.23, -1.3) (point14) {};
\node [dot=2pt, fill=cbBlue] at (1.57, -1.0) (point15) {};
\node [dot=2pt, fill=cbBlue] at (1.98, -0.25) (point16) {};
\node [dot=2pt, fill=cbBlue] at (2.01, -0.1) (point17) {};
\node [dot=2pt, fill=cbBlue] at (1.82, -0.02) (point18) {};
\node [dot=2pt, fill=cbBlue] at (2.08, -0.0) (point19) {};
\node [dot=2pt, fill=cbBlue] at (2.25, -0.01) (point20) {};

% Draw the dashed line first (visible)
\path[draw=cbBlue!60, line width=1.0pt] plot[smooth cycle, tension=1.4] coordinates{(2.5,0) (0,-2) (1.5, 0) (0,2) (1.7, 1.7)};

\draw[myptr] (starcoord) to [out=300,in=150,looseness=1.0] node [below] {} (p1);
\draw[myptr] (p1) to [out=300,in=150,looseness=1.0] node [below] {} (X);
% \node[dot=2pt, fill=black] at (p1) (dot1) {};

\end{tikzpicture}
}
\subcaptionbox{Diffiner~[\citenum{sawata2023diffiner}]\label{fig:diffiner}}[0.195\textwidth]{
    \centering
    \raisebox{-5mm}{%
        \begin{tikzpicture}[
    scale=0.5,
    font=\small,
    dot/.style = {circle, fill, minimum size=#1, inner sep=0pt, outer sep=0pt},
    myptr/.style={
    line width=0.8pt,
    -{Stealth[length=2.5pt,width=3pt]}
}
]

% Input
\coordinate (y) at (-0.8, -0.4);

\node[dot=2pt, fill=black] at (y) {};
\node[left] at (y) {$Y$};

\coordinate (meanpt) at (1.3, -0.1);

% Different levels of the distribution

% Draw the first line (visible)
\path[draw=cbBlue!10, line width=0.5pt, fill=white] plot[smooth cycle, tension=1.4] coordinates{(2.20,-0.10) (0,-1.8) (-0.5, 0) (-1,1.95) (1.7, 1.6)};

% Draw the second line (visible)
\path[draw=cbBlue!20, line width=0.5pt, fill=white] plot[smooth cycle, tension=1.4] coordinates{(2.2,-0.05) (-0.1,-1.85) (0.2, 0) (-0.5, 2.0) (1.7, 1.62)};

% Draw the third line (visible)
\path[draw=cbBlue!30, line width=0.5pt, fill=white] plot[smooth cycle, tension=1.4] coordinates{(2.20,0.0) (-0.08,-1.93) (0.9, 0) (-0.2,2.0) (1.65, 1.65)};

\node[text=cbBlue, font=\tiny] at (0.0,2.7)
{$\mathcal N(0, \sigma^2\!-\!\textcolor{cbOrange}{\hat{\sigma}_N^2})$};

\node [dot=2pt, fill=black] at (-0.2,0.5) (sam1) {};
\node [dot=0pt, fill=black] at (0.5,0.16) (sam2) {};
\node [dot=0pt, fill=black] at (1.2,-0.23) (sam3) {};
\node [dot=0pt, fill=black] at (1.9,-0.5) (sam4) {};

\draw[myptr] (sam1) to [out=310,in=170] node [below] {} (sam2);
\draw[myptr] (sam2) to [out=310,in=170] node [below] {} (sam3);
\draw[myptr] (sam3) to [out=310,in=170] node [below] {} (sam4);

% Invisible  point 
\node [dot=0pt, fill=blue!0] at (1.9, -0.6) (X) {};

% Label for the distribution
\node at (2.4, 2.0) [text=cbBlue] {$p_{S|Y}$};

% Draw the dashed line first (visible)
\path[draw=cbBlue!60, line width=0.0pt, fill=white] plot[smooth cycle, tension=1.4] coordinates{(2.5,0) (0,-2) (1.5, 0) (0,2) (1.7, 1.7)};

% Then apply the clip path (invisible)
\path[clip] plot[smooth cycle, tension=1.4] coordinates{(2.5,0) (0,-2) (1.5, 0) (0,2) (1.7, 1.7)};

% First cluster (thesisblue, semi-transparent)
\shade[inner color=cbBlue!60, outer color=blue!0, opacity=0.5, shading=radial] 
    (2.0, 0.2) circle [radius=1.2cm];

% Second cluster (thesisblue, semi-transparent)
\shade[inner color=cbBlue!30, outer color=blue!0, opacity=0.2, shading=radial] 
    (1.2, -1.4) circle [radius=0.6cm];

% Third cluster (thesisblue, semi-transparent)
\shade[inner color=cbBlue!50, outer color=blue!0, opacity=0.2, shading=radial] 
    (0.8, 1.7) circle [radius=0.9cm];

% Draw the dashed line first (visible)
\path[draw=cbBlue!60, line width=1.0pt] plot[smooth cycle, tension=1.4] coordinates{(2.5,0) (0,-2) (1.5, 0) (0,2) (1.7, 1.7)};

% Points (same as before)
\node [dot=2pt, fill=cbBlue] at (1.0,1.5) (point1) {};
\node [dot=2pt, fill=cbBlue] at (1.2, 1.3) (point2) {};
\node [dot=2pt, fill=cbBlue] at (1.5, 1.0) (point3) {};
\node [dot=2pt, fill=cbBlue] at (1.8, 0.8) (point5) {};
\node [dot=2pt, fill=cbBlue] at (1.9, 0.7) (point6) {};
\node [dot=2pt, fill=cbBlue] at (1.7, 0.7) (point7) {};
\node [dot=2pt, fill=cbBlue] at (1.9, 0.2) (point8) {};
\node [dot=2pt, fill=cbBlue] at (1.95, 0.1) (point9) {};
\node [dot=2pt, fill=cbBlue] at (1.95, 0.02) (point10) {};
\node [dot=2pt, fill=cbBlue] at (2.1, 0.0) (point11) {};
\node [dot=2pt, fill=cbBlue] at (2.2, 0.01) (point12) {};
\node [dot=2pt, fill=cbBlue] at (1.12,-1.4) (point13) {};
\node [dot=2pt, fill=cbBlue] at (1.23, -1.3) (point14) {};
\node [dot=2pt, fill=cbBlue] at (1.57, -1.0) (point15) {};
\node [dot=2pt, fill=cbBlue] at (1.98, -0.25) (point16) {};
\node [dot=2pt, fill=cbBlue] at (2.01, -0.1) (point17) {};
\node [dot=2pt, fill=cbBlue] at (1.82, -0.02) (point18) {};
\node [dot=2pt, fill=cbBlue] at (2.08, -0.0) (point19) {};
\node [dot=2pt, fill=cbBlue] at (2.25, -0.01) (point20) {};

\node [dot=0pt, fill=cbBlue] at (1.9,-0.5) (init) {};
\draw[myptr] (sam3) to [out=310,in=170] node [below] {} (sam4);

\end{tikzpicture}%
    }%
}
\subcaptionbox{SIPS (ours)\label{fig:sips}}[0.195\textwidth]{
    \centering
    \begin{tikzpicture}[
    scale=0.5,
    font=\small,
    dot/.style = {circle, fill, minimum size=#1, inner sep=0pt, outer sep=0pt},
    myptr/.style={
    line width=0.8pt,
    -{Stealth[length=2.5pt,width=3pt]}
}
]

% Gaussian density / orbit symbol
\newcommand{\gaussorbit}[2][]{%
  \begin{scope}[shift={(#2)}, rotate=190, #1]
    % \fill[cyan!90!blue] (0,0) circle (2.0pt);
    \draw[cbBlue!50, line width=0.5pt] (0,0) ellipse (0.07 and 0.21);
    \draw[cbBlue!30, line width=0.5pt] (0,0) ellipse (0.14 and 0.42);    
  \end{scope}
}

\newcommand{\gaussorbitsmall}[2][]{%
  \begin{scope}[shift={(#2)}, rotate=190, #1]
    % \fill[cyan!90!blue] (0,0) circle (2.0pt);
    \draw[cbBlue!50, line width=0.5pt] (0,0) ellipse (0.06 and 0.19);
    \draw[cbBlue!30, line width=0.5pt] (0,0) ellipse (0.12 and 0.38);
  \end{scope}
}

\newcommand{\gaussorbitbig}[2][]{%
  \begin{scope}[shift={(#2)}, rotate=190, #1]
    % \fill[cyan!90!blue] (0,0) circle (2.0pt);
    \draw[cbBlue!50, line width=0.5pt] (0,0) ellipse (0.09 and 0.3);
    \draw[cbBlue!30, line width=0.5pt] (0,0) ellipse (0.18 and 0.60);
  \end{scope}
}

\newcommand{\orbitpoint}[2][]{%
  \begin{scope}[shift={(#2)}, rotate=190, #1]
    \fill[cbBlue] (0,0) circle (2.0pt);
  \end{scope}
}

% Input
\coordinate (y) at (-0.8, -0.4);

\node[dot=2pt, fill=black] at (y) {};
\node[left] at (y) {$Y$};

\coordinate (meanpt) at (1.3, -0.1);

% Mean    
\node [
  star,
  fill=cbOrange,
  minimum size=4pt,
  inner sep=0pt,
  star point ratio=2.6
] at (1.3, -0.1) (mean) {};

\node[above left=0pt of mean, xshift=3pt, yshift=0pt, text=cbOrange] {$\mathbb{E}[S|Y]$}; 
% $\mathbb{E}[S|Y]$

% Arrow (corrected)
\draw[myptr, draw=black!50, dash pattern=on 0.8pt off 0.8pt, line width=0.6pt] (y) to (mean);

\path (y) -- (mean) coordinate[pos=0] (start) coordinate[pos=1] (end);

% Base points on the arrow
\path (y) -- (mean)
  coordinate[pos=1/3] (p13)
  coordinate[pos=2/3] (p23)
  coordinate[pos=1]   (p33);

% Shifted versions
\coordinate (p13a) at ($(p13)+(3pt,-10pt)$);
\coordinate (p23a) at ($(p23)+(3pt,-10pt)$);

\coordinate (p23b) at ($(p23)+(5pt,-24pt)$);
\coordinate (p33b) at ($(p33)+(5pt,-24pt)$);

\gaussorbit{p13}
\gaussorbitbig{p23a}
\gaussorbitsmall{p33b}

% Draw arrows
\draw[myptr] (start) -- ($(start)!1/3!(end)$);
\draw[myptr] (p13a) -- (p23a);
\draw[myptr] (p23b) -- (p33b);

% Invisible  point 
\node [dot=0pt, fill=blue!0] at (1.9, -0.6) (X) {};

% \draw[myptr] (mean) to [out=270,in=180] node [below] {} (X);

% Label for the distribution
\node at (2.4, 2.0) [text=cbBlue] {$p_{S|Y}$};

% Then apply the clip path (invisible)
\path[clip] plot[smooth cycle, tension=1.4] coordinates{(2.5,0) (0,-2) (1.5, 0) (0,2) (1.7, 1.7)};

% First cluster (thesisblue, semi-transparent)
\shade[inner color=cbBlue!60, outer color=blue!0, opacity=0.5, shading=radial] 
    (2.0, 0.2) circle [radius=1.2cm];

% Second cluster (thesisblue, semi-transparent)
\shade[inner color=cbBlue!30, outer color=blue!0, opacity=0.2, shading=radial] 
    (1.2, -1.4) circle [radius=0.6cm];

% Third cluster (thesisblue, semi-transparent)
\shade[inner color=cbBlue!50, outer color=blue!0, opacity=0.2, shading=radial] 
    (0.8, 1.7) circle [radius=0.9cm];

\gaussorbitsmall{p33b}

% Draw the dashed line first (visible)
\path[draw=cbBlue!60, line width=1.0pt] plot[smooth cycle, tension=1.4] coordinates{(2.5,0) (0,-2) (1.5, 0) (0,2) (1.7, 1.7)};

% Points (same as before)
\node [dot=2pt, fill=cbBlue] at (1.0,1.5) (point1) {};
\node [dot=2pt, fill=cbBlue] at (1.2, 1.3) (point2) {};
\node [dot=2pt, fill=cbBlue] at (1.5, 1.0) (point3) {};
\node [dot=2pt, fill=cbBlue] at (1.8, 0.8) (point5) {};
\node [dot=2pt, fill=cbBlue] at (1.9, 0.7) (point6) {};
\node [dot=2pt, fill=cbBlue] at (1.7, 0.7) (point7) {};
\node [dot=2pt, fill=cbBlue] at (1.9, 0.2) (point8) {};
\node [dot=2pt, fill=cbBlue] at (1.95, 0.1) (point9) {};
\node [dot=2pt, fill=cbBlue] at (1.95, 0.02) (point10) {};
\node [dot=2pt, fill=cbBlue] at (2.1, 0.0) (point11) {};
\node [dot=2pt, fill=cbBlue] at (2.2, 0.01) (point12) {};
\node [dot=2pt, fill=cbBlue] at (1.12,-1.4) (point13) {};
\node [dot=2pt, fill=cbBlue] at (1.23, -1.3) (point14) {};
\node [dot=2pt, fill=cbBlue] at (1.57, -1.0) (point15) {};
\node [dot=2pt, fill=cbBlue] at (1.98, -0.25) (point16) {};
\node [dot=2pt, fill=cbBlue] at (2.01, -0.1) (point17) {};
\node [dot=2pt, fill=cbBlue] at (1.82, -0.02) (point18) {};
\node [dot=2pt, fill=cbBlue] at (2.08, -0.0) (point19) {};
\node [dot=2pt, fill=cbBlue] at (2.25, -0.01) (point20) {};

\draw[myptr] (p23b) -- (p33b);

% \orbitpoint{p33b} 

% \draw[myptr] (mean) to [out=270,in=180] node [below] {} (X);

\end{tikzpicture}
}

\caption{
Overview of different modeling approaches for speech enhancement:
(a) Predictive models learn a deterministic mapping that approximates the posterior mean $\mathbb{E}[S | Y]$ under squared-error regression;
(b) SGMSE models the posterior distribution $p_{S  | Y}$ by conditioning on the noisy observation $Y$;
(c) StoRM conditions a diffusion-based regeneration process on a predictive estimate;
(d) Diffiner uses a clean speech diffusion prior and estimates the noise variance $\hat{\sigma}_N^2$ from the predictive output for DDRM-based refinement;
(e) SIPS decomposes the sampling dynamics into a predictor-induced drift and a generative component.
}%
\label{fig:modeling_approaches}
\end{figure}

However, these hybrid approaches remain limited in flexibility and efficiency. StoRM requires the generative model to be trained jointly with, or explicitly conditioned on, the output of a specific predictor, restricting its ability to generalize across predictive backbones in a plug-and-play manner. 
Diffiner, on the other hand, relies on heuristic parameter choices and computationally expensive diffusion posterior sampling~\cite{kawar2022denoising}, typically requiring hundreds of reverse steps at inference (see Appendix~\ref{app:diffiner}). 
While Diffiner was originally introduced for speech enhancement, subsequent work extended the framework to speech separation through task-specific adaptations~\cite{hirano2026diffusion}.
In contrast, SIPS incorporates pretrained predictors directly into stochastic interpolant sampling through a deterministic drift, while retaining a reusable clean speech prior without predictor-specific conditioning during training.
This formulation enables plug-and-play integration across predictive backbones and additive degradation tasks with a small set of interpretable hyperparameters.

\section{Method}%
\label{sec:method}

\Acp{SGM}~\cite{song2021score} and flow-based models~\cite{lipman2023flow, liu2023flow, albergo2025stochastic} have emerged as two prominent frameworks for generative modeling. We adopt stochastic interpolants~\cite{albergo2025stochastic}, which provide a unified formulation of these approaches by constructing stochastic paths between variables via simple interpolation. 
Within this framework, we propose to combine a predictor-induced drift with an unconditional score model by decomposing the overall drift term in the \ac{SDE}-based sampling scheme in a mathematically principled manner.
This yields a coherent integration of predictive and generative models without ad hoc weighting rules, while maintaining a flexible formulation.

\paragraph{Notations.} In the following, uppercase letters (e.g., $X$) denote random variables, and lowercase letters (e.g., $x$) their realizations. 
All random variables are defined on a common probability space. 
For a random variable $X$, we write $p_X$ for its distribution (or density when it exists).
For clarity, we present the formulation in the one-dimensional setting. All results extend naturally to higher-dimensional representations such as spectrograms by applying the stochastic interpolant component-wise to each time--frequency bin. The neural networks used to estimate the drift components nevertheless operate on the entire spectrogram and can therefore model dependencies across time and frequency.

\subsection{Interpolant design and noise schedule}

Stochastic interpolants~\cite{albergo2025stochastic} define a continuous family of random variables that smoothly connect two distributions through controlled noise injection, enabling generative modeling using \acp{ODE} or \acp{SDE} for sampling.
A more detailed background is provided in Appendix~\ref{app:stochastic_interpolants}.

We define a linear stochastic interpolant between clean speech $S$ and corrupted speech $Y$ as
\begin{equation}
\label{eq:X_lin}
X_t 
= tS + (1-t)Y + \gamma(t) Z, 
\qquad t \in [0,1],
\end{equation}
where $(S, Y)\!\sim\!p_{S,Y}$ are jointly sampled from paired clean and corrupted speech signals, $Z\!\sim\!\mathcal{N}(0,1)$ denotes a standard Gaussian random variable independent of $(S, Y)$, and $\gamma(t)$ is a scalar noise schedule.
We parameterize the noise schedule as
\begin{equation}
\label{eq:gamma}
\gamma(t) = c \sin^2(\pi t),
\end{equation}
whose derivative with respect to $t$ is
\begin{equation}
\dot{\gamma}(t) = c\, \pi \sin(2\pi t),
\end{equation}
with hyperparameter $c \geq 0$. 
Appendix~\ref{app:hyperparameter} presents the small-scale hyperparameter search used to select $c = 0.5$ in Tables~\ref{tab:hyperparameter_vbdmd_valid} and~\ref{tab:hyperparameter_earswham_valid_with_postprocessing} and illustrates the noise schedule $\gamma(t)$ and its derivative $\dot{\gamma}(t)$ in Figure~\ref{fig:gamma}.

Since $\gamma(0)=\gamma(1)=0$, the Gaussian perturbation vanishes at the boundaries, yielding $X_0=Y$ and $X_1=S$.
Denoting the distribution of $X_t$ at time $t$ by $p_t$, we obtain $p_0 = p_Y$ and $p_1 = p_S$.
The additive term $\gamma(t)Z$ injects Gaussian noise along the path, broadening the intermediate distributions and enabling stochastic exploration during generation. 

\subsection{Inference via stochastic dynamics}
\label{sec:inference_stochastic_dynamics}

We begin by defining two key conditional expectations associated with the stochastic interpolant in~\eqref{eq:X_lin}. The first is the conditional velocity
\begin{align}
\label{eq:velocity_component}
v(t,x) &= \mathbb{E}[\partial_t(tS + (1-t)Y) | X_t = x] \\
&=  \mathbb{E}[S - Y | X_t = x], \label{eq:velocity_v}
\end{align}
which captures the expected time derivative of the interpolation path given the current state. The second is
\begin{equation}
\eta_z(t,x) = \mathbb{E}[Z | X_t = x],
\end{equation}
which represents the conditional expectation of the injected noise.

Following~\citet{albergo2025stochastic}, it can be shown that, for each $t \in [0,1]$, the distribution of the interpolant defined in~\eqref{eq:X_lin} coincides with that  of a stochastic process $(X_t^F)_{t\in[0,1]}$ evolving according to the forward \ac{SDE} (see Appendix~\ref{app:ode_sde} for a derivation)
\begin{equation}
\label{eq:sde}
\mathrm d X_t^F
=
\big(
v(t,X_t^F)
+
(\dot{\gamma}(t) - \kappa)\eta_z(t,X_t^F)
\big)\mathrm dt
+
\sqrt{2\kappa\gamma(t)}\mathrm dW_t,
\end{equation}
where $W_t$ is a standard Wiener process and $\kappa \geq 0$ is a noise scaling parameter. Notably, for $\kappa = 0$, the \ac{SDE} reduces to an \ac{ODE}.

Although the stochastic interpolant in~\eqref{eq:X_lin} and the process $X_t^F$ have identical time-marginal distributions, they generally do not induce the same probability measure on path space.
For $c>0$ and $\kappa>0$, $X_t^F$ is a Markov diffusion with a nonzero diffusion coefficient on $t\in(0,1)$, whereas for $c=0$ or $\kappa=0$, its evolution is deterministic given its initial state.
In contrast, the interpolant in~\eqref{eq:X_lin} is obtained by fixing $(S,Y,Z)$ and evaluating a deterministic interpolation path over time.
Thus, the equivalence is marginal in time rather than an equality in law of the full trajectories.

\subsection{Predictive–generative drift decomposition for inference}

During inference, only $Y$ is observed, so instead of taking the conditional expectation given $X_t$ in~\eqref{eq:velocity_v}, we consider the conditional expectation given $Y=y$, namely $\mathbb{E}[S - Y | Y = y]$.
This formulation resembles a predictive modeling problem, where the goal is to estimate the target $S$ from the input $Y$. Motivated by this perspective, we approximate the velocity field as
\vspace{1pt}
\begin{equation}
\hat v(y) = P_\phi(y) - y,
\end{equation}
where $P_\phi$ denotes a predictive model parameterized by $\phi$.

Given a pretrained predictor $P_\phi$, clean speech is obtained by numerically integrating an approximation of the \ac{SDE} in~\eqref{eq:sde}, starting from the initialization $X_0 = Y$.
During integration, the intractable term $\eta_z(t,x_t)$ is replaced by a learned denoiser $D_\theta(t,x_t)$. We discretize the \ac{SDE} using a time grid $\{t_i\}_{i=0}^{M}$ with $t_0\!=\!0$ and $t_M\!=\!1$.
In our experiments, we use a uniform grid with $M\!=\!15$ steps.
The full sampling procedure, which produces the clean speech estimate $\hat S$ from the observation $Y$, is summarized in Algorithm~\ref{alg:predictor_with_prior} in Appendix~\ref{app:sampling}.

The predictor can optionally be applied again after sampling for speech enhancement. 
Based on validation results, we use this post-processing step in the mismatched condition ($\kappa=0.4$), but omit it in the matched condition ($\kappa=0$).
Appendix~\ref{app:postprocessing} provides the analysis.

\subsection{Denoiser training}
\label{sec:denoiser_training}

We train an unconditional score model or denoiser independently of corrupted observations by injecting Gaussian noise into clean speech samples.
Since the interpolant noise level $\gamma(t)$ vanishes at the boundaries, we introduce a small constant offset $a \geq 0$ to control the minimum noise level during training.
The resulting objective, detailed in Appendix~\ref{app:regression_objectives}, is
\vspace{4pt}
\begin{equation}
\label{eq:J_D}
\min_\theta \, 
\mathbb{E}
[
\lVert
D_\theta (t, S + (a + \gamma(t)) Z) - Z
\rVert^2
],
\end{equation}
where $t \sim \mathrm{Uniform}[0,1]$, $S \sim p_S$, and
$Z \sim \mathcal{N}(0,1)$ is a standard Gaussian random variable.
Note that this training objective is closely related to denoising objectives arising from mirror interpolants~\cite{albergo2025stochastic}, with an additional noise offset $a$.
We set $a = 0.1$ and use $c = 0.5$ for the noise schedule $\gamma(t)$ defined in~\eqref{eq:gamma}, selected via a small-scale hyperparameter search (see Tables~\ref{tab:hyperparameter_vbdmd_valid} and~\ref{tab:hyperparameter_earswham_valid_with_postprocessing} in Appendix~\ref{app:hyperparameter}).

While the objective in~\eqref{eq:J_D} enables degradation-agnostic training, it also introduces a mismatch between training and inference.
During training, the denoiser is exposed only to clean speech corrupted by Gaussian noise, whereas at inference time $X_t$ additionally contains environmental noise or interfering speakers inherited from the observation $Y$.
We hypothesize that, despite this mismatch, the denoiser learns to separate Gaussian perturbations from the underlying speech structure and thereby approximates the score of the perturbed clean speech distribution.
Consequently, it can still provide meaningful guidance toward regions of higher likelihood during inference, even in the presence of non-Gaussian or structured corruption.
Our experimental results support this hypothesis and further suggest that a single prior model can generalize across different tasks, such as speech enhancement and speech separation.

\section{Experimental Validation}

\subsection{Data representation}

We follow~\cite{richter2023speech} and operate in an amplitude-compressed \ac{STFT} domain using a hop length of 128, an FFT size and window length of 512, and a Hann window. 
The resulting complex-valued representation is converted into a two-channel real-valued tensor by stacking the real and imaginary components. 
Further details on the representation are provided in Appendix~\ref{app:representation}.

\subsection{Training data and implementation details}

To train our denoiser models, we use clean speech from the 28-speaker training set of the VoiceBank-DEMAND dataset~\cite{botinhao2016investigating}, with all audio signals downsampled to \SI{16}{\kilo\hertz}. We reserve the speakers \texttt{p226} and \texttt{p287} for the validation set. Our implementation is based on the EDM2SE repository~\cite{richter2026do}, which employs the EDM2~\cite{karras2024analyzing} framework with a magnitude-preserving network architecture. 
Further implementation and training details are provided in Appendix~\ref{app:training}.

\subsection{Evaluation metrics}

We use objective instrumental metrics to evaluate the performance of our framework.
SI-SDR~\cite{leroux2019sdr} measures scale-invariant signal-to-distortion ratio in the time domain. 
PESQ~\cite{rix2001perceptual} is an intrusive perceptual metric comparing enhanced speech to a clean reference. 
DNSMOS P.808~\cite{reddy2021dnsmos} as well as NISQA~\cite{mittag2021nisqa} and UTMOS~\cite{saeki2022utmos} are non-intrusive, learning-based metrics that predict perceived speech quality without requiring a reference.
We use QuartzNet15x5Base-En from the NVIDIA NeMo toolkit to compute the \ac{WER}. 
We report corpus-level \ac{WER}, computed as the total number of word errors divided by the total number of reference words over the test set. 

\subsection{Speech enhancement}

\paragraph{Test data.} 
We consider two test conditions. In the matched condition, we use the VoiceBank-DEMAND~\cite{botinhao2016investigating} test set. In the mismatched condition, we evaluate on the EARS-WHAM (v2) test set~\cite{richter2024ears}, for which we additionally created manual text transcriptions available via the EARS-WHAM project page\footnote{\scriptsize\url{https://sp-uhh.github.io/ears_dataset/}}. 
All audio is resampled to \SI{16}{\kilo\hertz}. Compared to VoiceBank-DEMAND, EARS-WHAM contains a lower \ac{SNR} range, making it a substantially more difficult test scenario.

\paragraph{Baselines.} All baselines are trained on the same training data.
We compare with SGMSE+~\cite{richter2023speech}, a purely generative diffusion-based method.
In addition, we include two hybrid methods, StoRM~\cite{lemercier2023storm} and Diffiner~\cite{sawata2023diffiner}. 
As predictors, we consider Conv-TasNet~\cite{luo2019conv}, NCSN++~\cite{song2021score,lemercier2023storm}, and SEMamba~\cite{chao2024investigation}. 
For StoRM, we use the original checkpoint with the NCSN++ predictor and train separate models for the Conv-TasNet and SEMamba predictors using the publicly available training code.
For all other methods, we use the official implementations and the provided checkpoints when available, except for Conv-TasNet, for which we use a third-party implementation\footnote{\scriptsize\url{https://github.com/kaituoxu/Conv-TasNet}} and train the model with an SDR loss such that it is not scale-invariant.

\begin{figure}[t]
    \centering
    \begin{subfigure}{0.48\columnwidth}
        \centering
        % \pgfplotstableread[row sep=\\]{
% kappa sisdr nisqa\\
% 0   16.628857 4.640973\\
% 0.1 15.760820 4.663833\\
% 0.2 15.068143 4.654137\\
% 0.3 14.518651 4.645731\\
% 0.4 14.032758 4.628975\\
% 0.5 13.583891 4.602111\\
% 0.6 13.214874 4.580248\\
% 0.7 12.886644 4.556058\\
% 0.8 12.589031 4.536120\\
% 0.9 12.340708 4.528184\\
% 1   12.086640 4.504237\\
% }\kappaMatched%

\pgfplotstableread[row sep=\\]{
kappa sisdr nisqa\\
0 17.190416 4.590389\\
0.1 16.438069 4.360039\\
0.2 15.798949 4.170691\\
0.3 15.241534 4.066256\\
0.4 14.749544 4.004096\\
0.5 14.311596 3.966268\\
0.6 13.919113 3.944003\\
0.7 13.564998 3.930081\\
0.8 13.242575 3.920304\\
0.9 12.946607 3.910198\\
1 12.672778 3.898444\\
}\kappaMatched%

% Plot dimensions
\pgfmathsetlengthmacro{\plotwidth}{3.3cm}
\pgfmathsetlengthmacro{\plotheight}{1.9cm}

\begin{tikzpicture}

% Left axis: SI-SDR
\begin{axis}[
    name=leftaxis,
    width=\plotwidth,
    height=\plotheight,
    font=\small,
    scale only axis,
    xlabel={$\kappa$},
    ylabel={SI-SDR [dB]},
    xmin=0.0, xmax=1.0,
    ymin=12, ymax=18,
    xtick={0,0.2,0.4,0.6,0.8,1},
    grid=major,
    grid style={dashed,gray!60},
    axis y line*=left,
    axis x line*=box,
    xtick pos=bottom,
    tick align=outside,
    legend style={
        at={(0.5,1.05)},
        anchor=south,
        draw=none,
        legend columns=2,
        /tikz/every even column/.append style={
            column sep=0.4cm
        }
    }
]

% SI-SDR curve
\addplot[
    cbBlue,
    thick,
    mark=*,
    mark size=1.5pt
] table[
    x=kappa,
    y=sisdr
] {\kappaMatched};

\addlegendentry{SI-SDR}

% Add NISQA to the same legend
\addlegendimage{
    cbOrange,
    thick,
    mark=diamond*,
    mark size=2.0pt
}
\addlegendentry{NISQA}

\end{axis}

% Right axis: NISQA
\begin{axis}[
    at={(leftaxis.south west)},
    anchor=south west,
    width=\plotwidth,
    height=\plotheight,
    font=\small,
    scale only axis,
    xmin=0.0, xmax=1.0,
    ymin=3.8, ymax=4.7,
    ylabel={NISQA},
    axis y line*=right,
    axis x line=none,
    ytick={3.8,4.1,4.4,4.7},
    tick align=outside
]

% NISQA curve
\addplot[
    cbOrange,
    thick,
    mark=diamond*,
    mark size=2.0pt
] table[
    x=kappa,
    y=nisqa
] {\kappaMatched};

\end{axis}

\end{tikzpicture}
        \vspace{-4pt}
        \caption{Matched condition}%
        \label{fig:kappa_matched}
    \end{subfigure}\hspace{1em}
    \begin{subfigure}{0.48\columnwidth}
        \centering
        % \pgfplotstableread[row sep=\\]{
% kappa sisdr nisqa\\
% 0 10.778978 3.687741\\
% 0.1 10.189226 3.805241\\
% 0.2 9.749525 3.858268\\
% 0.3 9.377704 3.871139\\
% 0.4 9.057904 3.882364\\
% 0.5 8.769870 3.885935\\
% 0.6 8.509200 3.884052\\
% 0.7 8.267917 3.878356\\
% 0.8 8.050279 3.854567\\
% 0.9 7.839185 3.860366\\
% 1 7.645140 3.852052\\
% }\kappaMismatched%

\pgfplotstableread[row sep=\\]{
kappa sisdr nisqa\\
0 10.762871 3.692526\\
0.1 10.310809 3.745746\\
0.2 9.992600 3.774303\\
0.3 9.695582 3.785557\\
0.4 9.413180 3.789242\\
0.5 9.158813 3.787457\\
0.6 8.924527 3.782929\\
0.7 8.706387 3.775361\\
0.8 8.499160 3.764570\\
0.9 8.306033 3.752718\\
1 8.123729 3.738257\\
}\kappaMismatched%

% Plot dimensions
\pgfmathsetlengthmacro{\plotwidth}{3.3cm}
\pgfmathsetlengthmacro{\plotheight}{1.9cm}

\begin{tikzpicture}

% Left axis: SI-SDR
\begin{axis}[
    name=leftaxis,
    width=\plotwidth,
    height=\plotheight,
    font=\small,
    scale only axis,
    xlabel={$\kappa$},
    ylabel={SI-SDR [dB]},
    xmin=0.0, xmax=1.0,
    ymin=6, ymax=12,
    xtick={0,0.2,0.4,0.6,0.8,1},
    grid=major,
    grid style={dashed,gray!60},
    axis y line*=left,
    axis x line*=box,
    xtick pos=bottom,
    tick align=outside,
    legend style={
        at={(0.5,1.05)},
        anchor=south,
        draw=none,
        legend columns=2,
        /tikz/every even column/.append style={
            column sep=0.4cm
        }
    }
]

% SI-SDR curve
\addplot[
    cbBlue,
    thick,
    mark=*,
    mark size=1.5pt
] table[
    x=kappa,
    y=sisdr
] {\kappaMismatched};

\addlegendentry{SI-SDR}

% Add NISQA to the same legend
\addlegendimage{
    cbOrange,
    thick,
    mark=diamond*,
    mark size=2.0pt
}
\addlegendentry{NISQA}

\end{axis}

% Right axis: NISQA
\begin{axis}[
    at={(leftaxis.south west)},
    anchor=south west,
    width=\plotwidth,
    height=\plotheight,
    font=\small,
    scale only axis,
    xmin=0.0, xmax=1.0,
    ymin=3.6, ymax=3.9,
    ylabel={NISQA},
    axis y line*=right,
    axis x line=none,
    ytick={3.6,3.7,3.8,3.9},
    tick align=outside
]

% NISQA curve
\addplot[
    cbOrange,
    thick,
    mark=diamond*,
    mark size=2.0pt
] table[
    x=kappa,
    y=nisqa
] {\kappaMismatched};

\end{axis}

\end{tikzpicture}
        \vspace{-4pt}
        \caption{Mismatched condition}%
        \label{fig:kappa_mismatched}
    \end{subfigure}
        \caption{SI-SDR and NISQA of SEMamba + SIPS over the noise scaling parameter $\kappa$ on the matched VoiceBank-DEMAND and mismatched EARS-WHAM (v2) validation sets. Predictor post-processing is disabled in (a) and enabled in (b).}%
    \label{fig:metrics_k_plot}
\end{figure}

\paragraph{Noise scaling and predictor post-processing.}
Figure~\ref{fig:metrics_k_plot} evaluates SEMamba + SIPS over $\kappa$, without predictor post-processing in the matched condition and with it in the mismatched condition.
Matched SI-SDR and NISQA peak at $\kappa=0$.
With post-processing enabled, mismatched NISQA peaks at $\kappa=0.4$, at the cost of lower SI-SDR.
At this value, the ablation in Appendix~\ref{app:postprocessing} shows that post-processing improves DNSMOS and NISQA for NCSN++ and SEMamba.
Together, these validation results support using $\kappa=0$ without post-processing for matched enhancement and $\kappa=0.4$ with post-processing for mismatched enhancement.
We use $M=15$ sampling steps in all main SIPS experiments, based on an ablation study presented in Appendix~\ref{app:num_steps}.

\begin{table}[!t]\centering
\caption{Speech enhancement results on the matched VoiceBank-DEMAND test set (mean values). Best values within each predictor group, including the predictor itself, are shown in bold. The underlying predictors are highlighted with a gray background. The separate Diffiner row uses the official checkpoint with released DCUnet predictions, provided only for VoiceBank-DEMAND.}%
\label{tab:vbdmd}
\begingroup
\setlength{\tabcolsep}{4pt}
\begin{tabular}{lS[table-format=2.2]S[table-format=1.2]S[table-format=1.2]S[table-format=1.2]S[table-format=1.2]S[table-format=2.2]}
\toprule
 & {SI-SDR [dB] $\uparrow$} & {PESQ $\uparrow$} & {DNSMOS $\uparrow$} & {NISQA $\uparrow$} & {UTMOS $\uparrow$} & {WER [\%] $\downarrow$} \\
\midrule
Clean & \multicolumn{1}{c}{-} & 4.64 & 3.55 & 4.50 & 4.09 & 6.96 \\
Noisy & 8.44 & 1.97 & 3.09 & 3.03 & 3.11 & 11.01 \\
\midrule
SGMSE+~\cite{richter2023speech} & 17.35 & 2.93 & 3.56 & 4.51 & 3.97 & 11.00 \\
Diffiner~\cite{sawata2023diffiner} & 17.55 & 2.67 & 3.49 & 4.79 & 3.99 & 14.65 \\
\midrule\noalign{\vskip-\belowrulesep}\rowcolor{gray!15}\rule{0pt}{2.0ex}%
Conv-TasNet~\cite{luo2019conv} & 18.57 & 2.51 & 3.31 & 3.45 & 3.61 & \bfseries 11.87 \\
\trimgrayrow
$\;$ + StoRM & 17.92 & \bfseries 2.67 & \bfseries 3.54 & 4.57 & 3.92 & 13.32 \\
$\;$ + Diffiner & 16.22 & 2.59 & 3.47 & \bfseries 4.74 & \bfseries 3.96 & 15.99 \\
$\;$ + SIPS (ours) & \bfseries 18.79 & 2.64 & 3.43 & 4.28 & 3.85 & 12.86 \\
\midrule\noalign{\vskip-\belowrulesep}\rowcolor{gray!15}\rule{0pt}{2.0ex}%
NCSN++~\cite{lemercier2023storm} & \bfseries 19.57 & 2.83 & \bfseries 3.59 & 4.64 & 3.93 & \bfseries 9.58 \\
\trimgrayrow
$\;$ + StoRM~\cite{lemercier2023storm} & 18.49 & \bfseries 2.89 & 3.56 & 4.52 & 3.92 & 10.21 \\
$\;$ + Diffiner & 16.11 & 2.83 & 3.52 & \bfseries 4.80 & \bfseries 4.02 & 13.03 \\
$\;$ + SIPS (ours) & 19.20 & 2.88 & 3.56 & 4.72 & 3.97 & 10.00 \\
\midrule\noalign{\vskip-\belowrulesep}\rowcolor{gray!15}\rule{0pt}{2.0ex}%
SEMamba~\cite{chao2024investigation} & \bfseries 19.72 & \bfseries 3.56 & \bfseries 3.58 & 4.60 & 4.07 & 8.87 \\
\trimgrayrow
$\;$ + StoRM & 18.89 & 3.17 & 3.56 & 4.54 & 4.01 & 9.37 \\
$\;$ + Diffiner & 16.51 & 2.87 & 3.53 & \bfseries 4.81 & 4.04 & 12.43 \\
$\;$ + SIPS (ours) & 19.63 & 3.43 & 3.57 & 4.73 & \bfseries 4.09 & \bfseries 8.81 \\
\bottomrule
\end{tabular}%
\endgroup
\end{table}

\begin{figure}[!t]
\centering
\input{figures/vbdmd_per_snr.tex}
\vspace{-1.0em}
\caption{Per-SNR results for SEMamba and its generative refiners on the matched VoiceBank-DEMAND test set. Each panel reports one metric over the four input-SNR bins.}
\label{fig:vbdmd_per_snr}
\end{figure}

\paragraph{Matched-condition results.}
\Cref{tab:vbdmd} reports the matched-condition results compared to the baselines. 
Across all predictors, SIPS consistently improves the non-intrusive perceptual metrics NISQA and UTMOS, while maintaining comparable SI-SDR and PESQ scores. 
Compared to existing hybrid approaches, SIPS generally exhibits smaller degradations in these reference-based metrics.
For example, with SEMamba, SIPS preserves a high PESQ of 3.43 while still improving perceptual quality. 
Notably, SIPS also improves WER for SEMamba, whereas other hybrid methods degrade WER, potentially due to generative hallucinations introduced during sampling.
The StoRM results in Table~\ref{tab:vbdmd} are obtained using separate checkpoints trained for each predictor. 
This predictor-specific training is essential, as applying the original StoRM checkpoint with NCSN++ as predictor to Conv-TasNet or SEMamba leads to substantial performance degradation, as shown in Appendix~\ref{app:storm}. 
These findings indicate that, unlike SIPS and Diffiner, StoRM is not inherently plug-and-play.
Although Diffiner achieves high NISQA scores, it substantially degrades the signal-level metrics, likely because posterior sampling introduces generative artifacts. 
In contrast, our plug-and-play framework yields only minor degradation in reference-based metrics while providing consistent improvements in perceptual quality. 
See Appendix~\ref{app:vbdmd_std} for mean improvements over each predictor and the standard deviations of these improvements.

\paragraph{Mismatched-condition results.}
In the mismatched condition (\cref{tab:earswham}), we observe similar trends but substantially higher \acp{WER} due to the more challenging \ac{SNR} conditions and domain mismatch. 
SIPS consistently improves NISQA and UTMOS while maintaining comparable signal-level metrics.
The elevated WERs across hybrid approaches suggest that generative refinement may affect linguistic content or speech intelligibility.
Although StoRM performs well under mismatch, it requires predictor-specific training and is therefore not plug-and-play. 
In contrast, Diffiner and SIPS are predictor-agnostic, though Diffiner struggles under mismatched conditions, particularly for Conv-TasNet and NCSN++. 
Interestingly, SGMSE+ achieves the highest DNSMOS and UTMOS scores, albeit with a substantially higher WER.

\begin{table}[!t]\centering
\caption{Speech enhancement results on the mismatched EARS-WHAM (v2) test set (mean values). Best values within each predictor group, including the predictor itself, are shown in bold. The underlying predictors are highlighted with a gray background.}%
\label{tab:earswham}
\begingroup
\setlength{\tabcolsep}{4pt}
\begin{tabular}{lS[table-format=-2.2]S[table-format=1.2]S[table-format=1.2]S[table-format=1.2]S[table-format=1.2]S[table-format=2.2]}
\toprule
 & {SI-SDR [dB] $\uparrow$} & {PESQ $\uparrow$} & {DNSMOS $\uparrow$} & {NISQA $\uparrow$} & {UTMOS $\uparrow$} & {WER [\%] $\downarrow$} \\
\midrule
Clean & \multicolumn{1}{c}{-} & 4.64 & 3.89 & 4.09 & 3.68 & 8.95 \\
Noisy & 5.36 & 1.24 & 2.73 & 1.95 & 1.68 & 32.87 \\
\midrule
SGMSE+~\cite{richter2023speech} & 11.64 & 1.86 & 3.86 & 4.09 & 3.10 & 37.60 \\
\midrule\noalign{\vskip-\belowrulesep}\rowcolor{gray!15}\rule{0pt}{2.0ex}%
Conv-TasNet~\cite{luo2019conv} & \bfseries 3.95 & \bfseries 1.35 & 2.90 & 1.43 & 1.66 & \bfseries 53.87 \\
\trimgrayrow
$\;$ + StoRM & 3.24 & 1.29 & \bfseries 3.44 & \bfseries 3.33 & \bfseries 2.21 & 64.56 \\
$\;$ + Diffiner & -2.31 & 1.16 & 2.99 & 3.02 & 1.97 & 87.79 \\
$\;$ + SIPS (ours) &  3.86 & 1.32 & 3.09 & 1.98 & 1.73 & 59.10 \\
\midrule\noalign{\vskip-\belowrulesep}\rowcolor{gray!15}\rule{0pt}{2.0ex}%
NCSN++~\cite{lemercier2023storm} & \bfseries 13.24 & 1.81 & 3.72 & 3.84 & 2.75 & \bfseries 29.15 \\
\trimgrayrow
$\;$ + StoRM~\cite{lemercier2023storm} &  12.49 & \bfseries 1.90 & \bfseries 3.83 & \bfseries 4.12 & \bfseries 2.86 & 31.74 \\
$\;$ + Diffiner & -0.11 & 1.30 & 3.50 & 3.40 & 2.22 & 69.05 \\
$\;$ + SIPS (ours) & 12.28 & 1.73 & 3.68 & 3.87 & 2.77 & 30.98 \\
\midrule\noalign{\vskip-\belowrulesep}\rowcolor{gray!15}\rule{0pt}{2.0ex}%
SEMamba~\cite{chao2024investigation} & \bfseries 11.36 & \bfseries 2.19 & 3.71 & 3.57 & 2.93 & \bfseries 28.08 \\
\trimgrayrow
$\;$ + StoRM & 10.92 & 2.01 & 3.71 & 3.72 & 2.88 & 29.63 \\
$\;$ + Diffiner & 4.88 & 1.43 & 3.44 & 3.67 & 2.53 & 64.21 \\
$\;$ + SIPS (ours) & 11.28 & 2.05 & \bfseries 3.74 & \bfseries 3.82 & \bfseries 2.97 &  29.27 \\
\bottomrule
\end{tabular}%
\endgroup
\end{table}

\begin{figure}[!t]
    \centering
    % System: clean
\pgfplotstableread[row sep=\\]{
snr sisdr pesq nisqa utmos vqscore dnsmos wer\\
0 nan 4.643888 4.053954 3.683301 0.708022 3.872628 9.105079\\
5 nan 4.643888 4.073906 3.666092 0.709587 3.887752 9.318726\\
10 nan 4.643888 4.162889 3.706027 0.710448 3.902541 8.488197\\
15 nan 4.643888 4.072530 3.645635 0.708805 3.885024 8.880309\\
}\earswhamClean

% System: noisy
\pgfplotstableread[row sep=\\]{
snr sisdr pesq nisqa utmos vqscore dnsmos wer\\
0 -1.987743 1.071482 1.223731 1.331226 0.588923 2.394548 63.241979\\
5 3.024680 1.112861 1.714125 1.294709 0.606626 2.638709 35.524493\\
10 7.959207 1.251844 2.284152 1.736009 0.629896 2.848801 20.140633\\
15 12.912497 1.520231 2.633430 2.386216 0.651579 3.054907 13.127413\\
}\earswhamNoisy

% System: convtasnet
\pgfplotstableread[row sep=\\]{
snr sisdr pesq nisqa utmos vqscore dnsmos wer\\
0 -1.737669 1.114265 1.005245 1.247356 0.656727 2.570087 86.556696\\
5 3.181656 1.203145 1.177950 1.346933 0.678698 2.772216 64.916257\\
10 6.095816 1.419034 1.579341 1.816017 0.695434 3.040610 39.929684\\
15 8.584799 1.696082 1.993199 2.267501 0.704781 3.250520 24.465894\\
}\earswhamConvtasnet

% System: diffiner_convtasnet
\pgfplotstableread[row sep=\\]{
snr sisdr pesq nisqa utmos vqscore dnsmos wer\\
0 -8.516784 1.112153 2.396262 1.804395 0.659332 2.581161 98.519288\\
5 -2.756315 1.093837 2.981244 1.830622 0.674060 2.899804 94.887294\\
10 0.028011 1.157088 3.312080 2.042613 0.688979 3.172721 86.175289\\
15 2.352552 1.263775 3.442073 2.228268 0.695904 3.335714 71.570142\\
}\earswhamDiffinerConvtasnet

% System: diffiner_ncsnpp
\pgfplotstableread[row sep=\\]{
snr sisdr pesq nisqa utmos vqscore dnsmos wer\\
0 -1.674287 1.143723 3.302116 1.969687 0.692702 3.336160 84.309651\\
5 0.422523 1.238634 3.437505 2.176389 0.701068 3.511013 72.711245\\
10 0.733076 1.385098 3.516085 2.354459 0.703433 3.598248 61.614766\\
15 0.152941 1.456044 3.353256 2.376462 0.699558 3.549712 57.824968\\
}\earswhamDiffinerNcsnpp

% System: diffiner_semamba
\pgfplotstableread[row sep=\\]{
snr sisdr pesq nisqa utmos vqscore dnsmos wer\\
0 -0.393379 1.133523 3.337766 1.998932 0.681831 3.146949 89.751916\\
5 3.581928 1.245128 3.568091 2.312116 0.696308 3.390043 74.272762\\
10 6.863188 1.520020 3.852472 2.802576 0.705137 3.580771 51.757911\\
15 9.774249 1.847230 3.928934 3.026417 0.706596 3.638251 41.389961\\
}\earswhamDiffinerSemamba

% System: ncsnpp
\pgfplotstableread[row sep=\\]{
snr sisdr pesq nisqa utmos vqscore dnsmos wer\\
0 6.378215 1.219937 3.039360 2.056241 0.709256 3.444310 54.747370\\
5 11.650337 1.519407 3.758706 2.591488 0.717637 3.679689 30.827352\\
10 15.997897 2.030632 4.243585 3.100899 0.721579 3.848886 18.445505\\
15 19.328041 2.530011 4.344994 3.301918 0.719401 3.925577 13.063063\\
}\earswhamNcsnpp

% System: semamba
\pgfplotstableread[row sep=\\]{
snr sisdr pesq nisqa utmos vqscore dnsmos wer\\
0 5.007628 1.436163 2.686972 2.180542 0.700292 3.434991 53.422522\\
5 9.801851 1.912069 3.463623 2.775632 0.708630 3.695738 29.719179\\
10 13.872221 2.504312 4.027058 3.306137 0.710737 3.849839 16.662481\\
15 17.120806 2.953585 4.152332 3.476484 0.709222 3.893358 12.985843\\
}\earswhamSemamba

% System: sgmsep
\pgfplotstableread[row sep=\\]{
snr sisdr pesq nisqa utmos vqscore dnsmos wer\\
0 4.900051 1.270924 3.930322 2.574816 0.714164 3.781245 68.320561\\
5 10.140407 1.586884 4.041088 2.998678 0.718348 3.860682 42.652059\\
10 14.403015 2.099652 4.205997 3.360606 0.718822 3.913386 24.070819\\
15 17.522489 2.523377 4.196424 3.478240 0.715154 3.905950 15.868726\\
}\earswhamSgmsep

% System: sips_convtasnet
\pgfplotstableread[row sep=\\]{
snr sisdr pesq nisqa utmos vqscore dnsmos wer\\
0 -1.948610 1.103383 1.405764 1.278815 0.663911 2.717591 90.089622\\
5 3.055519 1.179338 1.675082 1.429727 0.685415 2.962158 71.703816\\
10 6.051944 1.382726 2.223038 1.913933 0.701197 3.257866 46.195379\\
15 8.616974 1.650827 2.649386 2.332990 0.709789 3.462837 28.751609\\
}\earswhamSipsConvtasnet

% System: sips_ncsnpp
\pgfplotstableread[row sep=\\]{
snr sisdr pesq nisqa utmos vqscore dnsmos wer\\
0 5.780075 1.200891 3.157428 2.082488 0.701017 3.373613 56.968437\\
5 10.706926 1.452702 3.788954 2.607672 0.708425 3.632511 32.741468\\
10 14.871278 1.906724 4.217518 3.115881 0.712295 3.828139 20.266198\\
15 18.135051 2.388101 4.344018 3.303284 0.711651 3.905392 14.414414\\
}\earswhamSipsNcsnpp

% System: sips_semamba
\pgfplotstableread[row sep=\\]{
snr sisdr pesq nisqa utmos vqscore dnsmos wer\\
0 4.987302 1.372302 3.203837 2.249582 0.694802 3.478527 54.942200\\
5 9.743981 1.779181 3.730457 2.848304 0.704488 3.719870 31.331067\\
10 13.749570 2.324579 4.156666 3.337024 0.707720 3.874066 17.629332\\
15 16.998184 2.780969 4.231842 3.486899 0.706201 3.918222 13.667954\\
}\earswhamSipsSemamba

% System: storm_convtasnet
\pgfplotstableread[row sep=\\]{
snr sisdr pesq nisqa utmos vqscore dnsmos wer\\
0 -2.404750 1.100911 2.785266 1.597891 0.707602 3.154934 91.687232\\
5 2.481231 1.156703 3.219439 1.980702 0.710009 3.356755 77.546908\\
10 5.384112 1.330408 3.614503 2.481027 0.714300 3.579167 53.955299\\
15 7.826257 1.599271 3.719700 2.799001 0.715478 3.692747 35.263835\\
}\earswhamStormConvtasnet

% System: storm_ncsnpp
\pgfplotstableread[row sep=\\]{
snr sisdr pesq nisqa utmos vqscore dnsmos wer\\
0 5.808482 1.281783 3.741290 2.152264 0.717224 3.702753 59.085596\\
5 11.017098 1.617456 4.100647 2.720076 0.724244 3.829448 34.416320\\
10 15.173984 2.139914 4.337523 3.218347 0.725222 3.892425 20.203415\\
15 18.346493 2.602488 4.312716 3.397572 0.721624 3.917498 13.732304\\
}\earswhamStormNcsnpp

% System: storm_semamba
\pgfplotstableread[row sep=\\]{
snr sisdr pesq nisqa utmos vqscore dnsmos wer\\
0 4.640769 1.344400 3.226489 2.138737 0.707523 3.466776 56.228082\\
5 9.431619 1.748112 3.643393 2.740920 0.714849 3.698765 31.872560\\
10 13.435897 2.275352 4.002006 3.262024 0.718466 3.825946 17.516323\\
15 16.563321 2.696852 4.045037 3.434370 0.717130 3.861864 13.410553\\
}\earswhamStormSemamba

\pgfmathsetlengthmacro{\earswhamLeftMargin}{2.3em}
\pgfmathsetlengthmacro{\earswhamRightMargin}{1.3em}
\pgfmathsetlengthmacro{\earswhamPanelGap}{4.2em}
\pgfmathsetlengthmacro{\earswhamPlotWidth}{
    (\columnwidth-\earswhamLeftMargin-\earswhamRightMargin-2*\earswhamPanelGap)/3}
\pgfmathsetlengthmacro{\earswhamPlotHeight}{2cm}
\pgfmathsetlengthmacro{\earswhamColumnSep}{\earswhamPlotWidth+\earswhamPanelGap}
\pgfmathsetlengthmacro{\earswhamRowSep}{3.75cm}

\pgfplotsset{
    earswham per snr axis/.style={
        width=\earswhamPlotWidth,
        height=\earswhamPlotHeight,
        font=\small,
        scale only axis,
        xmin=0, xmax=15,
        xtick={0,5,10,15},
        xticklabels={Very Low, Low, Medium, High},
        tick label style={font=\small},
        label style={font=\small},
        yticklabel pos=left,
        ytick pos=left,
        grid=major,
        grid style={dashed,gray!50},
        tick align=outside,
        xtick pos=bottom,
        title style={yshift=-4pt},
        legend cell align=left,
        legend style={
            draw=none,
            font=\small,
            legend columns=4,
            /tikz/every even column/.append style={column sep=0.28cm}
        },
        every axis plot/.append style={thick, mark size=1.4pt}
    }
}

\newcommand{\plotEarswhamSemambaCurves}[1]{%
    \addplot[black, mark=*, densely dashed]
        table[x=snr,y=#1] {\earswhamSemamba};
    \addplot[cbBlue, mark=square*]
        table[x=snr,y=#1] {\earswhamStormSemamba};
    \addplot[cbGreen, mark=triangle*]
        table[x=snr,y=#1] {\earswhamDiffinerSemamba};
    \addplot[cbOrange, mark=diamond*]
        table[x=snr,y=#1] {\earswhamSipsSemamba};
}

% Discard spaces emitted by the data and style definitions.
\loop\ifdim\lastskip>0pt\unskip\repeat%
\begin{tikzpicture}
\begin{axis}[
    earswham per snr axis,
    name=pesq,
    title={PESQ $\uparrow$},
    ymin=1.0, ymax=3.0,
    ytick={1.0,1.5,2.0,2.5,3.0},
    legend to name=earswhamPerSnrLegend
]
\plotEarswhamSemambaCurves{pesq}
\addlegendentry{SEMamba}
\addlegendentry{+ StoRM}
\addlegendentry{+ Diffiner}
\addlegendentry{+ SIPS (ours)}
\end{axis}

\begin{axis}[
    earswham per snr axis,
    at={($(pesq.south west)+(\earswhamColumnSep,0)$)},
    anchor=south west,
    name=nisqa,
    title={NISQA $\uparrow$},
    ymin=2.5, ymax=4.5,
    ytick={2.5,3.0,3.5,4.0,4.5}
]
\plotEarswhamSemambaCurves{nisqa}
\end{axis}

\begin{axis}[
    earswham per snr axis,
    at={($(nisqa.south west)+(\earswhamColumnSep,0)$)},
    anchor=south west,
    name=wer,
    title={WER [\%] $\downarrow$},
    ymin=0, ymax=100,
    ytick={0,25,50,75,100}
]
\plotEarswhamSemambaCurves{wer}
\end{axis}

% Optional second row: SI-SDR, DNSMOS, and UTMOS.
% \begin{axis}[
%     earswham per snr axis,
%     at={($(pesq.south west)+(0,-\earswhamRowSep)$)},
%     anchor=south west,
%     name=sisdr,
%     title={SI-SDR [dB]},
%     ymin=-1, ymax=18,
%     ytick={0,5,10,15}
% ]
% \plotEarswhamSemambaCurves{sisdr}
% \end{axis}
%
% \begin{axis}[
%     earswham per snr axis,
%     at={($(sisdr.south west)+(\earswhamColumnSep,0)$)},
%     anchor=south west,
%     name=dnsmos,
%     title={DNSMOS},
%     ymin=3.0, ymax=4.0,
%     ytick={3.0,3.25,3.5,3.75,4.0}
% ]
% \plotEarswhamSemambaCurves{dnsmos}
% \end{axis}
%
% \begin{axis}[
%     earswham per snr axis,
%     at={($(dnsmos.south west)+(\earswhamColumnSep,0)$)},
%     anchor=south west,
%     name=utmos,
%     title={UTMOS},
%     ymin=1.8, ymax=3.7,
%     ytick={2.0,2.5,3.0,3.5}
% ]
% \plotEarswhamSemambaCurves{utmos}
% \end{axis}

\node[anchor=south] at ($(pesq.north east)!0.5!(wer.north west)+(0,0.65cm)$)
    {\pgfplotslegendfromname{earswhamPerSnrLegend}};
\end{tikzpicture}%
    \caption{Per-SNR results for SEMamba and its generative refiners on the mismatched EARS-WHAM (v2) test set. Each panel reports one metric over the four input SNR bins.}
    \label{fig:earswham_per_snr}
\end{figure}

\paragraph{Per-SNR analysis.}
The benefits of generative refinement become more pronounced in challenging conditions, such as low input \ac{SNR}.
We examine this effect in both the matched VoiceBank-DEMAND and mismatched EARS-WHAM settings by evaluating SEMamba and its generative refiners in four input-\ac{SNR} bins, labeled Very Low, Low, Medium, and High in Figures~\ref{fig:vbdmd_per_snr} and~\ref{fig:earswham_per_snr}.
For VoiceBank-DEMAND, these bins are centered at 2.5, 7.5, 12.5, and 17.5 dB, respectively.
For EARS-WHAM, they correspond to the original benchmark ranges [-2.5,2.5], (2.5,7.5], (7.5,12.5], and (12.5,17.5] dB~\cite{richter2024ears}.

In the matched setting (Figure~\ref{fig:vbdmd_per_snr}), Diffiner and SIPS improve NISQA over SEMamba in all four regimes, with Diffiner achieving the highest scores throughout.
StoRM also improves NISQA in the Very Low regime, but falls below SEMamba in the other three bins.
Diffiner's perceptual gains come at the cost of recognition performance, particularly at very low input \ac{SNR}, where its \ac{WER} increases to 22.15\% compared with 13.00\% for SEMamba.
Among the refiners, SIPS best preserves PESQ in every bin while keeping \ac{WER} close to SEMamba.

In the mismatched setting (Figure~\ref{fig:earswham_per_snr}), the NISQA gain from SIPS over SEMamba increases as the input \ac{SNR} decreases, from 0.08 in the High regime to 0.52 in the Very Low regime (Table~\ref{tab:earswham_per_snr_2} in Appendix~\ref{app:per_snr}).
For the lowest-\ac{SNR} bin, this NISQA improvement has a 95\% \ac{CI} of [0.47,0.57], indicating that SIPS is particularly beneficial when the predictive estimate is degraded by severe noise.
The other refiners exhibit different trade-offs: StoRM also improves NISQA in the lowest bin, while Diffiner achieves the highest NISQA there.
However, as seen in Figure~\ref{fig:earswham_per_snr}, Diffiner does so at the cost of a much higher \ac{WER} in the same bin (89.75\%), whereas SIPS keeps \ac{WER} close to SEMamba (54.94\% vs.\ 53.42\%) while still improving NISQA substantially.
Numerical results for both settings, covering all combinations of predictors and generative refiners across all evaluation metrics, are provided in Appendix~\ref{app:per_snr}.

\paragraph{Subjective listening test.}
We also conducted a \ac{CCR} listening test comparing SEMamba with SEMamba + SIPS on the mismatched EARS-WHAM setting.
Positive \ac{CCR} ratings indicate preference for SEMamba + SIPS, negative ratings indicate preference for SEMamba, and zero denotes no preference.
Across 4,210 accepted ratings, listeners preferred SEMamba + SIPS overall, with a mean CCR of 0.234 and a 95\% bootstrap \ac{CI} of [0.152, 0.323]. 44.2\% of ratings favored SEMamba + SIPS, compared with 29.2\% favoring SEMamba and 26.6\% indicating no preference.
The per-\ac{SNR} trend is consistent with the objective analysis, as shown in Figure~\ref{fig:mean_ccr_snr}. 
The strongest preference occurs in the lowest-\ac{SNR} bin and decreases as the input \ac{SNR} increases.
Appendix~\ref{app:ccr_listening_test} provides the listening-test protocol, detailed statistics, and the overall and per-\ac{SNR} preference distributions (Figures~\ref{fig:ccr_preference_by_snr} and~\ref{fig:ccr_preference_overall}). 

\begin{figure}[t]
    \centering
    \begin{tikzpicture}
\begin{axis}[
    width=6.5cm,
    height=3.8cm,
    ymin=0,
    ymax=0.50,
    ytick={0,0.1,0.2,0.3,0.4,0.5},
    xmin=0.5,
    xmax=4.5,
    xtick={1,2,3,4},
    xticklabels={Very Low, Low, Medium, High},
    font=\small,
    tick label style={font=\small},
    label style={font=\small},
    title={CMOS with 95\% CI},
    tick align=outside,
    xtick pos=bottom,
    yticklabel pos=left,
    ytick pos=left,
    grid=major,
    grid style={dashed,gray!50},
    error bars/y dir=both,
    error bars/y explicit,
]
\addplot+[
    cbBlue,
    line width=1pt,
    mark=diamond*,
    mark size=1.4pt,
    mark options={
        fill=cbBlue,
        draw=cbBlue
    },
    error bars/.cd,
        y dir=both,
        y explicit,
        error bar style={line width=0.8pt},
        error mark=-,
        error mark options={
            line width=0.8pt,
            mark size=3pt,
            rotate=90
        },
] table[
    x=bin,
    y=mean,
    y error plus expr=\thisrow{upper}-\thisrow{mean},
    y error minus expr=\thisrow{mean}-\thisrow{lower}
] {figures/listening_test/mean_ccr_by_snr_95ci.dat};
\end{axis}
\end{tikzpicture}
    \caption{\Ac{CMOS} by SNR bin with 95\% \ac{CI} for the mismatched EARS-WHAM setting.
    The positive preference for SEMamba + SIPS is strongest in the lowest-SNR bin and decreases as the SNR increases.}%
    \label{fig:mean_ccr_snr}
\end{figure}

\subsection{Speech separation}

\paragraph{Test data.} We use the monaural noisy reverberant subset of the WHAMR! dataset~\cite{maciejewski2020} in the 16~kHz \texttt{max} configuration.
Its mixtures are generated by convolving simulated room impulse responses with dry speech signals from the WSJ0-2mix~\cite{hershey2016} dataset and adding background noise.

\paragraph{Baselines.}
We consider two representative predictor models: a time-domain approach, SepFormer~\cite{subakan2021attention}, and a time-frequency-domain one, FlexIO~\cite{masuyama2026flexio}.
For SepFormer, we use the official checkpoint that was trained on the WHAMR! dataset~\footnote{\url{https://huggingface.co/speechbrain/sepformer-whamr16k}}.
Since the official model was trained with the SI-SDR loss~\cite{luo2019conv}, the scale of the separated signals can deviate from the true signals. 
This needs to be addressed in order to combine SepFormer with SIPS, which requires the predictor estimate to be at the correct scale.
Difficulties in reproducing the official recipe prevented us from re-training the model with SNR loss, so we rescaled each SepFormer output to minimize the mean squared error to the ground-truth signal after resolving the speaker permutation based on SI-SDR. 
The rescaled signals were used both for baseline evaluation and as inputs to SIPS, without additional scaling after SIPS.
While this procedure is not applicable in practice, it provides insight into the achievable performance of the combination with SepFormer.
There is no such issue with FlexIO, which is trained on a combination of multiple speech enhancement and separation datasets using the vanilla SNR loss~\cite{masuyama2026flexio}.
We use its large configuration, yielding the best performance on the WHAMR! dataset in the original paper.
To measure WERs, we use QuartzNet15x5Base-En from the NVIDIA NeMo toolkit similarly to our enhancement experiments.

\paragraph{Results.} Table~\ref{tab:whamr} summarizes separation performance on the WHAMR! dataset. 
For SIPS, we set $\kappa\!=\!0$ and disable predictor post-processing.
We observe a trend similar to speech enhancement, i.e., the combination with SIPS improves non-intrusive metrics (DNSMOS, NISQA and UTMOS) with comparable SI-SDR. 
These perceptual improvements come at the cost of lower PESQ and higher WER for both predictors, highlighting a trade-off between perceptual quality and speech fidelity. 
We also applied the original Diffiner implementation to the outputs of both separation models but observed substantial performance degradation. We therefore omit these results from \cref{tab:whamr}. 
While subsequent work extends Diffiner to speech separation~\cite{hirano2026diffusion}, the corresponding implementation was unavailable to us, preventing a direct comparison.

Figure~\ref{fig:separation} presents the metric distributions with and without our prior.
The left panel demonstrates that our method with FlexIO results in relatively small changes in SI-SDR, regardless of the performance of the initial separation.
In contrast, the UTMOS score generally improves in the right panel.
An additional evaluation on real-recorded two-speaker mixtures from MC-WSJ-AV~\cite{lincoln2005mcwsjav} shows improvements in all three non-intrusive quality metrics with a slight increase in \ac{WER}.
Appendix~\ref{app:mcwsjav} provides the details and results.

\begin{table}[t]
\caption{Speech separation results on WHAMR!.
For the mixture, SI-SDR and PESQ are computed with respect to each speaker's utterance and averaged. Best values within each predictor group, including the predictor itself, are shown in bold. The underlying predictors are highlighted with a gray background.}
\label{tab:whamr}
\centering
\begingroup
\setlength{\tabcolsep}{4.6pt}
\begin{tabular}{lS[table-format=-1.2]S[table-format=1.2]S[table-format=1.2]S[table-format=1.2]S[table-format=1.2]S[table-format=2.2]}
\toprule
Model
& {SI-SDR [dB] $\uparrow$}
& {PESQ $\uparrow$}
& {DNSMOS $\uparrow$}
& {NISQA $\uparrow$}
& {UTMOS $\uparrow$}
& {WER [\%] $\downarrow$} \\
\midrule
Mixture & -7.20 & 1.08 & 2.53 & 1.21 & 1.35 & 92.62 \\
\midrule\noalign{\vskip-\belowrulesep}\rowcolor{gray!15}\rule{0pt}{2.0ex}%
SepFormer~[\citenum{subakan2021attention}] & \bfseries 6.99 & \bfseries 1.55 & 3.13 & 2.06 & 2.34 & \bfseries 32.22 \\
\trimgrayrow
% $\;$ + Diffiner~[\citenum{sawata2023diffiner}] & -23.76  & 1.09  &   &  1.68 &  1.92 &  \\
$\;$ + SIPS (ours) & 6.79 & 1.41 & \bfseries 3.36 & \bfseries 3.01 & \bfseries 2.60 & 34.29 \\
\midrule\noalign{\vskip-\belowrulesep}\rowcolor{gray!15}\rule{0pt}{2.0ex}%
FlexIO~[\citenum{masuyama2026flexio}] & 8.45 & \bfseries 1.76 & 3.62 & 3.54 & 2.79 & \bfseries 21.50 \\
\trimgrayrow
% $\;$ + Diffiner~[\citenum{sawata2023diffiner}] & -25.41  & 1.08  &  & 1.64   & 1.95 &  \\
$\;$ + SIPS (ours) & \bfseries 8.51 & 1.56 & \bfseries 3.68 & \bfseries 4.01 & \bfseries 3.01 & 23.43 \\
\bottomrule
\end{tabular}%
\endgroup
\end{table}

\begin{figure}[t]
    \centering
    \begingroup
% Regenerate the compact plot data with: python3 figures/separation_improvement/prepare.py

\pgfplotsset{
    separation axis/.style={
        scale only axis,
        width=0.24\columnwidth,
        height=0.14\columnwidth,
        font=\small,
        xlabel={FlexIO},
        ylabel={SIPS improvement},
        title style={yshift=-0.5em},
        tick align=outside,
        xtick pos=bottom,
        ytick pos=left,
        scaled ticks=false,
        grid=major,
        grid style={dashed,gray!30},
        set layers=standard,
        mark layer=main,
    },
    separation points/.style={
        on layer=main,
        mark layer=main,
        color=cbBlue!80,
        only marks,
        % mark=*,
        % mark size=0.9pt,
        % mark options={draw=none,draw opacity=0,fill=cbBlue!50,fill opacity=0.65},
        mark=star,
        mark size=1pt,
        mark options={draw opacity=0.8,line width=0.2pt},
    },
}

    \begin{tikzpicture}
        \begin{axis}[
            separation axis,
            title={SI-SDR [dB]},
            xmin=-15, xmax=20,
            ymin=-2, ymax=2,
            xtick={-10,0,10,20},
            ytick={-2,-1,0,1,2},
        ]
            \addplot+[separation points] table[
                col sep=comma,
                x=flexio, y=improvement
            ] {figures/separation_improvement/sisdr_points.csv};
            \begin{pgfonlayer}{main}
                \input{figures/separation_improvement/sisdr_density.tex}
            \end{pgfonlayer}
            \addplot[black,mark=none,on layer=axis foreground]
                coordinates {(-15,0) (20,0)};
        \end{axis}
    \end{tikzpicture}%
\hspace{5em}%
    \begin{tikzpicture}
        \begin{axis}[
            separation axis,
            title={UTMOS},
            xmin=1, xmax=5,
            ymin=-0.75, ymax=1.25,
            xtick={1,2,3,4,5},
            ytick={-0.5,0,0.5,1},
        ]
            \addplot+[separation points] table[
                col sep=comma,
                x=flexio, y=improvement
            ] {figures/separation_improvement/utmos_points.csv};
            \begin{pgfonlayer}{main}
                \input{figures/separation_improvement/utmos_density.tex}
            \end{pgfonlayer}
            \addplot[black,mark=none,on layer=axis foreground]
                coordinates {(1,0) (5,0)};
        \end{axis}
    \end{tikzpicture}
\endgroup
    \caption{SI-SDR and UTMOS improvement from adding SIPS to FlexIO on WHAMR!, plotted against the corresponding FlexIO score. Improvement is the score with SIPS minus the FlexIO score. Individual test utterances in sparse regions are shown as points, with dense regions visualized as a density cloud. The horizontal line at zero indicates equal performance.}
    \label{fig:separation}
\end{figure}

\section{Conclusion}

In this work, we presented SIPS (Stochastic Interpolant Prior for Speech), a plug-and-play framework for speech enhancement and separation that augments existing predictors with a generative clean speech prior.
Built on stochastic interpolants, SIPS combines a predictor-induced drift with a score-based generative model trained exclusively on clean speech, enabling integration across predictors and additive degradation tasks.
Experiments with recent predictors, including SEMamba and FlexIO, demonstrate consistent improvements in non-intrusive speech quality metrics while maintaining competitive performance on reference-based distortion measures and \ac{WER}.
A per-SNR analysis reveals particularly pronounced perceptual quality improvements under challenging low-SNR conditions, which are further supported by a subjective listening test conducted for speech enhancement.
Overall, these findings highlight the potential of combining predictive and generative modeling to improve perceptual quality while maintaining reconstruction fidelity across speech restoration tasks.

\begin{ack}
This work was supported by Mitsubishi Electric Research Laboratories.

\textbf{Competing interests.}
The authors declare no competing interests beyond their disclosed affiliations.
\end{ack}

\addtolength{\bibsep}{-2.0pt} % Reduce spacing between references
\bibliographystyle{plainnat}
\bibliography{mybib}

\clearpage

\appendix

\section{Background on Stochastic Interpolants}
\label{app:stochastic_interpolants}

Stochastic interpolants~\cite{albergo2023building, albergo2025stochastic} provide a continuous-time formulation for constructing transport maps between probability distributions. 
They offer a unified view of flow-based models~\cite{lipman2023flow,liu2023flow}, score-based diffusion models~\cite{song2021score}, and related transport-based generative modeling methods such as the Schrödinger bridge~\cite{de2021diffusion}.
In this appendix, we briefly summarize the main concepts used in Section~\ref{sec:method}.

\subsection{Dynamical transport of probability measures}

Let $p_0$ and $p_1$ denote two probability densities on $\mathbb{R}^d$.
Generative modeling can be viewed as the problem of constructing a transport from a base distribution $p_0$ to a target distribution $p_1$.
Given a differentiable random path $\{X_t\}_{t\in[0,1]}$ with $X_0\sim p_0$ and $X_1\sim p_1$, we define the velocity field
\begin{equation}
    \label{eq:app_transport_velocity}
    b(t,x)
    :=
    \mathbb E[\dot X_t | X_t=x],
\end{equation}
which averages the path velocities conditional on the current state.
Under suitable regularity assumptions, the velocity field defines an associated ODE process $X_t^{\mathrm{ODE}}$, which evolves according to
\begin{equation}
    \label{eq:ode}
    \frac{\mathrm d}{\mathrm dt}X_t^{\mathrm{ODE}}
    = b(t,X_t^{\mathrm{ODE}}),
    \qquad X_0^{\mathrm{ODE}}\sim p_0.
\end{equation}
If $X_t$ has density $p(t,\cdot)$ at time $t$, then $p$ evolves according to the transport equation
\begin{equation}
    \label{eq:transport_equation}
    \partial_t p(t,x)
    +
    \nabla \cdot \bigl(b(t,x)p(t,x)\bigr)
    = 0.
\end{equation}
Appendix~\ref{app:proof_transport_equation} proves this relation for the stochastic interpolants defined below.
Consequently, the ODE initialized with density $p_0$ has the same time-marginal densities as the random path and transports $p_0$ to $p_1$.

\subsection{Definition of stochastic interpolants}

Given two endpoint variables $X_0 \sim p_0$ and $X_1 \sim p_1$, a stochastic interpolant is a family of random variables
\begin{equation}
    X_t = f(t,X_0,X_1) + \gamma(t)Z,
    \qquad t\in[0,1],
\end{equation}
where $Z\sim\mathcal N(0,I)$ is independent of $(X_0,X_1)$,
$f : [0,1]\times\mathbb R^d\times\mathbb R^d \rightarrow \mathbb R^d$
is a sufficiently smooth interpolation function, and $\gamma:[0,1]\rightarrow\mathbb R_{\geq 0}$ is a scalar noise schedule.
The interpolation function satisfies the endpoint conditions
\begin{equation}
    f(0,X_0,X_1)=X_0,
    \qquad
    f(1,X_0,X_1)=X_1,
\end{equation}
and the noise schedule satisfies
\begin{equation}
    \gamma(0)=\gamma(1)=0,
    \qquad
    \gamma(t)>0 \quad \text{for } t\in(0,1).
\end{equation}
Consequently, $X_t$ has endpoint distributions $p_0$ and $p_1$ by construction.

A common example is the linear stochastic interpolant
\begin{equation}
    X_t = (1-t)X_0 + tX_1 + \gamma(t)Z.
\end{equation}
In this paper, we use this form with $X_0=Y$ denoting corrupted speech and $X_1=S$ denoting clean speech. 
Defining the residual $N:=Y-S$ in the working representation, the interpolant can be rewritten as
\begin{equation}
    X_t = S + (1-t)N + \gamma(t)Z,
\end{equation}
which describes a gradual reduction of the residual while introducing a controlled Gaussian perturbation.

\subsection{Velocity field and drift decomposition}

The time derivative of the interpolant is
\begin{equation}
    \dot X_t
    =
    \partial_t f(t,X_0,X_1) + \dot\gamma(t)Z .
\end{equation}
Substituting this derivative into~\eqref{eq:app_transport_velocity} gives
\begin{equation}
    b(t,x)
    =
    \mathbb E[\partial_t f(t,X_0,X_1)+\dot\gamma(t)Z | X_t=x].
\end{equation}

For the linear interpolant
\begin{equation}
    X_t = (1-t)X_0+tX_1+\gamma(t)Z,
\end{equation}
we obtain
\begin{equation}
    \dot X_t = X_1-X_0+\dot\gamma(t)Z,
\end{equation}
and therefore
\begin{equation}
    b(t,x)
    =
    \mathbb E[X_1-X_0+\dot\gamma(t)Z  | X_t=x].
\end{equation}
In this work, we exploit the fact that this velocity can be written as
\begin{equation}
    b(t,x)
    =
    v(t,x)+\dot\gamma(t)\eta_z(t,x),
\end{equation}
where
\begin{equation}
    v(t,x)
    =
    \mathbb E[X_1-X_0 | X_t=x],
    \qquad
    \eta_z(t,x)
    =
    \mathbb E[Z | X_t=x].
\end{equation}

\subsection{ODE and SDE samplers}
\label{app:ode_sde}

The transport equation above induces the deterministic probability flow ODE
\begin{equation}
    \frac{\mathrm d}{\mathrm dt} X_t^{\mathrm{ODE}} = b(t,X_t^{\mathrm{ODE}}),
\end{equation}
as defined in~\eqref{eq:ode}, with initial density $p_0$.
The same time marginals $p(t,\cdot)$ can also be realized by a family of stochastic differential equations. 
In particular, adding a diffusion term with variance schedule $\epsilon(t)\geq 0$ requires a corresponding correction of the drift to preserve the same marginals, as shown by the Fokker--Planck equation (see Appendix~\ref{app:ode_sde_proof}). 
For any diffusion schedule $\epsilon(t)\geq 0$, define
\begin{equation}
    \label{eq:b_F}
    b_F(t,x)
    =
    b(t,x)+\epsilon(t)\nabla_x\log p(t,x).
\end{equation}
Then the forward SDE
\begin{equation}
    \label{eq:forward_sde}
    \mathrm d X_t^F
    =
    b_F(t,X_t^F)\,\mathrm dt
    +
    \sqrt{2\epsilon(t)}\,\mathrm dW_t
\end{equation}
has the same marginal density $p(t,\cdot)$ as the interpolant and the probability flow ODE, provided it is initialized with density $p_0$.
For $t\!\in\!(0,1)$, where $\gamma(t)\!>\!0$, the de\-noi\-ser–score relation gives
\begin{equation}
    \nabla_x \log p(t,x)
    =
    -\gamma(t)^{-1}\eta_z(t,x).
\end{equation}
Substituting this identity into \eqref{eq:b_F}, we obtain
\begin{equation}
    b_F(t,x)
    =
    b(t,x)-\epsilon(t)\gamma(t)^{-1}\eta_z(t,x).
\end{equation}

For the decomposition
\begin{equation}
    b(t,x)=v(t,x)+\dot\gamma(t)\eta_z(t,x),
\end{equation}
the forward drift becomes
\begin{equation}
    b_F(t,x)
    =
    v(t,x)
    +
    \left(
        \dot\gamma(t)-\frac{\epsilon(t)}{\gamma(t)}
    \right)
    \eta_z(t,x).
\end{equation}
The sampler used in the main text corresponds to the choice
\begin{equation}
    \epsilon(t)=\kappa\gamma(t),
\end{equation}
which yields
\begin{equation}
    \mathrm dX_t^F
    =
    \left[
        v(t,X_t^F)
        +
        \bigl(\dot\gamma(t)-\kappa\bigr)
        \eta_z(t,X_t^F)
    \right]\mathrm dt
    +
    \sqrt{2\kappa\gamma(t)}\,\mathrm dW_t .
\end{equation}
This is the SDE formulation used in Section~\ref{sec:method}. 
When $\kappa=0$, the stochastic term vanishes and the sampler reduces to the corresponding probability flow ODE.

\subsection{Denoiser regression objective}
\label{app:regression_objectives}

In SIPS, the velocity component $v(t,x)$ is approximated by $\hat v(y)=P_\phi(y)-y$ using a fixed pretrained predictor.
Only the denoiser $D_\theta$ is trained.
For clean speech $S\sim p_S$ and independent Gaussian noise $Z\sim\mathcal{N}(0,1)$, define the denoiser training input
\begin{equation}
    \tilde X_t = S+(a+\gamma(t))Z,
\end{equation}
where $a \ge 0$ is the noise offset.
The tilde distinguishes this perturbed clean speech sample from the stochastic interpolant $X_t$ in~\eqref{eq:X_lin}.
The denoiser objective in~\eqref{eq:J_D} can then be written as
\begin{equation}
    \mathcal{L}_D(\theta)
    := \int_0^1
    \mathbb{E}_{S,Z}
    \Big[
        \big\lVert
            D_\theta(t,\tilde X_t)-Z
        \big\rVert^2
    \Big] \mathrm dt.
\end{equation}
Its minimizer over unrestricted denoising functions is the conditional expectation
\begin{equation}
    D^*(t,x)=\mathbb{E}[Z | \tilde X_t=x].
\end{equation}
This objective uses only clean speech and Gaussian noise, independently of the predictor and the degradation process.
To make the distinction explicit, the stochastic interpolant in~\eqref{eq:X_lin} can be rewritten as $X_t=S+(1-t)(Y-S)+\gamma(t)Z$.
The training input $\tilde X_t$ omits the observation-dependent residual $(1-t)(Y-S)$ and uses the additional noise offset $a$.
Consequently, $D^*(t,x)$ and the interpolant denoising term $\eta_z(t,x)=\mathbb{E}[Z | X_t=x]$ generally differ.
SIPS uses the learned clean speech denoiser as a surrogate for this term, together with the predictor-based velocity approximation.
These substitutions do not in general preserve the exact time-marginal equivalence of the ideal SDE. 
However, their effectiveness is assessed empirically.
This training--inference mismatch is also discussed in Section~\ref{sec:denoiser_training}.

\section{Proofs}
\label{app:proofs}

\subsection{Proof that the interpolant density satisfies the transport equation}
\label{app:proof_transport_equation}

We follow the derivations in Appendix~B of \citet{albergo2025stochastic}. Let
\begin{equation}
    X_t = f(t,X_0,X_1) + \gamma(t)Z,
\end{equation}
with time derivative
\begin{equation}
    \dot X_t
    =
    \partial_t f(t,X_0,X_1) + \dot\gamma(t)Z.
\end{equation}
We define the velocity field
\begin{equation}
    b(t,x)
    :=
    \mathbb E[\dot X_t | X_t=x].
\end{equation}
Let $p(t,x)$ denote the density of $X_t$. We show that $p$ satisfies
\begin{equation}
    \partial_t p(t,x)
    +
    \nabla_x\cdot\bigl(b(t,x)p(t,x)\bigr)
    =0.
\end{equation}

Consider the characteristic function of $X_t$,
\begin{equation}
    \varphi_{X_t}(t,k)
    :=
    \mathbb E\left[e^{i k\cdot X_t}\right]
    =
    \int_{\mathbb R^d} e^{i k\cdot x}p(t,x)\,dx .
\end{equation}
Differentiating with respect to time gives
\begin{equation}
    \partial_t \varphi_{X_t}(t,k)
    =
    \mathbb E\left[
        i k\cdot \dot X_t \,
        e^{i k\cdot X_t}
    \right]
    =
    i k\cdot m(t,k),
\end{equation}
where
\begin{equation}
    m(t,k)
    :=
    \mathbb E\left[
        \dot X_t e^{i k\cdot X_t}
    \right].
\end{equation}
Using the law of total expectation, we can write
\begin{align}
    m(t,k)
    &=
    \mathbb E\left[
        \mathbb E[\dot X_t e^{i k\cdot X_t} | X_t]
    \right] \notag \\
    &=
    \mathbb E\left[
        e^{i k\cdot X_t}
        \mathbb E[\dot X_t | X_t]
    \right] \notag \\
    &=
    \int_{\mathbb R^d}
        e^{i k\cdot x}
        b(t,x)
        p(t,x)
    \,dx .
\end{align}
Therefore,
\begin{equation}
    \partial_t \varphi_{X_t}(t,k)
    =
    i k\cdot
    \int_{\mathbb R^d}
        e^{i k\cdot x}
        b(t,x)p(t,x)
    \,dx .
\end{equation}
On the other hand, differentiating the density representation gives
\begin{equation}
    \partial_t \varphi_{X_t}(t,k)
    =
    \int_{\mathbb R^d}
        e^{i k\cdot x}
        \partial_t p(t,x)
    \,dx .
\end{equation}
Thus, for all $k\in\mathbb R^d$,
\begin{equation}
    \int_{\mathbb R^d}
        e^{i k\cdot x}
        \partial_t p(t,x)
    \,dx
    =
    i k\cdot
    \int_{\mathbb R^d}
        e^{i k\cdot x}
        b(t,x)p(t,x)
    \,dx .
\end{equation}
Since
\begin{equation}
     \int_{\mathbb R^d}
        e^{i k\cdot x}
        \nabla\cdot\bigl(b(t,x)p(t,x)\bigr)
    \,dx
    =
    - i k\cdot
    \int_{\mathbb R^d}
        e^{i k\cdot x}
        b(t,x)p(t,x)
    \,dx ,
\end{equation}
where we used the Fourier transform identity
\(
\mathcal{F}[\nabla \cdot g](k) = - i k \cdot \mathcal{F}[g](k),
\)
we obtain
\begin{equation}
    \int_{\mathbb R^d}
        e^{i k\cdot x}
        \left[
            \partial_t p(t,x)
            +
            \nabla\cdot\bigl(b(t,x)p(t,x)\bigr)
        \right]
    \,dx
    =
    0 .
\end{equation}
The left-hand side is the Fourier transform of 
\(
\partial_t p(t,x)
+
\nabla\cdot\bigl(b(t,x)p(t,x)\bigr).
\)
Since this transform vanishes for all $k \in \mathbb R^d$, and the Fourier transform is injective on integrable functions, the integrand itself must vanish almost everywhere. Hence,
\begin{equation}
    \partial_t p(t,x)
    +
    \nabla\cdot\bigl(b(t,x)p(t,x)\bigr)
    =
    0.
\end{equation}
This proves that the interpolant density satisfies the transport equation.

\subsection{Proof of the equivalent SDE marginals}
\label{app:ode_sde_proof}

We show that the forward SDE in~\eqref{eq:forward_sde} has the same time marginals as the interpolant. 
The interpolant density $p(t,x)$ satisfies the transport equation
\begin{equation}
    \partial_t p(t,x)
    +
    \nabla_x \cdot \bigl(b(t,x)p(t,x)\bigr)
    =
    0 .
\end{equation}
The Fokker--Planck equation associated with the forward SDE is
\begin{equation}
    \partial_t p(t,x)
    =
    - \nabla_x \cdot \bigl(b_F(t,x)p(t,x)\bigr)
    +
    \epsilon(t)\Delta_x p(t,x).
\end{equation}
Substituting 
\begin{equation}
    b_F(t,x)=b(t,x)+\epsilon(t)\nabla_x\log p(t,x)
\end{equation}
gives
\begin{align}
    \partial_t p(t,x)
    &=
    - \nabla_x \cdot \bigl(b(t,x)p(t,x)\bigr)
    - \epsilon(t)\nabla_x \cdot 
    \bigl(p(t,x)\nabla_x\log p(t,x)\bigr)
    +
    \epsilon(t)\Delta_x p(t,x) \\
    &=
    - \nabla_x \cdot \bigl(b(t,x)p(t,x)\bigr)
    - \epsilon(t)\Delta_x p(t,x)
    +
    \epsilon(t)\Delta_x p(t,x) \\
    &=
    - \nabla_x \cdot \bigl(b(t,x)p(t,x)\bigr).
\end{align}
Thus, the Fokker--Planck equation reduces to the original transport equation.
With the same initial density $p_0$ and under regularity conditions ensuring uniqueness of the density evolution, the forward SDE and the probability flow ODE therefore share the same time-marginal densities $p(t,\cdot)$.

\section{Implementation Details and Ablations}
\label{app:ablations}

\subsection{Diffiner with reduced sampling steps}
\label{app:diffiner}

To compare computational efficiency under a matched sampling budget, we additionally evaluate Diffiner using $M=15$ sampling steps. 
The corresponding results on the VoiceBank-DEMAND dataset are reported in Table~\ref{tab:diffiner}. Reducing Diffiner from its default configuration of 200 steps to 15 steps results in a substantial degradation across all metrics, whereas SIPS maintains performance close to the predictive model at the same 15-step budget. 
One possible explanation is that the two methods incorporate the predictive estimate differently. Diffiner converts the predictor residual into a per-bin noise-scale estimate that controls its noise-level-dependent posterior updates. 
Consequently, reducing the number of diffusion steps coarsens both the reverse dynamics and the switching between these updates. 
In contrast, SIPS directly uses the full complex-valued predictor residual as a deterministic drift. Since its explicit cumulative contribution satisfies $\sum_{i=0}^{M-1}\hat{v}(y)\Delta t_i=\hat{v}(y)$ for $\sum_i\Delta t_i=1$, this contribution is not attenuated when fewer integration steps are used, although the overall dynamics are discretized more coarsely. 
These results therefore suggest that SIPS is more robust to step reduction and can more effectively preserve the predictor's signal and linguistic content under a small inference budget.
The effect of the number of sampling steps on SIPS is further examined in Appendix~\ref{app:num_steps}.

\begin{table}\centering
\caption{Speech enhancement results on the matched VoiceBank-DEMAND test set. Best values among the refiners for each underlying predictor are shown in bold; predictor rows are highlighted with a gray background.}
\label{tab:diffiner}
\begingroup
\setlength{\tabcolsep}{2pt}
\begin{tabular}{lS[table-format=2.2]S[table-format=1.2]S[table-format=1.2]S[table-format=1.2]S[table-format=1.2]S[table-format=2.2]}
\toprule
 & {SI-SDR [dB] $\uparrow$} & {PESQ $\uparrow$} & {DNSMOS $\uparrow$} & {NISQA $\uparrow$} & {UTMOS $\uparrow$} & {WER [\%] $\downarrow$} \\
\midrule
Noisy & 8.44 & 1.97 & 3.09 & 3.03 & 3.11 & 11.01 \\
\midrule\noalign{\vskip-\belowrulesep}\rowcolor{gray!15}\rule{0pt}{2.0ex}%
Conv-TasNet~\cite{luo2019conv} & 18.57 & 2.51 & 3.31 & 3.45 & 3.61 & 11.87 \\
\trimgrayrow
$\;$ + Diffiner~\cite{sawata2023diffiner} (M=200) & 16.22 & 2.59 & \bfseries 3.47 & \bfseries 4.74 & \bfseries 3.96 & 15.99 \\
$\;$ + Diffiner~\cite{sawata2023diffiner} (M=15) & 13.05 & 2.07 & 3.22 & 4.43 & 3.64 & 31.41 \\
$\;$ + SIPS (M=15) & \bfseries 18.79 & \bfseries 2.64 & 3.43 & 4.28 & 3.85 & \bfseries 12.86 \\
\midrule\noalign{\vskip-\belowrulesep}\rowcolor{gray!15}\rule{0pt}{2.0ex}%
NCSN++~\cite{lemercier2023storm} & 19.57 & 2.83 & 3.59 & 4.64 & 3.93 & 9.58 \\
\trimgrayrow
$\;$ + Diffiner~\cite{sawata2023diffiner} (M=200) & 16.11 & 2.83 & 3.52 & \bfseries 4.80 & \bfseries 4.02 & 13.03 \\
$\;$ + Diffiner~\cite{sawata2023diffiner} (M=15) & 13.05 & 2.16 & 3.26 & 4.52 & 3.69 & 29.59 \\
$\;$ + SIPS (M=15) & \bfseries 19.20 & \bfseries 2.88 & \bfseries 3.56 & 4.72 & 3.97 & \bfseries 10.00 \\
\midrule\noalign{\vskip-\belowrulesep}\rowcolor{gray!15}\rule{0pt}{2.0ex}%
SEMamba~\cite{chao2024investigation} & 19.72 & 3.56 & 3.58 & 4.60 & 4.07 & 8.87 \\
\trimgrayrow
$\;$ + Diffiner~\cite{sawata2023diffiner} (M=200) & 16.51 & 2.87 & 3.53 & \bfseries 4.81 & 4.04 & 12.43 \\
$\;$ + Diffiner~\cite{sawata2023diffiner} (M=15) & 13.18 & 2.19 & 3.29 & 4.56 & 3.72 & 26.23 \\
$\;$ + SIPS (M=15) & \bfseries 19.63 & \bfseries 3.43 & \bfseries 3.57 & 4.73 & \bfseries 4.09 & \bfseries 8.81 \\
\bottomrule
\end{tabular}
\endgroup
\end{table}

\subsection{Hyperparameter Selection for the Noise Schedule}
\label{app:hyperparameter}

We study the influence of the hyperparameters $a$ and $c$ on speech enhancement performance. 
The hyperparameter $a$ controls the Gaussian noise offset during denoiser training, while $c$ controls the amplitude of the interpolation noise schedule $\gamma(t)$.
The noise scale $\gamma(t)$ and its derivative $\dot{\gamma}(t)$ are visualized in Figure~\ref{fig:gamma} for $t \in [0,1]$.

To determine suitable values, we evaluate a grid of configurations on the VoiceBank-DEMAND and EARS-WHAM validation sets, keeping all other training and inference settings fixed within each condition.
We use $\kappa=0$ without predictor post-processing for the matched condition and $\kappa=0.4$ with post-processing for the mismatched condition.
Tables~\ref{tab:hyperparameter_vbdmd_valid} and~\ref{tab:hyperparameter_earswham_valid_with_postprocessing} report these validation results, which are used to select $a$ and $c$.

\begin{figure}
    \centering
    \begin{tikzpicture}
\begin{axis}[
    width=6cm,
    height=4cm,
    font=\small,
    tick label style={font=\small},
    label style={font=\small},
    xlabel={$t$},
    xmin=0, xmax=1,
    ymin=-2, ymax=2,
    xtick={0,0.25,0.5,0.75,1},
    xticklabels={$0$,$1/4$,$1/2$,$3/4$,$1$},
    minor x tick num=1,
    ytick={-2,-1,0,1,2},
    minor y tick num=1,
    tick align=outside,
    xtick pos=bottom,
    ytick pos=left,
    grid=major,
    xminorgrids=true,
    yminorgrids=true,
    grid style={dashed,gray!50},
    legend style={
        at={(0.5,1.05)},
        anchor=south,
        draw=none,
        legend columns=3,
        /tikz/every even column/.append style={column sep=0.3cm}
    },
    domain=0:1,
    samples=101
]

\addplot[cbBlue,thick] {0.5*sin(deg(pi*x))^2};
\addlegendentry{$\gamma(t)$}

\addplot[cbBlue,thick,dashed] {0.5*sin(deg(pi*x))^2 + 0.1};
\addlegendentry{$\gamma(t)+a$}

\addplot[cbOrange,thick] {0.5*pi*sin(deg(2*pi*x))};
\addlegendentry{$\dot{\gamma}(t)$}

\end{axis}
\end{tikzpicture}
    \caption{Interpolation noise schedule $\gamma(t)$, training noise level $\gamma(t)+a$, and derivative $\dot{\gamma}(t)$ for $c=0.5$ and $a=0.1$.}
    \label{fig:gamma}
\end{figure}

The results show that the choice of noise schedule parameters has a noticeable impact on both perceptual quality and distortion-based metrics. 
Positive offsets can improve perceptual quality, with the effect depending on the setting and the schedule amplitude $c$. Overly large values of $c$ tend to degrade SI-SDR, indicating reduced reconstruction fidelity. Based on the validation trade-off between distortion-based and perceptual metrics, we select $a=0.1$ and $c=0.5$ for all subsequent experiments.

\paragraph{Predictor-only limit and signal processing.}
For the matched setting in Table~\ref{tab:hyperparameter_vbdmd_valid}, $\kappa=0$ and predictor post-processing is disabled.
When $c=0$, both $\gamma(t)$ and $\dot{\gamma}(t)$ vanish, so the denoiser and stochastic terms in Algorithm~\ref{alg:predictor_with_prior} have zero weight.
The sampler therefore reduces to
\begin{equation}
    x_1 = y + \sum_{i=0}^{M-1}\bigl(P_\phi(y)-y\bigr)\Delta t_i = P_\phi(y),
\end{equation}
since $\sum_{i=0}^{M-1}\Delta t_i=1$.
Thus, the matched $c=0$ rows reproduce the predictor output within the SIPS processing pipeline, independently of $a$.
The standalone SEMamba baseline processes the original waveform directly, whereas SEMamba within SIPS receives an \ac{STFT}-reconstructed, padded waveform.
Its output then undergoes an additional spectrogram encoding and decoding cycle followed by cropping.
These signal-processing differences explain why the matched $c=0$ rows can differ from the standalone SEMamba row, even though no learned denoising contribution is active.
Such differences should therefore be attributed to the processing pipeline, rather than to generative refinement.
The contribution of each processing operation has not been isolated.
For the mismatched setting in Table~\ref{tab:hyperparameter_earswham_valid_with_postprocessing}, $\kappa=0.4$, so the denoiser coefficient $\dot{\gamma}(t)-\kappa=-\kappa$ remains nonzero when $c=0$.
The mismatched $c=0$ rows therefore retain a denoising contribution and can vary with $a$.
These rows also include predictor post-processing, $\hat{s}\leftarrow P_\phi(x_1)$, which can further alter the output and evaluation scores.
Their differences from the standalone SEMamba baseline thus reflect the combined effects of sampling, signal processing, and predictor post-processing. 
The individual contributions cannot be isolated from this table.
    
\begin{table}\centering
\caption{Speech enhancement results on the matched VoiceBank-DEMAND validation set for different noise schedule hyperparameters $a$ and $c$. SEMamba + SIPS uses 15 sampling steps and $\kappa=0$, without optional predictor post-processing.}
\label{tab:hyperparameter_vbdmd_valid}
\begingroup
\setlength{\tabcolsep}{6.1pt}
\begin{tabular}{lccS[table-format=-2.2]S[table-format=1.2]S[table-format=1.2]S[table-format=1.2]S[table-format=1.2]}
\toprule
 & $a$ & $c$ & {SI-SDR [dB] $\uparrow$} & {PESQ $\uparrow$} & {DNSMOS $\uparrow$} & {NISQA $\uparrow$} & {UTMOS $\uparrow$} \\
\midrule
Clean & - & - & \multicolumn{1}{c}{-} & 4.64 & 3.68 & 4.46 & 4.05 \\
Noisy & - & - & 6.02 & 1.57 & 2.85 & 2.14 & 2.31 \\
\midrule\noalign{\vskip-\belowrulesep}\rowcolor{gray!15}\rule{0pt}{2.0ex}%
SEMamba & - & - & 17.61 & 3.44 & 3.65 & 4.29 & 3.98 \\
\trimgrayrow
$\;$ + Run 1 & 0.0 & 0.3 & 17.26 & 3.17 & 3.67 & 4.51 & 3.99 \\
$\;$ + Run 2 & 0.0 & 0.5 & 17.24 & 3.05 & 3.68 & 4.55 & 3.99 \\
$\;$ + Run 3 & 0.001 & 0.1 & 17.26 & 3.36 & 3.63 & 4.32 & 4.00 \\
$\;$ + Run 4 & 0.001 & 0.5 & 17.26 & 3.05 & 3.69 & 4.56 & 3.99 \\
$\;$ + Run 5 & 0.01 & 2.0 & 15.31 & 2.63 & 3.71 & 4.70 & 3.94 \\
$\;$ + Run 6 & 0.05 & 0.0 & 17.30 & 3.40 & 3.65 & 4.32 & 4.00 \\
$\;$ + Run 7 & 0.05 & 0.5 & 17.26 & 3.06 & 3.67 & 4.56 & 4.00 \\
$\;$ + Run 8 & 0.05 & 1.0 & 16.56 & 2.85 & 3.70 & 4.64 & 3.99 \\
$\;$ + Run 9 & 0.05 & 2.0 & 15.08 & 2.58 & 3.68 & 4.67 & 3.95 \\
$\;$ + Run 10 & 0.05 & 3.0 & 13.82 & 2.19 & 3.59 & 4.55 & 3.85 \\
$\;$ + Run 11 & 0.05 & 4.0 & 12.01 & 1.85 & 3.49 & 4.39 & 3.66 \\
$\;$ + Run 12 & 0.1 & 0.0 & 17.30 & 3.40 & 3.65 & 4.32 & 4.00 \\
$\;$ + Run 13 & 0.1 & 0.1 & 17.34 & 3.39 & 3.65 & 4.42 & 4.02 \\
$\;$ + Run 14 & 0.1 & 0.3 & 17.34 & 3.23 & 3.66 & 4.53 & 4.02 \\
$\;$ + Run 15 & 0.1 & 0.5 & 17.19 & 3.10 & 3.67 & 4.59 & 4.02 \\
$\;$ + Run 16 & 0.1 & 1.0 & 16.93 & 2.84 & 3.67 & 4.63 & 3.98 \\
$\;$ + Run 17 & 0.1 & 2.0 & 14.51 & 2.37 & 3.68 & 4.68 & 3.93 \\
$\;$ + Run 18 & 0.2 & 0.0 & 17.30 & 3.40 & 3.65 & 4.32 & 4.00 \\
$\;$ + Run 19 & 0.2 & 0.1 & 17.36 & 3.38 & 3.65 & 4.41 & 4.00 \\
$\;$ + Run 20 & 0.2 & 0.3 & 17.29 & 3.23 & 3.66 & 4.48 & 3.99 \\
\bottomrule
\end{tabular}
\endgroup
\end{table}

\begin{table}\centering
\caption{Speech enhancement results on the mismatched EARS-WHAM (v2) validation set for different noise schedule hyperparameters $a$ and $c$. SEMamba + SIPS uses 15 sampling steps with $\kappa=0.4$ and predictor post-processing.}
\label{tab:hyperparameter_earswham_valid_with_postprocessing}
\begingroup
\setlength{\tabcolsep}{6.1pt}
\begin{tabular}{lccS[table-format=-2.2]S[table-format=1.2]S[table-format=1.2]S[table-format=1.2]S[table-format=1.2]}
\toprule
 & $a$ & $c$ & {SI-SDR [dB] $\uparrow$} & {PESQ $\uparrow$} & {DNSMOS $\uparrow$} & {NISQA $\uparrow$} & {UTMOS $\uparrow$} \\
\midrule
Clean & - & - & \multicolumn{1}{c}{-} & 4.64 & 3.77 & 3.81 & 3.49 \\
Noisy & - & - & 5.05 & 1.29 & 2.70 & 1.92 & 1.65 \\
\midrule\noalign{\vskip-\belowrulesep}\rowcolor{gray!15}\rule{0pt}{2.0ex}%
SEMamba & - & - & 10.85 & 2.09 & 3.61 & 3.35 & 2.89 \\
\trimgrayrow
$\;$ + Run 1 & 0.0 & 0.3 & 9.61 & 1.88 & 3.71 & 3.80 & 2.90 \\
$\;$ + Run 2 & 0.0 & 0.5 & 9.31 & 1.85 & 3.73 & 3.88 & 2.93 \\
$\;$ + Run 3 & 0.001 & 0.1 & 9.51 & 1.77 & 3.65 & 3.40 & 2.86 \\
$\;$ + Run 4 & 0.001 & 0.5 & 9.33 & 1.85 & 3.73 & 3.87 & 2.91 \\
$\;$ + Run 5 & 0.01 & 2.0 & 7.38 & 1.58 & 3.73 & 3.85 & 2.71 \\
$\;$ + Run 6 & 0.05 & 0.0 & 10.73 & 1.99 & 3.63 & 3.48 & 2.82 \\
$\;$ + Run 7 & 0.05 & 0.5 & 9.53 & 1.90 & 3.68 & 3.83 & 2.92 \\
$\;$ + Run 8 & 0.05 & 1.0 & 8.77 & 1.80 & 3.71 & 3.90 & 2.95 \\
$\;$ + Run 9 & 0.05 & 2.0 & 7.54 & 1.62 & 3.70 & 3.86 & 2.81 \\
$\;$ + Run 10 & 0.05 & 3.0 & 6.25 & 1.44 & 3.64 & 3.75 & 2.61 \\
$\;$ + Run 11 & 0.05 & 4.0 & 4.76 & 1.31 & 3.55 & 3.59 & 2.39 \\
$\;$ + Run 12 & 0.1 & 0.0 & 10.62 & 1.92 & 3.58 & 3.44 & 2.79 \\
$\;$ + Run 13 & 0.1 & 0.1 & 10.26 & 1.93 & 3.62 & 3.57 & 2.84 \\
$\;$ + Run 14 & 0.1 & 0.3 & 9.82 & 1.91 & 3.64 & 3.72 & 2.89 \\
$\;$ + Run 15 & 0.1 & 0.5 & 9.40 & 1.88 & 3.67 & 3.79 & 2.94 \\
$\;$ + Run 16 & 0.1 & 1.0 & 8.91 & 1.81 & 3.68 & 3.83 & 2.90 \\
$\;$ + Run 17 & 0.1 & 2.0 & 7.33 & 1.56 & 3.72 & 3.91 & 2.80 \\
$\;$ + Run 18 & 0.2 & 0.0 & 10.35 & 1.83 & 3.52 & 3.35 & 2.75 \\
$\;$ + Run 19 & 0.2 & 0.1 & 10.05 & 1.88 & 3.55 & 3.48 & 2.81 \\
$\;$ + Run 20 & 0.2 & 0.3 & 9.57 & 1.86 & 3.59 & 3.61 & 2.87 \\
\bottomrule
\end{tabular}
\endgroup
\end{table}

\subsection{Sampling Procedure}
\label{app:sampling}

Algorithm~\ref{alg:predictor_with_prior} summarizes the sampling procedure, and Figure~\ref{fig:schmatic} illustrates the corresponding processing pipeline, including the optional post-processing step in which the predictor is applied again after sampling. 

{\renewcommand{\baselinestretch}{1.1}\selectfont
\begin{algorithm}[t]
\caption{Sampling with SIPS}
\label{alg:predictor_with_prior}
\begin{algorithmic}[1]
\Require Observation $y$, predictor $P_\phi$, denoiser $D_\theta$, noise scaling parameter $\kappa \geq 0$, number of steps $M$, arbitrary time schedule $\{t_i\}_{i=0}^{M}$ with $t_0=0$, $t_M=1$, and step sizes $\{\Delta t_i\}_{i=0}^{M-1}$ with $\Delta t_i = t_{i+1} -t_i$
\Ensure Clean speech estimate $\hat{s}$
\Statex \textbf{Initialization}
\State $\hat{v} = P_\phi(y) - y$
% \State $z \sim \mathcal{N}(0,I)$
\State $x_0 = y$
\Statex \textbf{Euler--Maruyama Steps}
\For{$i = 0$ to $M-1$}
    \State $\hat{z} = D_\theta(t_i, x_{t_i})$
    \State $z \sim \mathcal{N}(0,I)$
   \State $x_{t_{i+1}} = x_{t_i}
    + \big(\hat{v} + (\dot{\gamma} (t_i) - \kappa )\hat{z} \big)\Delta t_i  + \sqrt{2 \Delta t_i \kappa\gamma(t_{i})}\, z$ \label{line:6}
\EndFor
\State $\hat{s} \leftarrow x_1$
\Statex \textbf{Post-Processing} (optional)
\State $\hat{s} \leftarrow P_\phi(\hat{s})$

\State \Return $\hat{s}$
\end{algorithmic}
\end{algorithm}
}

\begin{figure}[t]
\centering
\begin{tikzpicture}[
  font=\small,
  >=Stealth,
  line/.style={-{Stealth[length=1.4mm, width=1.4mm]}, semithick},
  block/.style={draw, thick, rounded corners, align=center, minimum height=6mm, minimum width=8mm},
  den/.style={draw, thick, rounded corners, align=center, minimum height=6mm, minimum width=7mm, fill=cbBlue!15},
  circ/.style={draw, thick, circle, inner sep=0pt, minimum size=8mm},
  pred/.style={block, fill=cbOrange!15, minimum width=7mm},
  sum/.style={draw, semithick, circle, inner sep=0pt, minimum size=4mm, font=\footnotesize},
  stepblock/.style={block, fill=black!5, minimum height=8mm},
  noise/.style={circ, fill=black!10}
]

% --- Central state x0 ---
\node[stepblock] (x0) {Init \\[-3pt]  \tiny \texttt{Alg:2}};

% --- Denoiser before first step ---
\node[den, right=9mm of x0] (D0) {$D_\theta$};
\draw[line] (x0) -- node[pos=0.5,above] {$x^{(0)}$} (D0.west);
\node[above=4mm of D0] (t0) {\scriptsize $t_0$};
\draw[line] (t0.south) -- (D0.north);

% --- First Step ---
\node[stepblock, right=5mm of D0] (S0) {Step\\[-3pt]  \tiny \texttt{Alg:6}};
\draw[line] (D0.east) -- (S0.west);

\node[above=7mm of S0] (E0) {$z^{(1)}$};
\draw[line] (E0.south) -- (S0.north);

% --- Dots indicating more steps ---
\node[right=7mm of S0] (dots) {$\cdots$};

% --- Denoiser before final step ---
\node[den, right=10mm of dots] (DN) {$D_\theta$};
\draw[line] (S0.east) -- node[pos=0.5,above] {$x^{(1)}$} (dots);
\draw[line] (dots) -- node[pos=0.4,above] {$x^{(M-1)}$} (DN.west);
\node[above=4mm of DN] (tN) {\scriptsize $t_{M-1}$};
\draw[line] (tN.south) -- (DN.north);

% --- Final Step ---
\node[stepblock, right=5mm of DN] (SN) {Step\\[-3pt]  \tiny \texttt{Alg:6}};
\draw[line] (DN.east) -- (SN.west);

\node[above=7mm of SN] (EN) {$z^{(M)}$};
\draw[line] (EN.south) -- (SN.north);

% --- Final output ---
\node[pred, right=8mm of SN] (D2) {$P_\phi$};
\draw[line] (SN.east) -- node[pos=0.5,above] {$x^{(M)}$} (D2.west);
\draw[line] (D2.east) -- node[pos=0.35,above] {$\hat{s}$} ++(6mm,0);
\node[below=0.3mm of D2, font=\footnotesize] {(optional)};

% --- Predictor and y ---
\node[below=6mm of x0] (y) {$y$};
\node[pred, right=4.0mm of y] (P) {$P_\phi$};
\draw[line] (y) -- (P.west);

% --- Single summation node (with signed inputs) ---
\node[sum, right=6.0mm of P] (sum) {\footnotesize \textbf{$\Sigma$}}; % or leave empty: {}

% Connect P output to sum with a "-" sign on the input
\draw[line] (P.east) -- (sum.west);
\node[anchor=south east, font=\tiny, inner sep=0.0em] () at (sum.155) {$+$};

% Tap from y down into the sum with a "+" sign on the input
\path (y) -- (y.south) coordinate[pos=0.35] (ytap);
\draw[line] (ytap) -- ++(0,-4mm) -| (sum.south);
\node[anchor=north east, font=\tiny, inner sep=0.0em] () at (sum.249) {$-$};

% v-hat bus feeding Step blocks from the sum output
\draw[line] (sum.east) -| node[pos=0.75, right] {$\hat{v}$} (SN.south);
\draw[line] (sum.east) -| node[pos=0.75, right] {$\hat{v}$} (S0.south);

% y contributes to x0
\draw[line] (y) -- (x0);

% initial noise into x0
% \node[above=7mm of x0] (Einit) {$z^{(0)}$};
% \draw[line] (Einit.south) -- (x0);

\end{tikzpicture}
% \end{document}
\caption{Schematic of the proposed framework.
The predictor $P_\phi$ defines a deterministic drift $\hat{v} = P_\phi(y) - y$ from the observation $y$, which steers the sampling dynamics at every step.
Concurrently, the denoiser $D_\theta$ guides the trajectory toward regions of higher likelihood.
After the sampling procedure, the predictor is optionally applied again as a post-processing step.}
\label{fig:schmatic}
\end{figure}

\subsection{Effect of Post-Processing After Generative Sampling}
\label{app:postprocessing}

After generative sampling, the predictor can optionally be applied again to the clean speech estimate as $\hat{s} \leftarrow P_\phi(\hat{s})$.
Tables~\ref{tab:vbdmd_post_valid} and~\ref{tab:earswham_post_valid} report validation results with and without this step for the matched condition ($\kappa=0$) and mismatched condition ($\kappa=0.4$), respectively.
In the matched condition, post-processing reduces SI-SDR for all three predictors, with limited or inconsistent gains in the other metrics.
For example, with SEMamba, SI-SDR decreases from 17.19 to 16.74~dB, while NISQA increases only from 4.59 to 4.65 and UTMOS remains unchanged.
We therefore omit post-processing in the matched condition to preserve signal fidelity.

In the mismatched condition, post-processing improves DNSMOS for all three predictors and substantially increases NISQA for NCSN++ and SEMamba.
For SEMamba, DNSMOS increases from 3.50 to 3.67, NISQA from 3.32 to 3.79, and UTMOS from 2.89 to 2.94, with a smaller SI-SDR reduction from 9.51 to 9.40~dB.
The gains are not uniform. 
PESQ decreases for all three predictors, and NISQA and UTMOS decrease for Conv-TasNet.
Figure~\ref{fig:kappa_mismatched} complements this comparison by varying $\kappa$ with post-processing enabled throughout. 
For SEMamba, NISQA increases from 3.69 at $\kappa=0$ to 3.79 at $\kappa=0.4$, while SI-SDR decreases from 10.76 to 9.40~dB.
Together, the two ablations support using $\kappa=0.4$ with predictor post-processing in the mismatched condition, prioritizing the observed quality gains while accepting the reductions in signal fidelity.

\begin{table}\centering
\caption{Speech enhancement results with and without post-processing on the matched VoiceBank-DEMAND validation set (mean scores). SIPS uses 15 sampling steps with $\kappa=0$. Best values among the two SIPS configurations for each underlying predictor are shown in bold. Predictor rows are highlighted in gray.}
\label{tab:vbdmd_post_valid}
\begingroup
\setlength{\tabcolsep}{3.7pt}
\begin{tabular}{lS[table-format=-2.2]S[table-format=1.2]S[table-format=1.2]S[table-format=1.2]S[table-format=1.2]}
\toprule
 & {SI-SDR [dB] $\uparrow$} & {PESQ $\uparrow$} & {DNSMOS $\uparrow$} & {NISQA $\uparrow$} & {UTMOS $\uparrow$} \\
\midrule
Noisy & 6.02 & 1.57 & 2.85 & 2.14 & 2.31 \\
\midrule\noalign{\vskip-\belowrulesep}\rowcolor{gray!15}\rule{0pt}{2.0ex}%
Conv-TasNet~\cite{luo2019conv} & 15.35 & 2.15 & 3.19 & 2.75 & 3.22 \\
\trimgrayrow
$\;$ + SIPS (w post) & 14.92 & 2.11 & \bfseries 3.38 & 3.54 & 3.50 \\
$\;$ + SIPS (w/o post) & \bfseries 15.28 & \bfseries 2.22 & \bfseries 3.38 & \bfseries 3.83 & \bfseries 3.54 \\
\midrule\noalign{\vskip-\belowrulesep}\rowcolor{gray!15}\rule{0pt}{2.0ex}%
NCSN++~\cite{lemercier2023storm} & 16.32 & 2.33 & 3.59 & 4.54 & 3.65 \\
\trimgrayrow
$\;$ + SIPS (w post) & 15.75 & 2.18 & \bfseries 3.58 & 4.47 & 3.71 \\
$\;$ + SIPS (w/o post) & \bfseries 16.21 & \bfseries 2.30 & 3.57 & \bfseries 4.48 & \bfseries 3.74 \\
\midrule\noalign{\vskip-\belowrulesep}\rowcolor{gray!15}\rule{0pt}{2.0ex}%
SEMamba~\cite{chao2024investigation} & 17.61 & 3.44 & 3.65 & 4.29 & 3.98 \\
\trimgrayrow
$\;$ + SIPS (w post) & 16.74 & \bfseries 3.11 & \bfseries 3.69 & \bfseries 4.65 & \bfseries 4.02 \\
$\;$ + SIPS (w/o post) & \bfseries 17.19 & 3.10 & 3.67 & 4.59 & \bfseries 4.02 \\
\bottomrule
\end{tabular}
\endgroup
\end{table}

\begin{table}\centering
\caption{Speech enhancement results with and without post-processing on the mismatched EARS-WHAM (v2) validation set (mean scores). SIPS uses 15 sampling steps with $\kappa=0.4$. Best values among the two SIPS configurations for each underlying predictor are shown in bold. Predictor rows are highlighted in gray.}
\label{tab:earswham_post_valid}
\begingroup
\setlength{\tabcolsep}{3.7pt}
\begin{tabular}{lS[table-format=-2.2]S[table-format=1.2]S[table-format=1.2]S[table-format=1.2]S[table-format=1.2]}
\toprule
 & {SI-SDR [dB] $\uparrow$} & {PESQ $\uparrow$} & {DNSMOS $\uparrow$} & {NISQA $\uparrow$} & {UTMOS $\uparrow$} \\
\midrule
Noisy & 5.05 & 1.29 & 2.70 & 1.92 & 1.65 \\
\midrule\noalign{\vskip-\belowrulesep}\rowcolor{gray!15}\rule{0pt}{2.0ex}%
Conv-TasNet~\cite{luo2019conv} & 4.49 & 1.40 & 2.90 & 1.48 & 1.74 \\
\trimgrayrow
$\;$ + SIPS (w post) & 3.80 & 1.36 & \bfseries 3.17 & 2.40 & 1.96 \\
$\;$ + SIPS (w/o post) & \bfseries 4.01 & \bfseries 1.41 & 3.13 & \bfseries 2.59 & \bfseries 2.00 \\
\midrule\noalign{\vskip-\belowrulesep}\rowcolor{gray!15}\rule{0pt}{2.0ex}%
NCSN++~\cite{lemercier2023storm} & 11.32 & 1.64 & 3.59 & 3.58 & 2.60 \\
\trimgrayrow
$\;$ + SIPS (w post) & 10.31 & 1.59 & \bfseries 3.62 & \bfseries 3.83 & 2.70 \\
$\;$ + SIPS (w/o post) & \bfseries 10.54 & \bfseries 1.75 & 3.52 & 3.42 & \bfseries 2.75 \\
\midrule\noalign{\vskip-\belowrulesep}\rowcolor{gray!15}\rule{0pt}{2.0ex}%
SEMamba~\cite{chao2024investigation} & 10.85 & 2.09 & 3.61 & 3.35 & 2.89 \\
\trimgrayrow
$\;$ + SIPS (w post) & 9.40 & 1.88 & \bfseries 3.67 & \bfseries 3.79 & \bfseries 2.94 \\
$\;$ + SIPS (w/o post) & \bfseries 9.51 & \bfseries 1.91 & 3.50 & 3.32 & 2.89 \\
\bottomrule
\end{tabular}
\endgroup
\end{table}

\subsection{Data representation}
\label{app:representation}

Following~\cite{richter2023speech}, we operate in an amplitude-compressed time–frequency domain using the \ac{STFT} with a hop length of 128, an FFT size and window length of 512, and a Hann window.
Let $F$ denote the number of frequency bins and $K$ the number of time frames.
For each coefficient $\tilde{x}_{fk} \in \mathbb{C}$ of the \ac{STFT}, where $f \in \{1,\dots,F\}$ and $k \in \{ 1,\dots,K\}$, we define the representation as
\begin{equation}
   x_{fk} = b\,|\tilde{x}_{fk}|^{p} e^{i\angle{\tilde{x}_{fk}}},
\end{equation}
where $p \in (0,1]$ denotes a compression exponent (set to $p=0.5$ in our experiments),
$b \in \mathbb{R}_+$ is a scaling factor used to normalize amplitudes (set to $b=0.15$),
and $\angle{(\cdot)}$ denotes the phase of a complex number.
The real and imaginary components are separated and stacked along a channel dimension, resulting in a tensor representation in $\mathbb{R}^{C \times F \times K}$ with $C=2$. This representation naturally supports processing with two-dimensional convolutional architectures over time and frequency.

\subsection{Implementation and Training Details}
\label{app:training}

Our implementation is based on the repository\footnote{\scriptsize\url{https://github.com/sp-uhh/edm2se}} of EDM2SE~\cite{richter2026do}, which employs the EDM2~\cite{karras2024analyzing} framework with a magnitude-preserving network architecture.
All denoiser models are trained on two NVIDIA A40 GPUs with a total batch size of 16 (2$\times$8) for approximately \num{4.2}M training samples of about \qty{2}{\second} of clean speech, taking roughly two to three days.
We use Adam with $\beta_1=0.9$ and $\beta_2=0.99$ and a base learning rate of \num{2.5e-3}.
The learning rate is multiplied by a linear warmup factor over the first \num{1}M training examples and an inverse-square-root decay factor starting at \num{0.48}M examples.
Both factors apply multiplicatively during their overlap, with training examples counted cumulatively across batches.

\subsection{Effect of the Number of Sampling Steps}
\label{app:num_steps}

We analyze the impact of the number of sampling steps $M$ using the validation sets. Figure~\ref{fig:steps_matched} presents the speech enhancement performance under matched conditions, while Figure~\ref{fig:steps_mismatched} reports the corresponding results for mismatched noise conditions. Overall, performance improves with increasing $M$ before saturating, indicating a trade-off between inference cost and enhancement quality. Based on these results, we use $M=15$ sampling steps as the default setting in all subsequent experiments.

\begin{figure}[t]
    \begin{subfigure}{0.49\columnwidth}
        \centering
        % \pgfplotstableread[row sep=\\]{
% M sisdr nisqa\\
% 3 16.295427 4.572443\\
% 4 16.377473 4.578515\\
% 5 16.476559 4.598142\\
% 6 16.534907 4.614109\\
% 7 16.544769 4.625033\\
% 8 16.566802 4.631033\\
% 9 16.590360 4.635553\\
% 10 16.604784 4.638158\\
% 11 16.614217 4.639321\\
% 12 16.619096 4.640154\\
% 13 16.624024 4.640570\\
% 14 16.627071 4.640594\\
% 15 16.628857 4.640997\\
% 16 16.630318 4.641091\\
% 17 16.631653 4.641026\\
% 18 16.632078 4.640790\\
% 19 16.632975 4.640668\\
% 20 16.632484 4.640241\\
% }\stepsMatched%

\pgfplotstableread[row sep=\\]{
M sisdr nisqa\\
3 16.884397 4.500149\\
4 16.941755 4.471459\\
5 17.060902 4.468192\\
6 17.144142 4.498631\\
7 17.179778 4.531518\\
8 17.193909 4.556371\\
9 17.198001 4.570742\\
10 17.198391 4.579079\\
11 17.197601 4.583810\\
12 17.196108 4.586901\\
13 17.194333 4.588107\\
14 17.192414 4.589489\\
15 17.190416 4.590404\\
16 17.188482 4.590819\\
17 17.186668 4.590785\\
18 17.184919 4.590618\\
19 17.183286 4.590418\\
20 17.181745 4.590076\\
}\stepsMatched%

\pgfmathsetlengthmacro{\plotwidth}{3.7cm}
\pgfmathsetlengthmacro{\plotheight}{2.0cm}

\begin{tikzpicture}
\begin{axis}[
    name=leftaxis,
    width=\plotwidth,
    height=\plotheight,
    font=\small,
    scale only axis,
    xlabel={$M$},
    ylabel={SI-SDR [dB]},
    xmin=2.5, xmax=21,
    ymin=16.8, ymax=17.4,
    xtick={5,10,15,20},
    ytick={16.8,17.0,17.2,17.4},
    grid=major,
    grid style={dashed,gray!60},
    axis y line*=left,
    axis x line*=box,
    xtick pos=bottom,
    tick align=outside,
    legend style={
        at={(0.5,1.05)},
        anchor=south,
        draw=none,
        legend columns=2,
        /tikz/every even column/.append style={column sep=0.4cm}
    }
]
\addplot[cbBlue,thick,mark=*,mark size=1.5pt]
    table[x=M,y=sisdr] {\stepsMatched};
\addlegendentry{SI-SDR}
\addlegendimage{cbOrange,thick,mark=diamond*,mark size=2.0pt}
\addlegendentry{NISQA}
\end{axis}

\begin{axis}[
    at={(leftaxis.south west)},
    anchor=south west,
    width=\plotwidth,
    height=\plotheight,
    font=\small,
    scale only axis,
    xmin=2.5, xmax=21,
    ymin=4.45, ymax=4.60,
    ylabel={NISQA},
    axis y line*=right,
    axis x line=none,
    ytick={4.45,4.50,4.55,4.60},
    tick align=outside
]
\addplot[cbOrange,thick,mark=diamond*,mark size=2.0pt]
    table[x=M,y=nisqa] {\stepsMatched};
\end{axis}
\end{tikzpicture}
        \caption{Matched condition ($\kappa=0$)}%
        \label{fig:steps_matched}
    \end{subfigure}\hfill%
    \begin{subfigure}{0.49\columnwidth}
        \centering
        \pgfplotstableread[row sep=\\]{
M sisdr nisqa\\
1 9.944326 3.370844\\
2 8.101243 3.145249\\
3 8.508284 3.428937\\
4 8.721379 3.530804\\
5 8.970312 3.644392\\
6 9.165191 3.741827\\
7 9.271526 3.790989\\
8 9.332460 3.811201\\
9 9.361420 3.804959\\
10 9.379272 3.807145\\
11 9.393901 3.802065\\
12 9.399311 3.805117\\
13 9.406161 3.800817\\
14 9.412000 3.796791\\
15 9.416775 3.793962\\
16 9.417939 3.794879\\
17 9.412883 3.792299\\
18 9.415317 3.786344\\
19 9.414396 3.782819\\
20 9.415385 3.785021\\
}\stepsMismatched%

\pgfmathsetlengthmacro{\plotwidth}{3.9cm}
\pgfmathsetlengthmacro{\plotheight}{2.0cm}

\begin{tikzpicture}
\begin{axis}[
    name=leftaxis,
    width=\plotwidth,
    height=\plotheight,
    font=\small,
    scale only axis,
    xlabel={$M$},
    ylabel={SI-SDR [dB]},
    xmin=2.5, xmax=21,
    ymin=8, ymax=11,
    xtick={5,10,15,20},
    ytick={8,9,10,11},
    grid=major,
    grid style={dashed,gray!60},
    axis y line*=left,
    axis x line*=box,
    xtick pos=bottom,
    tick align=outside,
    legend style={
        at={(0.5,1.05)},
        anchor=south,
        draw=none,
        legend columns=2,
        /tikz/every even column/.append style={column sep=0.4cm}
    }
]
\addplot[cbBlue,thick,mark=*,mark size=1.5pt]
    table[x=M,y=sisdr] {\stepsMismatched};
\addlegendentry{SI-SDR}
\addlegendimage{cbOrange,thick,mark=diamond*,mark size=2.0pt}
\addlegendentry{NISQA}
\end{axis}

\begin{axis}[
    at={(leftaxis.south west)},
    anchor=south west,
    width=\plotwidth,
    height=\plotheight,
    font=\small,
    scale only axis,
    xmin=2.5, xmax=21,
    ymin=3.0, ymax=3.9,
    ylabel={NISQA},
    axis y line*=right,
    axis x line=none,
    ytick={3.0,3.3,3.6,3.9},
    tick align=outside
]
\addplot[cbOrange,thick,mark=diamond*,mark size=2.0pt]
    table[x=M,y=nisqa] {\stepsMismatched};
\end{axis}
\end{tikzpicture}
        \caption{Mismatched condition ($\kappa=0.4$)}
        \label{fig:steps_mismatched}
    \end{subfigure}
    \caption{Speech enhancement performance of SEMamba + SIPS over the number of sampling steps $M$ on the matched and mismatched validation sets. Predictor post-processing is disabled in (a) and enabled in (b).}%
    \label{fig:steps_comparison}
\end{figure}

\subsection{StoRM under predictor mismatch}
\label{app:storm}

StoRM trains its score model conditional on the output distribution of a predictor.
To test whether this conditioning transfers across predictors, we evaluate the original StoRM checkpoint, whose score model was trained on NCSN++ predictor outputs, after replacing the predictor at inference time with Conv-TasNet or SEMamba.
We compare these mismatched runs with StoRM checkpoints trained on the corresponding predictor outputs.
The results in Tables~\ref{tab:storm_vbdmd} and~\ref{tab:storm_earswham} show that swapping the predictor causes large degradations, especially in SI-SDR and WER.
We therefore report only predictor-specific StoRM checkpoints in the main tables. 
Thus, the three StoRM entries in Tables~\ref{tab:vbdmd} and~\ref{tab:earswham} correspond to three different checkpoints.

\begin{table}[t]\centering
\caption{StoRM under predictor mismatch on the matched VoiceBank-DEMAND test set (mean values). ``Mismatched'' denotes the original NCSN++-conditioned StoRM checkpoint used with a different predictor at inference time; ``matched'' denotes a StoRM checkpoint trained on outputs of the corresponding predictor.}%
\label{tab:storm_vbdmd}
\begingroup
\setlength{\tabcolsep}{2pt}
\begin{tabular}{lS[table-format=2.2]S[table-format=1.2]S[table-format=1.2]S[table-format=1.2]S[table-format=1.2]S[table-format=2.2]}
\toprule
 & {SI-SDR [dB] $\uparrow$} & {PESQ $\uparrow$} & {DNSMOS $\uparrow$} & {NISQA $\uparrow$} & {UTMOS $\uparrow$} & {WER [\%] $\downarrow$} \\
\midrule
Clean & \multicolumn{1}{c}{-} & 4.64 & 3.55 & 4.50 & 4.09 & 6.96 \\
Noisy & 8.44 & 1.97 & 3.09 & 3.03 & 3.11 & 11.01 \\
\midrule\noalign{\vskip-\belowrulesep}\rowcolor{gray!15}\rule{0pt}{2.0ex}%
Conv-TasNet~\cite{luo2019conv} & 18.57 & 2.51 & 3.31 & 3.45 & 3.61 & 11.87 \\
\trimgrayrow
$\;$ + StoRM (mismatched) & 11.99 & 2.31 & 3.36 & 3.93 & 3.65 & 16.87 \\
$\;$ + StoRM (matched) & \bfseries 17.92 & \bfseries 2.67 & \bfseries 3.54 & \bfseries 4.57 & \bfseries 3.92 & \bfseries 13.32 \\
\midrule\noalign{\vskip-\belowrulesep}\rowcolor{gray!15}\rule{0pt}{2.0ex}%
NCSN++~\cite{lemercier2023storm} & 19.57 & 2.83 & 3.59 & 4.64 & 3.93 & 9.58 \\
\trimgrayrow
$\;$ + StoRM (matched) & \bfseries 18.49 & \bfseries 2.89 & \bfseries 3.56 & \bfseries 4.52 & \bfseries 3.92 & \bfseries 10.21 \\
\midrule\noalign{\vskip-\belowrulesep}\rowcolor{gray!15}\rule{0pt}{2.0ex}%
SEMamba~\cite{chao2024investigation} & 19.72 & 3.56 & 3.58 & 4.60 & 4.07 & 8.87 \\
\trimgrayrow
$\;$ + StoRM (mismatched) & 12.49 & 2.79 & 3.54 & 4.42 & 3.91 & 11.11 \\
$\;$ + StoRM (matched) & \bfseries 18.89 & \bfseries 3.17 & \bfseries 3.56 & \bfseries 4.54 & \bfseries 4.01 & \bfseries 9.37 \\
\bottomrule
\end{tabular}%
\endgroup
\end{table}

\begin{table}[t]\centering
\caption{StoRM under predictor mismatch on the mismatched EARS-WHAM (v2) test set (mean values). ``Mismatched'' denotes the original NCSN++-conditioned StoRM checkpoint used with a different predictor at inference time; ``matched'' denotes a StoRM checkpoint trained on outputs of the corresponding predictor.}%
\label{tab:storm_earswham}
\begingroup
\setlength{\tabcolsep}{2pt}
\begin{tabular}{lS[table-format=-2.2]S[table-format=1.2]S[table-format=1.2]S[table-format=1.2]S[table-format=1.2]S[table-format=2.2]}
\toprule
 & {SI-SDR [dB] $\uparrow$} & {PESQ $\uparrow$} & {DNSMOS $\uparrow$} & {NISQA $\uparrow$} & {UTMOS $\uparrow$} & {WER [\%] $\downarrow$} \\
\midrule
Clean & \multicolumn{1}{c}{-} & 4.64 & 3.89 & 4.09 & 3.68 & 8.95 \\
Noisy & 5.36 & 1.24 & 2.73 & 1.95 & 1.68 & 32.87 \\
\midrule\noalign{\vskip-\belowrulesep}\rowcolor{gray!15}\rule{0pt}{2.0ex}%
Conv-TasNet~\cite{luo2019conv} & 3.95 & 1.35 & 2.90 & 1.43 & 1.66 & 53.87 \\
\trimgrayrow
$\;$ + StoRM (mismatched) & -2.69 & 1.15 & 3.22 & 2.14 & 1.50 & 80.60 \\
$\;$ + StoRM (matched) & \bfseries 3.24 & \bfseries 1.29 & \bfseries 3.44 & \bfseries 3.33 & \bfseries 2.21 & \bfseries 64.56 \\
\midrule\noalign{\vskip-\belowrulesep}\rowcolor{gray!15}\rule{0pt}{2.0ex}%
NCSN++~\cite{lemercier2023storm} & 13.24 & 1.81 & 3.72 & 3.84 & 2.75 & 29.15 \\
\trimgrayrow
$\;$ + StoRM (matched) & \bfseries 12.49 & \bfseries 1.90 & \bfseries 3.83 & \bfseries 4.12 & \bfseries 2.86 & \bfseries 31.74 \\
\midrule\noalign{\vskip-\belowrulesep}\rowcolor{gray!15}\rule{0pt}{2.0ex}%
SEMamba~\cite{chao2024investigation} & 11.36 & 2.19 & 3.71 & 3.57 & 2.93 & 28.08 \\
\trimgrayrow
$\;$ + StoRM (mismatched) & 1.40 & 1.38 & \bfseries 3.63 & 2.90 & 2.07 & 55.60 \\
$\;$ + StoRM (matched) & \bfseries 10.92 & \bfseries 2.01 & 3.71 & \bfseries 3.72 & \bfseries 2.88 & \bfseries 29.63 \\
\bottomrule
\end{tabular}%
\endgroup
\end{table}

\subsection{Speech separation on MC-WSJ-AV}
\label{app:mcwsjav}

To complement the evaluation on simulated WHAMR! mixtures, we evaluate FlexIO with and without SIPS on real-recorded two-speaker mixtures from MC-WSJ-AV~\cite{lincoln2005mcwsjav}.
The corpus contains read Wall Street Journal speech recorded in meeting rooms, including a scenario with two stationary overlapping speakers, and is distributed by the Linguistic Data Consortium as Multi-Channel WSJ Audio (LDC2014S03)~\cite{lincoln2014mcwsj}.
From the EVAL1 set of the overlapping scenario~\cite{drude2019integration}, which comprises utterances corresponding to text subsets 16--20, we select the first channel of Array 1, yielding 330 mixtures.
We use the same SIPS hyperparameters as in the WHAMR! experiments.
We report the non-intrusive quality metrics DNSMOS, NISQA, and UTMOS, together with \ac{WER}, in Table~\ref{tab:mcwsjav}.
Adding SIPS improves all three quality metrics.
DNSMOS increases from 3.80 to 3.82, NISQA from 2.37 to 2.46, and UTMOS from 2.87 to 3.01.
The \ac{WER} increases slightly, from 8.92\% to 9.36\%.
These results extend the observed perceptual-quality improvements to real-recorded overlapping speech, with a small trade-off in recognition performance.

\begin{table}[t]\centering
\caption{Speech separation results on real-recorded two-speaker mixtures from MC-WSJ-AV. Entries show absolute scores, with the best value in each metric in bold. The predictor has a gray background.}
\label{tab:mcwsjav}
\begingroup
\setlength{\tabcolsep}{5pt}
\begin{tabular}{lS[table-format=1.2]S[table-format=1.2]S[table-format=1.2]S[table-format=1.2]}
\toprule
Model & {DNSMOS $\uparrow$} & {NISQA $\uparrow$} & {UTMOS $\uparrow$} & {WER [\%] $\downarrow$} \\
\midrule\noalign{\vskip-\belowrulesep}\rowcolor{gray!15}\rule{0pt}{2.0ex}%
FlexIO~\cite{masuyama2026flexio} & 3.80 & 2.37 & 2.87 & \bfseries 8.92 \\
\trimgrayrow
$\;$ + SIPS (ours) & \bfseries 3.82 & \bfseries 2.46 & \bfseries 3.01 & 9.36 \\
\bottomrule
\end{tabular}
\endgroup
\end{table}

\subsection{Results with standard deviations}
\label{app:vbdmd_std}

Tables~\ref{tab:vbdmd_std} and~\ref{tab:earswham_std} report mean improvements over each predictor and the standard deviations of these improvements for VoiceBank-DEMAND and EARS-WHAM, respectively. 
Standard deviations are computed over per-utterance differences between refiner and predictor scores. 
Clean, noisy, and predictor results are shown as absolute mean values without standard deviations.

For \ac{WER}, the mean improvement is the difference between corpus-level error rates, each computed as the total number of word errors divided by the total number of reference words over the test set. 
The standard deviation is computed over per-utterance differences in \ac{WER}.

\begin{table}[t]\centering
\caption{Speech enhancement results on the matched VoiceBank-DEMAND test set (absolute means for clean, noisy, and predictors). Refiner entries show the mean improvement over the predictor, with its standard deviation in smaller text. Changes are computed from unrounded scores and rounded only for display. Green mean changes indicate improvement and red mean changes indicate degradation. Best mean values within each predictor group, including the predictor itself, are shown in bold. The underlying predictors are highlighted with a gray background.}%
\label{tab:vbdmd_std}
\begingroup
\setlength{\tabcolsep}{3.5pt}
\begin{tabular}{lS[table-format=+2.2,detect-weight=true,retain-explicit-plus=true]@{\hspace{\tabcolsep}}lS[table-format=+1.2,detect-weight=true,retain-explicit-plus=true]@{\hspace{\tabcolsep}}lS[table-format=+1.2,detect-weight=true,retain-explicit-plus=true]@{\hspace{\tabcolsep}}lS[table-format=+1.2,detect-weight=true,retain-explicit-plus=true]@{\hspace{\tabcolsep}}lS[table-format=+1.2,detect-weight=true,retain-explicit-plus=true]@{\hspace{\tabcolsep}}lS[table-format=+2.2,detect-weight=true,retain-explicit-plus=true]@{\hspace{\tabcolsep}}l}
\toprule
 & \multicolumn{2}{c}{\hspace{-0.5em}\makebox[0pt][c]{SI-SDR [dB] $\uparrow$}\hspace{0.5em}} & \multicolumn{2}{c}{PESQ $\uparrow$} & \multicolumn{2}{c}{DNSMOS $\uparrow$} & \multicolumn{2}{c}{NISQA $\uparrow$} & \multicolumn{2}{c}{UTMOS $\uparrow$} & \multicolumn{2}{c}{WER [\%] $\downarrow$} \\
\midrule
Clean & \multicolumn{1}{c}{-} &  & 4.64 &  & 3.55 &  & 4.50 &  & 4.09 &  & 6.96 &  \\
Noisy & 8.44 &  & 1.97 &  & 3.09 &  & 3.03 &  & 3.11 &  & 11.01 &  \\
\midrule\noalign{\vskip-\belowrulesep}\rowcolor{gray!15}\rule{0pt}{2.0ex}%
Conv-TasNet~\cite{luo2019conv} & 18.57 &  & 2.51 &  & 3.31 &  & 3.45 &  & 3.61 &  & \bfseries 11.87 &  \\
\trimgrayrow
$\;$ + StoRM~\cite{lemercier2023storm} & \color{red} -0.65 & \hspace{-2pt}{\tiny $\pm$ 0.61} & \color{green!50!black} \bfseries +0.16 & \hspace{-2pt}{\tiny $\pm$ 0.32} & \color{green!50!black} \bfseries+0.23 & \hspace{-2pt}{\tiny $\pm$ 0.20} & \color{green!50!black} +1.13 & \hspace{-2pt}{\tiny $\pm$ 0.79} & \color{green!50!black} +0.31 & \hspace{-2pt}{\tiny $\pm$ 0.29} & \color{red} +1.45 & \hspace{-2pt}{\tiny $\pm$ \tablenum[table-format=2.2]{11.46}} \\
$\;$ + Diffiner~\cite{sawata2023diffiner} & \color{red} -2.35 & \hspace{-2pt}{\tiny $\pm$ 1.24} & \color{green!50!black} +0.08 & \hspace{-2pt}{\tiny $\pm$ 0.36} & \color{green!50!black} +0.16 & \hspace{-2pt}{\tiny $\pm$ 0.17} & \color{green!50!black} \bfseries +1.30 & \hspace{-2pt}{\tiny $\pm$ 0.67} & \color{green!50!black} \bfseries +0.35 & \hspace{-2pt}{\tiny $\pm$ 0.29} & \color{red} +4.12 & \hspace{-2pt}{\tiny $\pm$ \tablenum[table-format=2.2]{16.00}} \\
$\;$ + SIPS (ours) & \color{green!50!black} \bfseries +0.22 & \hspace{-2pt}{\tiny $\pm$ 0.63} & \color{green!50!black} +0.13 & \hspace{-2pt}{\tiny $\pm$ 0.22} & \color{green!50!black} +0.12 & \hspace{-2pt}{\tiny $\pm$ 0.15} & \color{green!50!black} +0.83 & \hspace{-2pt}{\tiny $\pm$ 0.50} & \color{green!50!black} +0.24 & \hspace{-2pt}{\tiny $\pm$ 0.16} & \color{red} +1.00 & \hspace{-2pt}{\tiny $\pm$ \tablenum[table-format=2.2]{10.49}} \\
\midrule\noalign{\vskip-\belowrulesep}\rowcolor{gray!15}\rule{0pt}{2.0ex}%
NCSN++~\cite{lemercier2023storm} & \bfseries 19.57 &  & 2.83 &  & \bfseries 3.59 &  & 4.64 &  & 3.93 &  & \bfseries 9.58 &  \\
\trimgrayrow
$\;$ + StoRM~\cite{lemercier2023storm} & \color{red} -1.08 & \hspace{-2pt}{\tiny $\pm$ 0.73} & \color{green!50!black} \bfseries +0.06 & \hspace{-2pt}{\tiny $\pm$ 0.32} & \color{red} -0.03 & \hspace{-2pt}{\tiny $\pm$ 0.11} & \color{red} -0.12 & \hspace{-2pt}{\tiny $\pm$ 0.35} & \color{red} -0.01 & \hspace{-2pt}{\tiny $\pm$ 0.11} & \color{red} +0.63 & \hspace{-2pt}{\tiny $\pm$ \tablenum[table-format=2.2]{9.41}} \\
$\;$ + Diffiner~\cite{sawata2023diffiner} & \color{red} -3.46 & \hspace{-2pt}{\tiny $\pm$ 2.33} & +0.00 & \hspace{-2pt}{\tiny $\pm$ 0.34} & \color{red} -0.07 & \hspace{-2pt}{\tiny $\pm$ 0.12} & \color{green!50!black} \bfseries +0.15 & \hspace{-2pt}{\tiny $\pm$ 0.36} & \color{green!50!black} \bfseries +0.08 & \hspace{-2pt}{\tiny $\pm$ 0.14} & \color{red} +3.45 & \hspace{-2pt}{\tiny $\pm$ \tablenum[table-format=2.2]{13.26}} \\
$\;$ + SIPS (ours) & \color{red} -0.37 & \hspace{-2pt}{\tiny $\pm$ 1.08} & \color{green!50!black} +0.05 & \hspace{-2pt}{\tiny $\pm$ 0.30} & \color{red} -0.03 & \hspace{-2pt}{\tiny $\pm$ 0.13} & \color{green!50!black} +0.07 & \hspace{-2pt}{\tiny $\pm$ 0.31} & \color{green!50!black} +0.04 & \hspace{-2pt}{\tiny $\pm$ 0.08} & \color{red} +0.42 & \hspace{-2pt}{\tiny $\pm$ \tablenum[table-format=2.2]{8.62}} \\
\midrule\noalign{\vskip-\belowrulesep}\rowcolor{gray!15}\rule{0pt}{2.0ex}%
SEMamba~\cite{chao2024investigation} & \bfseries 19.72 &  & \bfseries 3.56 &  & \bfseries 3.58 &  & 4.60 &  & 4.07 &  & 8.87 &  \\
\trimgrayrow
$\;$ + StoRM~\cite{lemercier2023storm} & \color{red} -0.83 & \hspace{-2pt}{\tiny $\pm$ 0.62} & \color{red} -0.39 & \hspace{-2pt}{\tiny $\pm$ 0.23} & \color{red} -0.02 & \hspace{-2pt}{\tiny $\pm$ 0.06} &\color{red} -0.07 & \hspace{-2pt}{\tiny $\pm$ 0.25} & \color{red} -0.06 & \hspace{-2pt}{\tiny $\pm$ 0.07} & \color{red} +0.50 & \hspace{-2pt}{\tiny $\pm$ \tablenum[table-format=2.2]{6.46}} \\
$\;$ + Diffiner~\cite{sawata2023diffiner} & \color{red} -3.20 & \hspace{-2pt}{\tiny $\pm$ 1.62} & \color{red} -0.69 & \hspace{-2pt}{\tiny $\pm$ 0.37} & \color{red} -0.06 & \hspace{-2pt}{\tiny $\pm$ 0.12} & \color{green!50!black} \bfseries +0.21 & \hspace{-2pt}{\tiny $\pm$ 0.30} & \color{red} -0.03 & \hspace{-2pt}{\tiny $\pm$ 0.13} & \color{red} +3.56 & \hspace{-2pt}{\tiny $\pm$ \tablenum[table-format=2.2]{12.94}} \\
$\;$ + SIPS (ours) & \color{red} -0.08 & \hspace{-2pt}{\tiny $\pm$ 0.88} & \color{red} -0.12 & \hspace{-2pt}{\tiny $\pm$ 0.21} & \color{red} -0.01 & \hspace{-2pt}{\tiny $\pm$ 0.11} & \color{green!50!black} +0.13 & \hspace{-2pt}{\tiny $\pm$ 0.25} & \color{green!50!black} \bfseries +0.02 & \hspace{-2pt}{\tiny $\pm$ 0.05} & \color{green!50!black} \bfseries -0.06 & \hspace{-2pt}{\tiny $\pm$ \tablenum[table-format=2.2]{7.70}} \\
\bottomrule
\end{tabular}%
\endgroup
\end{table}

\begin{table}[t]\centering
\caption{Speech enhancement results on the mismatched EARS-WHAM (v2) test set (absolute means for clean, noisy, and predictors). Refiner entries show the mean improvement over the predictor, with its standard deviation in smaller text. Changes are computed from unrounded scores and rounded only for display. Green mean changes indicate improvement and red mean changes indicate degradation. Best mean values within each predictor group, including the predictor itself, are shown in bold. The underlying predictors are highlighted with a gray background.}%
\label{tab:earswham_std}
\begingroup
\setlength{\tabcolsep}{3.5pt}
\begin{tabular}{lS[table-format=+2.2,detect-weight=true,retain-explicit-plus=true]@{\hspace{\tabcolsep}}lS[table-format=+1.2,detect-weight=true,retain-explicit-plus=true]@{\hspace{\tabcolsep}}lS[table-format=+1.2,detect-weight=true,retain-explicit-plus=true]@{\hspace{\tabcolsep}}lS[table-format=+1.2,detect-weight=true,retain-explicit-plus=true]@{\hspace{\tabcolsep}}lS[table-format=+1.2,detect-weight=true,retain-explicit-plus=true]@{\hspace{\tabcolsep}}lS[table-format=+2.2,detect-weight=true,retain-explicit-plus=true]@{\hspace{\tabcolsep}}l}
\toprule
 & \multicolumn{2}{c}{\hspace{-0.5em}\makebox[0pt][c]{SI-SDR [dB] $\uparrow$}\hspace{0.5em}} & \multicolumn{2}{c}{PESQ $\uparrow$} & \multicolumn{2}{c}{DNSMOS $\uparrow$} & \multicolumn{2}{c}{NISQA $\uparrow$} & \multicolumn{2}{c}{UTMOS $\uparrow$} & \multicolumn{2}{c}{WER [\%] $\downarrow$} \\
\midrule
Clean & \multicolumn{1}{c}{-} &  & 4.64 &  & 3.89 &  & 4.09 &  & 3.68 &  & 8.95 &  \\
Noisy & 5.36 &  & 1.24 &  & 2.73 &  & 1.95 &  & 1.68 &  & 32.87 &  \\
\midrule\noalign{\vskip-\belowrulesep}\rowcolor{gray!15}\rule{0pt}{2.0ex}%
Conv-TasNet~\cite{luo2019conv} & \bfseries 3.95 &  & \bfseries 1.35 &  & 2.90 &  & 1.43 &  & 1.66 &  & \bfseries 53.87 &  \\
\trimgrayrow
$\;$ + StoRM~\cite{lemercier2023storm} & \color{red} -0.71 & \hspace{-2pt}{\tiny $\pm$ 0.86} & \color{red} -0.06 & \hspace{-2pt}{\tiny $\pm$ 0.13} & \color{green!50!black} \bfseries +0.54 & \hspace{-2pt}{\tiny $\pm$ 0.21} & \color{green!50!black} \bfseries +1.90 & \hspace{-2pt}{\tiny $\pm$ 0.65} & \color{green!50!black} \bfseries +0.54 & \hspace{-2pt}{\tiny $\pm$ 0.35} & \color{red} +10.69 & \hspace{-2pt}{\tiny $\pm$ \tablenum[table-format=2.2]{13.20}} \\
$\;$ + Diffiner~\cite{sawata2023diffiner} & \color{red} -6.26 & \hspace{-2pt}{\tiny $\pm$ 3.44} & \color{red} -0.20 & \hspace{-2pt}{\tiny $\pm$ 0.25} & \color{green!50!black} +0.09 & \hspace{-2pt}{\tiny$\pm$ 0.23} & \color{green!50!black} +1.59 & \hspace{-2pt}{\tiny $\pm$ 0.84} & \color{green!50!black} +0.31 & \hspace{-2pt}{\tiny $\pm$ 0.40} & \color{red} +33.93 & \hspace{-2pt}{\tiny $\pm$ \tablenum[table-format=2.2]{22.37}} \\
$\;$ + SIPS (ours) & \color{red} -0.09 & \hspace{-2pt}{\tiny $\pm$ 0.27} & \color{red} -0.03 & \hspace{-2pt}{\tiny $\pm$ 0.04} & \color{green!50!black} +0.19 & \hspace{-2pt}{\tiny $\pm$ 0.10} & \color{green!50!black} +0.55 & \hspace{-2pt}{\tiny $\pm$ 0.36} & \color{green!50!black} +0.07 & \hspace{-2pt}{\tiny $\pm$ 0.10} & \color{red} +5.24 & \hspace{-2pt}{\tiny $\pm$ \tablenum[table-format=2.2]{7.42}} \\
\midrule\noalign{\vskip-\belowrulesep}\rowcolor{gray!15}\rule{0pt}{2.0ex}%
NCSN++~\cite{lemercier2023storm} & \bfseries 13.24 &  & 1.81 &  & 3.72 &  & 3.84 &  & 2.75 &  & \bfseries 29.15 &  \\
\trimgrayrow
$\;$ + StoRM~\cite{lemercier2023storm} & \color{red} -0.75 & \hspace{-2pt}{\tiny $\pm$ 0.58} & \color{green!50!black} \bfseries +0.09 & \hspace{-2pt}{\tiny $\pm$ 0.10} & \color{green!50!black} \bfseries+0.11 & \hspace{-2pt}{\tiny $\pm$ 0.15} & \color{green!50!black} \bfseries +0.28 & \hspace{-2pt}{\tiny $\pm$ 0.41} & \color{green!50!black} \bfseries +0.11 & \hspace{-2pt}{\tiny $\pm$ 0.11} & \color{red} +2.58 & \hspace{-2pt}{\tiny $\pm$ \tablenum[table-format=2.2]{8.47}} \\
$\;$ + Diffiner~\cite{sawata2023diffiner} & \color{red} -13.34 & \hspace{-2pt}{\tiny $\pm$ 5.60} & \color{red} -0.51 & \hspace{-2pt}{\tiny $\pm$ 0.47} & \color{red} -0.22 & \hspace{-2pt}{\tiny $\pm$ 0.25} & \color{red} -0.43 & \hspace{-2pt}{\tiny $\pm$ 0.79} & \color{red} -0.54 & \hspace{-2pt}{\tiny $\pm$ 0.59} & \color{red} +39.90 & \hspace{-2pt}{\tiny $\pm$ \tablenum[table-format=2.2]{22.42}} \\
$\;$ + SIPS (ours) & \color{red} -0.96 & \hspace{-2pt}{\tiny $\pm$ 1.45} & \color{red} -0.09 & \hspace{-2pt}{\tiny $\pm$ 0.11} & \color{red} -0.04 & \hspace{-2pt}{\tiny $\pm$ 0.09} & \color{green!50!black} +0.03 & \hspace{-2pt}{\tiny $\pm$ 0.20} & \color{green!50!black} +0.01 & \hspace{-2pt}{\tiny $\pm$ 0.10} & \color{red} +1.83 & \hspace{-2pt}{\tiny $\pm$ \tablenum[table-format=2.2]{7.81}} \\
\midrule\noalign{\vskip-\belowrulesep}\rowcolor{gray!15}\rule{0pt}{2.0ex}%
SEMamba~\cite{chao2024investigation} & \bfseries 11.36 &  & \bfseries 2.19 &  & 3.71 &  & 3.57 &  & 2.93 &  & \bfseries 28.08 &  \\
\trimgrayrow
$\;$ + StoRM~\cite{lemercier2023storm} & \color{red} -0.43 & \hspace{-2pt}{\tiny $\pm$ 0.32} & \color{red} -0.18 & \hspace{-2pt}{\tiny $\pm$ 0.13} & +0.00 & \hspace{-2pt}{\tiny $\pm$ 0.09} & \color{green!50!black} +0.15 & \hspace{-2pt}{\tiny $\pm$ 0.38} & \color{red} -0.04 & \hspace{-2pt}{\tiny $\pm$ 0.10} & \color{red} +1.56 & \hspace{-2pt}{\tiny $\pm$ \tablenum[table-format=2.2]{5.85}} \\
$\;$ + Diffiner~\cite{sawata2023diffiner} & \color{red} -6.48 & \hspace{-2pt}{\tiny $\pm$ 2.20} & \color{red} -0.76 & \hspace{-2pt}{\tiny $\pm$ 0.42} & \color{red} -0.28 & \hspace{-2pt}{\tiny $\pm$ 0.19} & \color{green!50!black} +0.10 & \hspace{-2pt}{\tiny $\pm$ 0.65} & \color{red} -0.40 & \hspace{-2pt}{\tiny $\pm$ 0.41} & \color{red} +36.14 & \hspace{-2pt}{\tiny $\pm$ \tablenum[table-format=2.2]{18.65}} \\
$\;$ + SIPS (ours) & \color{red} -0.08 & \hspace{-2pt}{\tiny $\pm$ 0.23} & \color{red} -0.14 & \hspace{-2pt}{\tiny $\pm$ 0.08} & \color{green!50!black} \bfseries +0.03 & \hspace{-2pt}{\tiny $\pm$ 0.08} & \color{green!50!black} \bfseries +0.25 & \hspace{-2pt}{\tiny $\pm$ 0.30} & \color{green!50!black} \bfseries +0.05 & \hspace{-2pt}{\tiny $\pm$ 0.08} & \color{red} +1.20 & \hspace{-2pt}{\tiny $\pm$ \tablenum[table-format=2.2]{5.28}} \\
\bottomrule
\end{tabular}%
\endgroup
\end{table}

\subsection{Results per input SNR}
\label{app:per_snr}

\paragraph{Matched setting.}
We analyze the matched setting as a function of the input SNR by grouping the VoiceBank-DEMAND test utterances at the four input-SNR levels of 2.5, 7.5, 12.5, and 17.5 dB.
Tables~\ref{tab:vbdmd_per_snr_1}--\ref{tab:vbdmd_per_snr_3} report the per-SNR results for all predictors and their generative refiners, including the SEMamba results shown in Figure~\ref{fig:vbdmd_per_snr}.
Table~\ref{tab:vbdmd_per_snr_1} shows PESQ and SI-SDR, Table~\ref{tab:vbdmd_per_snr_2} shows DNSMOS and NISQA, and Table~\ref{tab:vbdmd_per_snr_3} shows UTMOS and WER.
Clean, noisy, and predictor entries report absolute means, whereas refiner entries report changes relative to the corresponding predictor. 
WER changes are expressed in percentage points.

\paragraph{Mismatched setting.}

We analyze the mismatched setting as a function of the input SNR.
For this analysis, we partition the test utterances according to the SNR of the original EARS-WHAM mixture, using the identical SNR partitioning as in the EARS-WHAM paper~\cite{richter2024ears}.
The four SNR ranges are shown in the column headers of the corresponding tables.
Tables~\ref{tab:earswham_per_snr_1}--\ref{tab:earswham_per_snr_3} report the per-SNR results for all metrics.
Table~\ref{tab:earswham_per_snr_1} shows the intrusive quality metrics PESQ and SI-SDR, Table~\ref{tab:earswham_per_snr_2} shows the non-intrusive quality metrics DNSMOS and NISQA, and Table~\ref{tab:earswham_per_snr_3} shows UTMOS and WER.
For the lowest-SNR bin, we compute the 95\% \ac{CI} for the mean paired utterance-level NISQA improvement using a two-sided Student's $t$ interval over 229 utterances.

\clearpage

\begin{table*}[t]\centering
\caption{\textbf{Results per input SNR (PESQ and SI-SDR [dB])}. Results on the matched VoiceBank-DEMAND test set for quality metrics, segmented into the four exact input-SNR levels specified in the column headers. Clean, noisy, and predictor entries show absolute means. Refiner entries show the mean change from their predictor, with the population standard deviation of paired per-utterance changes in smaller text. Changes are computed from unrounded scores and rounded only for display. Green changes indicate improvement and red changes indicate degradation. Best absolute means within each predictor group and SNR level, including the predictor, are bold. Predictors have a gray background.}
\label{tab:vbdmd_per_snr_1}
\begingroup
\setlength{\tabcolsep}{3.5pt}
\begin{tabular}{lS[table-format=+1.2,detect-weight=true,retain-explicit-plus=true]@{\hspace{\tabcolsep}}lS[table-format=+1.2,detect-weight=true,retain-explicit-plus=true]@{\hspace{\tabcolsep}}lS[table-format=+1.2,detect-weight=true,retain-explicit-plus=true]@{\hspace{\tabcolsep}}lS[table-format=+1.2,detect-weight=true,retain-explicit-plus=true]@{\hspace{\tabcolsep}}l}
\toprule
& \multicolumn{8}{c}{\textbf{PESQ} $\uparrow$} \\
\cmidrule(lr){2-9}
& \multicolumn{2}{c}{2.5 dB} & \multicolumn{2}{c}{7.5 dB} & \multicolumn{2}{c}{12.5 dB} & \multicolumn{2}{c}{17.5 dB} \\ \midrule
Clean & 4.64 &  & 4.64 &  & 4.64 &  & 4.64 &  \\
Noisy & 1.42 &  & 1.76 &  & 2.10 &  & 2.60 &  \\
\midrule\noalign{\vskip-\belowrulesep}\rowcolor{gray!15}\rule{0pt}{2.0ex}%
Conv-TasNet~\cite{luo2019conv} & 1.96 &  & 2.38 &  & 2.66 &  & 3.04 &  \\
\trimgrayrow
$\;$ + StoRM~\cite{lemercier2023storm} & \color{green!50!black} \bfseries +0.20 & \hspace{-2pt}{\tiny $\pm$ 0.29} & \color{green!50!black} \bfseries +0.21 & \hspace{-2pt}{\tiny $\pm$ 0.31} & \color{green!50!black} \bfseries +0.14 & \hspace{-2pt}{\tiny $\pm$ 0.30} & \color{green!50!black} +0.10 & \hspace{-2pt}{\tiny $\pm$ 0.35} \\
$\;$ + Diffiner~\cite{sawata2023diffiner} & \color{green!50!black} +0.09 & \hspace{-2pt}{\tiny $\pm$ 0.33} & \color{green!50!black} +0.11 & \hspace{-2pt}{\tiny $\pm$ 0.38} & \color{green!50!black} +0.06 & \hspace{-2pt}{\tiny $\pm$ 0.35} & \color{green!50!black} +0.07 & \hspace{-2pt}{\tiny $\pm$ 0.37} \\
$\;$ + SIPS (ours) & \color{green!50!black} +0.08 & \hspace{-2pt}{\tiny $\pm$ 0.21} & \color{green!50!black} +0.11 & \hspace{-2pt}{\tiny $\pm$ 0.21} & \color{green!50!black} +0.13 & \hspace{-2pt}{\tiny $\pm$ 0.22} & \color{green!50!black} \bfseries +0.20 & \hspace{-2pt}{\tiny $\pm$ 0.22} \\
\midrule\noalign{\vskip-\belowrulesep}\rowcolor{gray!15}\rule{0pt}{2.0ex}%
NCSN++~\cite{lemercier2023storm} & 2.25 &  & 2.71 &  & 2.98 &  & 3.37 &  \\
\trimgrayrow
$\;$ + StoRM~\cite{lemercier2023storm} & \color{green!50!black} \bfseries +0.13 & \hspace{-2pt}{\tiny $\pm$ 0.25} & \color{green!50!black} \bfseries +0.12 & \hspace{-2pt}{\tiny $\pm$ 0.30} & \color{green!50!black} +0.05 & \hspace{-2pt}{\tiny $\pm$ 0.33} & \color{red} -0.06 & \hspace{-2pt}{\tiny $\pm$ 0.34} \\
$\;$ + Diffiner~\cite{sawata2023diffiner} & +0.00 & \hspace{-2pt}{\tiny $\pm$ 0.29} & \color{green!50!black} +0.01 & \hspace{-2pt}{\tiny $\pm$ 0.33} & \color{green!50!black} +0.02 & \hspace{-2pt}{\tiny $\pm$ 0.38} & \color{red} -0.01 & \hspace{-2pt}{\tiny $\pm$ 0.34} \\
$\;$ + SIPS (ours) & \color{red} -0.04 & \hspace{-2pt}{\tiny $\pm$ 0.22} & \color{green!50!black} +0.03 &\hspace{-2pt}{\tiny $\pm$ 0.24} & \color{green!50!black} \bfseries +0.08 & \hspace{-2pt}{\tiny $\pm$ 0.33} & \color{green!50!black} \bfseries +0.15 & \hspace{-2pt}{\tiny $\pm$ 0.35} \\
\midrule\noalign{\vskip-\belowrulesep}\rowcolor{gray!15}\rule{0pt}{2.0ex}%
SEMamba~\cite{chao2024investigation} & \bfseries 3.10 &  & \bfseries 3.49 &  & \bfseries 3.70 &  & \bfseries 3.95 &  \\
\trimgrayrow
$\;$ + StoRM~\cite{lemercier2023storm} & \color{red} -0.36 & \hspace{-2pt}{\tiny $\pm$ 0.24} & \color{red} -0.37 & \hspace{-2pt}{\tiny $\pm$ 0.17} & \color{red} -0.40 & \hspace{-2pt}{\tiny $\pm$ 0.25} & \color{red} -0.44 & \hspace{-2pt}{\tiny $\pm$ 0.24} \\
$\;$ + Diffiner~\cite{sawata2023diffiner} & \color{red} -0.81 & \hspace{-2pt}{\tiny $\pm$ 0.38} & \color{red} -0.75 & \hspace{-2pt}{\tiny $\pm$ 0.37} & \color{red} -0.65 & \hspace{-2pt}{\tiny $\pm$ 0.35} & \color{red} -0.55 & \hspace{-2pt}{\tiny $\pm$ 0.32} \\
$\;$ + SIPS (ours) & \color{red} -0.21 & \hspace{-2pt}{\tiny $\pm$ 0.20} & \color{red} -0.16 & \hspace{-2pt}{\tiny $\pm$ 0.21} & \color{red} -0.09 & \hspace{-2pt}{\tiny $\pm$ 0.25} & \color{red} -0.04 & \hspace{-2pt}{\tiny $\pm$ 0.13} \\
\bottomrule
\end{tabular}
\vskip 2em
\begin{tabular}{lS[table-format=+2.2,detect-weight=true,retain-explicit-plus=true]@{\hspace{\tabcolsep}}lS[table-format=+2.2,detect-weight=true,retain-explicit-plus=true]@{\hspace{\tabcolsep}}lS[table-format=+2.2,detect-weight=true,retain-explicit-plus=true]@{\hspace{\tabcolsep}}lS[table-format=+2.2,detect-weight=true,retain-explicit-plus=true]@{\hspace{\tabcolsep}}l}
\toprule
& \multicolumn{8}{c}{\textbf{SI-SDR [dB]} $\uparrow$} \\
\cmidrule(lr){2-9}
& \multicolumn{2}{c}{2.5 dB} & \multicolumn{2}{c}{7.5 dB} & \multicolumn{2}{c}{12.5 dB} & \multicolumn{2}{c}{17.5 dB} \\ \midrule
Clean & \multicolumn{1}{c}{-} &  & \multicolumn{1}{c}{-} &  & \multicolumn{1}{c}{-} &  & \multicolumn{1}{c}{-} &  \\
Noisy & 0.95 &  & 6.03 &  & 10.92 &  & 15.97 &  \\
\midrule\noalign{\vskip-\belowrulesep}\rowcolor{gray!15}\rule{0pt}{2.0ex}%
Conv-TasNet~\cite{luo2019conv} & 15.24 &  & 17.86 &  & 19.63 &  & \bfseries 21.61 &  \\
\trimgrayrow
$\;$ + StoRM~\cite{lemercier2023storm} & \color{red} -0.25 & \hspace{-2pt}{\tiny $\pm$ 0.51} & \color{red} -0.49 & \hspace{-2pt}{\tiny $\pm$ 0.57} & \color{red} -0.77 & \hspace{-2pt}{\tiny $\pm$ 0.48} & \color{red} -1.10 & \hspace{-2pt}{\tiny $\pm$ 0.52} \\
$\;$ + Diffiner~\cite{sawata2023diffiner} & \color{red} -2.21 & \hspace{-2pt}{\tiny $\pm$ 1.42} & \color{red} -2.00 & \hspace{-2pt}{\tiny $\pm$ 1.32} & \color{red} -2.28 & \hspace{-2pt}{\tiny $\pm$ 1.03} & \color{red} -2.93 & \hspace{-2pt}{\tiny $\pm$ 0.89} \\
$\;$ + SIPS (ours) & \color{green!50!black} \bfseries +0.32 & \hspace{-2pt}{\tiny $\pm$ 0.45} & \color{green!50!black} \bfseries +0.38 & \hspace{-2pt}{\tiny $\pm$ 0.60} & \color{green!50!black} \bfseries +0.28 &\hspace{-2pt}{\tiny $\pm$ 0.52} & \color{red} -0.11 & \hspace{-2pt}{\tiny $\pm$ 0.77} \\
\midrule\noalign{\vskip-\belowrulesep}\rowcolor{gray!15}\rule{0pt}{2.0ex}%
NCSN++~\cite{lemercier2023storm} & \bfseries 16.51 &  & \bfseries 18.92 &  & \bfseries 20.45 &  & \bfseries 22.43 &  \\
\trimgrayrow
$\;$ + StoRM~\cite{lemercier2023storm} & \color{red} -0.95 & \hspace{-2pt}{\tiny $\pm$ 0.70} & \color{red} -1.07 & \hspace{-2pt}{\tiny $\pm$ 0.83} & \color{red} -1.09 & \hspace{-2pt}{\tiny $\pm$ 0.69} & \color{red} -1.22 & \hspace{-2pt}{\tiny $\pm$ 0.66} \\
$\;$ + Diffiner~\cite{sawata2023diffiner} & \color{red} -2.85 & \hspace{-2pt}{\tiny $\pm$ 1.98} & \color{red} -3.11 & \hspace{-2pt}{\tiny $\pm$ 2.01} & \color{red} -3.48 & \hspace{-2pt}{\tiny $\pm$ 2.39} & \color{red} -4.41 & \hspace{-2pt}{\tiny $\pm$ 2.60} \\
$\;$ + SIPS (ours) & \color{red} -0.10 & \hspace{-2pt}{\tiny $\pm$ 1.10} & \color{red} -0.24 & \hspace{-2pt}{\tiny $\pm$ 0.97} & \color{red} -0.32 & \hspace{-2pt}{\tiny $\pm$ 0.98} & \color{red} -0.81 & \hspace{-2pt}{\tiny $\pm$ 1.12} \\
\midrule\noalign{\vskip-\belowrulesep}\rowcolor{gray!15}\rule{0pt}{2.0ex}%
SEMamba~\cite{chao2024investigation} & \bfseries 16.98 &  & \bfseries 19.02 &  & 20.52 &  & \bfseries 22.38 &  \\
\trimgrayrow
$\;$ + StoRM~\cite{lemercier2023storm} & \color{red} -0.69 & \hspace{-2pt}{\tiny $\pm$ 0.52} & \color{red} -0.76 & \hspace{-2pt}{\tiny $\pm$ 0.58} & \color{red} -0.79 & \hspace{-2pt}{\tiny $\pm$ 0.54} & \color{red} -1.08 & \hspace{-2pt}{\tiny $\pm$ 0.75} \\
$\;$ + Diffiner~\cite{sawata2023diffiner} & \color{red} -3.34 & \hspace{-2pt}{\tiny $\pm$ 1.97} & \color{red} -3.01 & \hspace{-2pt}{\tiny $\pm$ 1.67} & \color{red} -2.97 & \hspace{-2pt}{\tiny $\pm$ 1.34} & \color{red} -3.50 & \hspace{-2pt}{\tiny $\pm$ 1.36} \\
$\;$ + SIPS (ours) & \color{red} -0.06 & \hspace{-2pt}{\tiny $\pm$ 0.81} & \color{red} -0.03 & \hspace{-2pt}{\tiny $\pm$ 0.75} & \color{green!50!black} \bfseries +0.02 & \hspace{-2pt}{\tiny $\pm$ 0.66} & \color{red} -0.26 & \hspace{-2pt}{\tiny $\pm$ 1.19} \\
\bottomrule
\end{tabular}
\endgroup
\vspace{-1em}
\end{table*}

\begin{table*}[t]\centering
\caption{\textbf{Results per input SNR (DNSMOS [P.808] and NISQA)}. Results on the matched VoiceBank-DEMAND test set for non-intrusive quality metrics, segmented into the four exact input-SNR levels specified in the column headers. Clean, noisy, and predictor entries show absolute means. Refiner entries show the mean change from their predictor, with the population standard deviation of paired per-utterance changes in smaller text. Changes are computed from unrounded scores and rounded only for display. Green changes indicate improvement and red changes indicate degradation. Best absolute means within each predictor group and SNR level, including the predictor, are bold. Predictors have a gray background.}
\label{tab:vbdmd_per_snr_2}
\begingroup
\setlength{\tabcolsep}{3.5pt}
\begin{tabular}{lS[table-format=+1.2,detect-weight=true,retain-explicit-plus=true]@{\hspace{\tabcolsep}}lS[table-format=+1.2,detect-weight=true,retain-explicit-plus=true]@{\hspace{\tabcolsep}}lS[table-format=+1.2,detect-weight=true,retain-explicit-plus=true]@{\hspace{\tabcolsep}}lS[table-format=+1.2,detect-weight=true,retain-explicit-plus=true]@{\hspace{\tabcolsep}}l}
\toprule
& \multicolumn{8}{c}{\textbf{DNSMOS [P.808]} $\uparrow$} \\
\cmidrule(lr){2-9}
& \multicolumn{2}{c}{2.5 dB} & \multicolumn{2}{c}{7.5 dB} & \multicolumn{2}{c}{12.5 dB} & \multicolumn{2}{c}{17.5 dB} \\ \midrule
Clean & 3.54 &  & 3.55 &  & 3.57 &  & 3.55 &  \\
Noisy & 2.81 &  & 3.03 &  & 3.19 &  & 3.33 &  \\
\midrule\noalign{\vskip-\belowrulesep}\rowcolor{gray!15}\rule{0pt}{2.0ex}%
Conv-TasNet~\cite{luo2019conv} & 3.12 &  & 3.30 &  & 3.40 &  & 3.44 &  \\
\trimgrayrow
$\;$ + StoRM~\cite{lemercier2023storm} & \color{green!50!black} \bfseries +0.39 & \hspace{-2pt}{\tiny $\pm$ 0.22} & \color{green!50!black} \bfseries +0.25 & \hspace{-2pt}{\tiny $\pm$ 0.17} & \color{green!50!black} \bfseries +0.16 & \hspace{-2pt}{\tiny $\pm$ 0.16} & \color{green!50!black} \bfseries +0.11 & \hspace{-2pt}{\tiny $\pm$ 0.13} \\
$\;$ + Diffiner~\cite{sawata2023diffiner} & \color{green!50!black} +0.26 & \hspace{-2pt}{\tiny $\pm$ 0.19} & \color{green!50!black} +0.17 & \hspace{-2pt}{\tiny $\pm$ 0.16} & \color{green!50!black} +0.11 & \hspace{-2pt}{\tiny $\pm$ 0.14} & \color{green!50!black} +0.09 & \hspace{-2pt}{\tiny $\pm$ 0.12} \\
$\;$ + SIPS (ours) & \color{green!50!black} +0.17 & \hspace{-2pt}{\tiny $\pm$ 0.17} & \color{green!50!black} +0.13 & \hspace{-2pt}{\tiny $\pm$ 0.15} & \color{green!50!black} +0.10 & \hspace{-2pt}{\tiny $\pm$ 0.13} & \color{green!50!black} +0.06 & \hspace{-2pt}{\tiny $\pm$ 0.13} \\
\midrule\noalign{\vskip-\belowrulesep}\rowcolor{gray!15}\rule{0pt}{2.0ex}%
NCSN++~\cite{lemercier2023storm} & 3.53 &  & \bfseries 3.60 &  & \bfseries 3.61 &  & \bfseries 3.60 &  \\
\trimgrayrow
$\;$ + StoRM~\cite{lemercier2023storm} & \color{green!50!black} \bfseries +0.01 & \hspace{-2pt}{\tiny $\pm$ 0.14} & \color{red} -0.03 & \hspace{-2pt}{\tiny $\pm$ 0.11} & \color{red} -0.04 & \hspace{-2pt}{\tiny $\pm$ 0.08} & \color{red} -0.06 & \hspace{-2pt}{\tiny $\pm$ 0.06} \\
$\;$ + Diffiner~\cite{sawata2023diffiner} & \color{red} -0.08 & \hspace{-2pt}{\tiny $\pm$ 0.14} & \color{red} -0.08 & \hspace{-2pt}{\tiny $\pm$ 0.12} & \color{red} -0.06 & \hspace{-2pt}{\tiny $\pm$ 0.10} & \color{red} -0.04 & \hspace{-2pt}{\tiny $\pm$ 0.10} \\
$\;$ + SIPS (ours) & \color{red} -0.06 & \hspace{-2pt}{\tiny $\pm$ 0.14} & \color{red} -0.04 & \hspace{-2pt}{\tiny $\pm$ 0.12} & \color{red} -0.01 & \hspace{-2pt}{\tiny $\pm$ 0.12} & \color{red} -0.02 & \hspace{-2pt}{\tiny $\pm$ 0.13} \\
\midrule\noalign{\vskip-\belowrulesep}\rowcolor{gray!15}\rule{0pt}{2.0ex}%
SEMamba~\cite{chao2024investigation} & \bfseries 3.54 &  & \bfseries 3.60 &  & \bfseries 3.61 &  & \bfseries 3.59 &  \\
\trimgrayrow
$\;$ + StoRM~\cite{lemercier2023storm} & \color{red} -0.02 & \hspace{-2pt}{\tiny $\pm$ 0.07} & \color{red} -0.02 & \hspace{-2pt}{\tiny $\pm$ 0.05} & \color{red} -0.02 & \hspace{-2pt}{\tiny $\pm$ 0.05} & \color{red} -0.03 & \hspace{-2pt}{\tiny $\pm$ 0.05} \\
$\;$ + Diffiner~\cite{sawata2023diffiner} & \color{red} -0.09 & \hspace{-2pt}{\tiny $\pm$ 0.15} & \color{red} -0.06 & \hspace{-2pt}{\tiny $\pm$ 0.12} & \color{red} -0.06 & \hspace{-2pt}{\tiny $\pm$ 0.10} & \color{red} -0.03 & \hspace{-2pt}{\tiny $\pm$ 0.10} \\
$\;$ + SIPS (ours) & \color{red} -0.02 & \hspace{-2pt}{\tiny $\pm$ 0.12} & \color{red} -0.02 & \hspace{-2pt}{\tiny $\pm$ 0.12} & \bfseries +0.00 & \hspace{-2pt}{\tiny $\pm$ 0.10} & \color{red} -0.01 & \hspace{-2pt}{\tiny $\pm$ 0.10} \\
\bottomrule
\end{tabular}
\vskip 2em
\begin{tabular}{lS[table-format=+1.2,detect-weight=true,retain-explicit-plus=true]@{\hspace{\tabcolsep}}lS[table-format=+1.2,detect-weight=true,retain-explicit-plus=true]@{\hspace{\tabcolsep}}lS[table-format=+1.2,detect-weight=true,retain-explicit-plus=true]@{\hspace{\tabcolsep}}lS[table-format=+1.2,detect-weight=true,retain-explicit-plus=true]@{\hspace{\tabcolsep}}l}
\toprule
& \multicolumn{8}{c}{\textbf{NISQA} $\uparrow$} \\
\cmidrule(lr){2-9}
& \multicolumn{2}{c}{2.5 dB} & \multicolumn{2}{c}{7.5 dB} & \multicolumn{2}{c}{12.5 dB} & \multicolumn{2}{c}{17.5 dB} \\ \midrule
Clean & 4.50 &  & 4.53 &  & 4.46 &  & 4.53 &  \\
Noisy & 2.29 &  & 2.91 &  & 3.28 &  & 3.65 &  \\
\midrule\noalign{\vskip-\belowrulesep}\rowcolor{gray!15}\rule{0pt}{2.0ex}%
Conv-TasNet~\cite{luo2019conv} & 2.74 &  & 3.41 &  & 3.69 &  & 3.94 &  \\
\trimgrayrow
$\;$ + StoRM~\cite{lemercier2023storm} & \color{green!50!black} +1.89 & \hspace{-2pt}{\tiny $\pm$ 0.70} &\color{green!50!black} +1.24 & \hspace{-2pt}{\tiny $\pm$ 0.63} & \color{green!50!black} +0.82 & \hspace{-2pt}{\tiny $\pm$ 0.60} & \color{green!50!black} +0.55 & \hspace{-2pt}{\tiny $\pm$ 0.49} \\
$\;$ + Diffiner~\cite{sawata2023diffiner} & \color{green!50!black} \bfseries +1.92 & \hspace{-2pt}{\tiny $\pm$ 0.61} & \color{green!50!black} \bfseries +1.37 & \hspace{-2pt}{\tiny $\pm$ 0.52} & \color{green!50!black} \bfseries +1.07 & \hspace{-2pt}{\tiny $\pm$ 0.51} & \color{green!50!black} \bfseries +0.83 & \hspace{-2pt}{\tiny $\pm$ 0.47} \\
$\;$ + SIPS (ours) & \color{green!50!black} +1.18 & \hspace{-2pt}{\tiny $\pm$ 0.55} & \color{green!50!black} +0.89 & \hspace{-2pt}{\tiny $\pm$ 0.47} & \color{green!50!black} +0.70 & \hspace{-2pt}{\tiny $\pm$ 0.35} & \color{green!50!black} +0.54 & \hspace{-2pt}{\tiny $\pm$ 0.35} \\
\midrule\noalign{\vskip-\belowrulesep}\rowcolor{gray!15}\rule{0pt}{2.0ex}%
NCSN++~\cite{lemercier2023storm} & 4.66 &  & 4.70 &  & 4.61 &  & 4.61 &  \\
\trimgrayrow
$\;$ + StoRM~\cite{lemercier2023storm} & \color{red} -0.03 & \hspace{-2pt}{\tiny $\pm$ 0.49} & \color{red} -0.10 & \hspace{-2pt}{\tiny $\pm$ 0.35} & \color{red} -0.18 & \hspace{-2pt}{\tiny $\pm$ 0.24} & \color{red} -0.18 & \hspace{-2pt}{\tiny $\pm$ 0.20} \\
$\;$ + Diffiner~\cite{sawata2023diffiner} & \color{green!50!black} \bfseries +0.14 & \hspace{-2pt}{\tiny $\pm$ 0.43} & \color{green!50!black} \bfseries +0.15 & \hspace{-2pt}{\tiny $\pm$ 0.41} & \color{green!50!black} \bfseries +0.16 & \hspace{-2pt}{\tiny $\pm$ 0.31} & \color{green!50!black} \bfseries +0.17 & \hspace{-2pt}{\tiny $\pm$ 0.26} \\
$\;$ + SIPS (ours) & \color{green!50!black} +0.02 & \hspace{-2pt}{\tiny $\pm$ 0.35} & \color{green!50!black} +0.05 & \hspace{-2pt}{\tiny $\pm$ 0.37} & \color{green!50!black} +0.10 & \hspace{-2pt}{\tiny $\pm$ 0.25} & \color{green!50!black} +0.13 & \hspace{-2pt}{\tiny $\pm$ 0.26} \\
\midrule\noalign{\vskip-\belowrulesep}\rowcolor{gray!15}\rule{0pt}{2.0ex}%
SEMamba~\cite{chao2024investigation} & 4.50 &  & 4.67 &  & 4.60 &  & 4.63 &  \\
\trimgrayrow
$\;$ + StoRM~\cite{lemercier2023storm} & \color{green!50!black} +0.05 & \hspace{-2pt}{\tiny $\pm$ 0.35} &\color{red} -0.10 & \hspace{-2pt}{\tiny $\pm$ 0.23} & \color{red} -0.10 & \hspace{-2pt}{\tiny $\pm$ 0.18}& \color{red} -0.11 & \hspace{-2pt}{\tiny $\pm$ 0.15} \\
$\;$ + Diffiner~\cite{sawata2023diffiner} & \color{green!50!black} \bfseries +0.30 & \hspace{-2pt}{\tiny $\pm$ 0.42} & \color{green!50!black} \bfseries +0.18 & \hspace{-2pt}{\tiny $\pm$ 0.25} & \color{green!50!black} \bfseries +0.18 & \hspace{-2pt}{\tiny $\pm$ 0.26} & \color{green!50!black} \bfseries +0.17 & \hspace{-2pt}{\tiny $\pm$ 0.19} \\
$\;$ + SIPS (ours) & \color{green!50!black} +0.23 & \hspace{-2pt}{\tiny $\pm$ 0.37} & \color{green!50!black} +0.13 & \hspace{-2pt}{\tiny $\pm$ 0.20} & \color{green!50!black} +0.09 & \hspace{-2pt}{\tiny $\pm$ 0.20} & \color{green!50!black} +0.08 & \hspace{-2pt}{\tiny $\pm$ 0.13} \\
\bottomrule
\end{tabular}
\endgroup
\vspace{-1em}
\end{table*}

\begin{table*}[t]\centering
\caption{\textbf{Results per input SNR (UTMOS and WER [\%])}. Results on the matched VoiceBank-DEMAND test set for speech quality and recognition metrics, segmented into the four exact input-SNR levels specified in the column headers. Clean, noisy, and predictor entries show absolute means. Refiner entries show the mean change from their predictor, with the population standard deviation of paired per-utterance changes in smaller text. Changes are computed from unrounded scores and rounded only for display. Green changes indicate improvement and red changes indicate degradation. Best absolute means within each predictor group and SNR level, including the predictor, are bold. Predictors have a gray background. WER means are corpus-level within each SNR level. Changes are in percentage points and their standard deviations use paired per-utterance WER differences.}
\label{tab:vbdmd_per_snr_3}
\begingroup
\setlength{\tabcolsep}{3.5pt}
\begin{tabular}{lS[table-format=+1.2,detect-weight=true,retain-explicit-plus=true]@{\hspace{\tabcolsep}}lS[table-format=+1.2,detect-weight=true,retain-explicit-plus=true]@{\hspace{\tabcolsep}}lS[table-format=+1.2,detect-weight=true,retain-explicit-plus=true]@{\hspace{\tabcolsep}}lS[table-format=+1.2,detect-weight=true,retain-explicit-plus=true]@{\hspace{\tabcolsep}}l}
\toprule
& \multicolumn{8}{c}{\textbf{UTMOS} $\uparrow$} \\
\cmidrule(lr){2-9}
& \multicolumn{2}{c}{2.5 dB} & \multicolumn{2}{c}{7.5 dB} & \multicolumn{2}{c}{12.5 dB} & \multicolumn{2}{c}{17.5 dB} \\ \midrule
Clean & 4.08 &  & 4.10 &  & 4.09 &  & 4.10 &  \\
Noisy & 2.37 &  & 3.00 &  & 3.40 &  & 3.65 &  \\
\midrule\noalign{\vskip-\belowrulesep}\rowcolor{gray!15}\rule{0pt}{2.0ex}%
Conv-TasNet~\cite{luo2019conv} & 3.19 &  & 3.59 &  & 3.77 &  & 3.90 &  \\
\trimgrayrow
$\;$ + StoRM~\cite{lemercier2023storm} & \color{green!50!black} +0.57 & \hspace{-2pt}{\tiny $\pm$ 0.33} &\color{green!50!black} +0.34 & \hspace{-2pt}{\tiny $\pm$ 0.24} & \color{green!50!black} +0.20 & \hspace{-2pt}{\tiny $\pm$ 0.18} & \color{green!50!black} +0.11 & \hspace{-2pt}{\tiny $\pm$ 0.15} \\
$\;$ + Diffiner~\cite{sawata2023diffiner} & \color{green!50!black} \bfseries +0.61 & \hspace{-2pt}{\tiny $\pm$ 0.38} & \color{green!50!black} \bfseries +0.37 & \hspace{-2pt}{\tiny $\pm$ 0.22} & \color{green!50!black} \bfseries +0.25 & \hspace{-2pt}{\tiny $\pm$ 0.16} & \color{green!50!black} \bfseries +0.18 & \hspace{-2pt}{\tiny $\pm$ 0.15} \\
$\;$ + SIPS (ours) & \color{green!50!black} +0.36 & \hspace{-2pt}{\tiny $\pm$ 0.20} & \color{green!50!black} +0.25 & \hspace{-2pt}{\tiny $\pm$ 0.14} & \color{green!50!black} +0.18 & \hspace{-2pt}{\tiny $\pm$ 0.11} & \color{green!50!black} +0.15 & \hspace{-2pt}{\tiny $\pm$ 0.09} \\
\midrule\noalign{\vskip-\belowrulesep}\rowcolor{gray!15}\rule{0pt}{2.0ex}%
NCSN++~\cite{lemercier2023storm} & 3.76 &  & 3.94 &  & 4.00 &  & 4.05 &  \\
\trimgrayrow
$\;$ + StoRM~\cite{lemercier2023storm} & \color{green!50!black} +0.04 & \hspace{-2pt}{\tiny $\pm$ 0.14} &+0.00 & \hspace{-2pt}{\tiny $\pm$ 0.10} & \color{red} -0.03 & \hspace{-2pt}{\tiny $\pm$ 0.09} & \color{red} -0.05 & \hspace{-2pt}{\tiny $\pm$ 0.08} \\
$\;$ + Diffiner~\cite{sawata2023diffiner} & \color{green!50!black} \bfseries +0.14 & \hspace{-2pt}{\tiny $\pm$ 0.19} & \color{green!50!black} \bfseries +0.08 & \hspace{-2pt}{\tiny $\pm$ 0.12} & \color{green!50!black} \bfseries +0.06 & \hspace{-2pt}{\tiny $\pm$ 0.09} & \color{green!50!black} \bfseries +0.05 & \hspace{-2pt}{\tiny $\pm$ 0.09} \\
$\;$ + SIPS (ours) & \color{green!50!black} +0.03 & \hspace{-2pt}{\tiny $\pm$ 0.09} & \color{green!50!black} +0.03 & \hspace{-2pt}{\tiny $\pm$ 0.08} & \color{green!50!black} +0.05 & \hspace{-2pt}{\tiny $\pm$ 0.08} & \color{green!50!black} \bfseries +0.05 & \hspace{-2pt}{\tiny $\pm$ 0.06} \\
\midrule\noalign{\vskip-\belowrulesep}\rowcolor{gray!15}\rule{0pt}{2.0ex}%
SEMamba~\cite{chao2024investigation} & 4.00 &  & 4.08 &  & 4.10 &  & 4.11 &  \\
\trimgrayrow
$\;$ + StoRM~\cite{lemercier2023storm} & \color{red} -0.07 & \hspace{-2pt}{\tiny $\pm$ 0.08} & \color{red} -0.06 & \hspace{-2pt}{\tiny $\pm$ 0.06} & \color{red} -0.06 & \hspace{-2pt}{\tiny $\pm$ 0.06} & \color{red} -0.05 & \hspace{-2pt}{\tiny $\pm$ 0.05} \\
$\;$ + Diffiner~\cite{sawata2023diffiner} & \color{red} -0.08 & \hspace{-2pt}{\tiny $\pm$ 0.20} & \color{red} -0.05 & \hspace{-2pt}{\tiny $\pm$ 0.12} & \color{red} -0.02 & \hspace{-2pt}{\tiny $\pm$ 0.09} & \color{green!50!black} +0.01 & \hspace{-2pt}{\tiny $\pm$ 0.07} \\
$\;$ + SIPS (ours) & \color{green!50!black} \bfseries +0.02 & \hspace{-2pt}{\tiny $\pm$ 0.08} & \color{green!50!black} \bfseries +0.01 & \hspace{-2pt}{\tiny $\pm$ 0.05} & \color{green!50!black} \bfseries +0.02 &\hspace{-2pt}{\tiny $\pm$ 0.04} & \color{green!50!black} \bfseries +0.03 & \hspace{-2pt}{\tiny $\pm$ 0.04} \\
\bottomrule
\end{tabular}
\vskip 2em
\begin{tabular}{lS[table-format=+2.2,detect-weight=true,retain-explicit-plus=true]@{\hspace{\tabcolsep}}lS[table-format=+2.2,detect-weight=true,retain-explicit-plus=true]@{\hspace{\tabcolsep}}lS[table-format=+2.2,detect-weight=true,retain-explicit-plus=true]@{\hspace{\tabcolsep}}lS[table-format=+2.2,detect-weight=true,retain-explicit-plus=true]@{\hspace{\tabcolsep}}l}
\toprule
& \multicolumn{8}{c}{\textbf{WER [\%]} $\downarrow$} \\
\cmidrule(lr){2-9}
& \multicolumn{2}{c}{2.5 dB} & \multicolumn{2}{c}{7.5 dB} & \multicolumn{2}{c}{12.5 dB} & \multicolumn{2}{c}{17.5 dB} \\ \midrule
Clean & 7.87 &  & 6.54 &  & 6.81 &  & 6.59 &  \\
Noisy & 17.80 &  & 10.42 &  & 8.04 &  & 7.70 &  \\
\midrule\noalign{\vskip-\belowrulesep}\rowcolor{gray!15}\rule{0pt}{2.0ex}%
Conv-TasNet~\cite{luo2019conv} & \bfseries 19.91 &  & \bfseries 11.37 &  & \bfseries 8.56 &  & \bfseries 7.50 &  \\
\trimgrayrow
$\;$ + StoRM~\cite{lemercier2023storm} & \color{red} +3.97 & \hspace{-2pt}{\tiny $\pm$ \tablenum[table-format=2.2]{16.00}} & \color{red} +1.27 & \hspace{-2pt}{\tiny $\pm$ \tablenum[table-format=2.2]{11.19}} & \color{red} +0.32 & \hspace{-2pt}{\tiny $\pm$ \tablenum[table-format=2.2]{8.77}} & \color{red} +0.20 & \hspace{-2pt}{\tiny $\pm$ \tablenum[table-format=2.2]{7.67}} \\
$\;$ + Diffiner~\cite{sawata2023diffiner} & \color{red} +10.18 & \hspace{-2pt}{\tiny $\pm$ \tablenum[table-format=2.2]{22.86}} & \color{red} +3.94 & \hspace{-2pt}{\tiny $\pm$ \tablenum[table-format=2.2]{15.72}} & \color{red} +1.36 & \hspace{-2pt}{\tiny $\pm$ \tablenum[table-format=2.2]{10.80}} & \color{red} +0.91 & \hspace{-2pt}{\tiny $\pm$ \tablenum[table-format=2.2]{9.50}} \\
$\;$ + SIPS (ours) & \color{red} +1.92 & \hspace{-2pt}{\tiny $\pm$ \tablenum[table-format=2.2]{13.79}} & \color{red} +1.40 & \hspace{-2pt}{\tiny $\pm$ \tablenum[table-format=2.2]{9.75}} & \color{red} +0.58 & \hspace{-2pt}{\tiny $\pm$ \tablenum[table-format=2.2]{9.63}} & \color{red} +0.07 & \hspace{-2pt}{\tiny $\pm$ \tablenum[table-format=2.2]{7.61}} \\
\midrule\noalign{\vskip-\belowrulesep}\rowcolor{gray!15}\rule{0pt}{2.0ex}%
NCSN++~\cite{lemercier2023storm} & \bfseries 14.79 &  & 9.34 &  & \bfseries 6.87 &  & 7.24 &  \\
\trimgrayrow
$\;$ + StoRM~\cite{lemercier2023storm} & \color{red} +2.18 & \hspace{-2pt}{\tiny $\pm$ \tablenum[table-format=2.2]{13.89}} & \color{green!50!black} -0.06 & \hspace{-2pt}{\tiny $\pm$ \tablenum[table-format=2.2]{7.95}} & \color{red} +1.04 & \hspace{-2pt}{\tiny $\pm$ \tablenum[table-format=2.2]{7.05}} & \color{green!50!black} \bfseries -0.65 & \hspace{-2pt}{\tiny $\pm$ \tablenum[table-format=2.2]{6.53}} \\
$\;$ + Diffiner~\cite{sawata2023diffiner} & \color{red} +8.96 & \hspace{-2pt}{\tiny $\pm$ \tablenum[table-format=2.2]{19.11}} & \color{red} +2.29 & \hspace{-2pt}{\tiny $\pm$ \tablenum[table-format=2.2]{12.50}} &\color{red} +2.01 & \hspace{-2pt}{\tiny $\pm$ \tablenum[table-format=2.2]{9.51}} & \color{red} +0.46 & \hspace{-2pt}{\tiny $\pm$ \tablenum[table-format=2.2]{7.51}} \\
$\;$ + SIPS (ours) & \color{red} +0.90 & \hspace{-2pt}{\tiny $\pm$ \tablenum[table-format=2.2]{11.25}} & \color{green!50!black} \bfseries -0.25 & \hspace{-2pt}{\tiny $\pm$ \tablenum[table-format=2.2]{8.39}} & \color{red} +1.10 & \hspace{-2pt}{\tiny $\pm$ \tablenum[table-format=2.2]{6.81}} & \color{green!50!black} -0.07 & \hspace{-2pt}{\tiny $\pm$ \tablenum[table-format=2.2]{7.20}} \\
\midrule\noalign{\vskip-\belowrulesep}\rowcolor{gray!15}\rule{0pt}{2.0ex}%
SEMamba~\cite{chao2024investigation} & 13.00 &  & 8.32 &  & \bfseries 7.13 &  & \bfseries 6.98 &  \\
\trimgrayrow
$\;$ + StoRM~\cite{lemercier2023storm} & \color{red} +1.15 & \hspace{-2pt}{\tiny $\pm$ \tablenum[table-format=2.2]{9.08}} & \color{red} +0.38 & \hspace{-2pt}{\tiny $\pm$ \tablenum[table-format=2.2]{6.28}} & \color{red} +0.32 & \hspace{-2pt}{\tiny $\pm$ \tablenum[table-format=2.2]{5.13}} & \color{red} +0.13 & \hspace{-2pt}{\tiny $\pm$ \tablenum[table-format=2.2]{4.27}} \\
$\;$ + Diffiner~\cite{sawata2023diffiner} & \color{red} +9.15 & \hspace{-2pt}{\tiny $\pm$ \tablenum[table-format=2.2]{17.35}} & \color{red} +3.62 & \hspace{-2pt}{\tiny $\pm$ \tablenum[table-format=2.2]{12.78}} &\color{red} +1.10 & \hspace{-2pt}{\tiny $\pm$ \tablenum[table-format=2.2]{8.49}} & \color{red} +0.26 & \hspace{-2pt}{\tiny $\pm$ \tablenum[table-format=2.2]{9.28}} \\
$\;$ + SIPS (ours) & \color{green!50!black} \bfseries -0.13 & \hspace{-2pt}{\tiny $\pm$ \tablenum[table-format=2.2]{12.33}} & \color{green!50!black} \bfseries -0.32 & \hspace{-2pt}{\tiny $\pm$ \tablenum[table-format=2.2]{6.00}} & \color{red} +0.06 & \hspace{-2pt}{\tiny $\pm$ \tablenum[table-format=2.2]{5.21}} & \color{red} +0.13 & \hspace{-2pt}{\tiny $\pm$ \tablenum[table-format=2.2]{4.61}} \\
\bottomrule
\end{tabular}
\endgroup
\vspace{-1em}
\end{table*}

\begin{table*}[t]\centering
\caption{\textbf{Results per input SNR (PESQ and SI-SDR [dB])}. Results on the EARS-WHAM (v2) test set for quality metrics, segmented into the four input-SNR bins specified in the column headers. Clean, noisy, and predictor entries show absolute means. Refiner entries show the mean change from their predictor, with the population standard deviation of paired per-utterance changes in smaller text. Changes are computed from unrounded scores and rounded only for display. Green changes indicate improvement and red changes indicate degradation. Best absolute means within each predictor group and SNR bin, including the predictor, are bold. Predictors have a gray background.}
\label{tab:earswham_per_snr_1}
\begingroup
\setlength{\tabcolsep}{3.5pt}
\begin{tabular}{lS[table-format=+1.2,detect-weight=true,retain-explicit-plus=true]@{\hspace{\tabcolsep}}lS[table-format=+1.2,detect-weight=true,retain-explicit-plus=true]@{\hspace{\tabcolsep}}lS[table-format=+1.2,detect-weight=true,retain-explicit-plus=true]@{\hspace{\tabcolsep}}lS[table-format=+1.2,detect-weight=true,retain-explicit-plus=true]@{\hspace{\tabcolsep}}l}
\toprule
& \multicolumn{8}{c}{\textbf{PESQ} $\uparrow$} \\
\cmidrule(lr){2-9}
& \multicolumn{2}{c}{[-2.5, 2.5] dB} & \multicolumn{2}{c}{(2.5, 7.5] dB} & \multicolumn{2}{c}{(7.5, 12.5] dB} & \multicolumn{2}{c}{(12.5, 17.5] dB} \\ \midrule
Clean & 4.64 &  & 4.64 &  & 4.64 &  & 4.64 &  \\
Noisy & 1.07 &  & 1.11 &  & 1.25 &  & 1.52 &  \\
\midrule\noalign{\vskip-\belowrulesep}\rowcolor{gray!15}\rule{0pt}{2.0ex}%
Conv-TasNet~\cite{luo2019conv} & \bfseries 1.11 &  & \bfseries 1.20 &  & \bfseries 1.42 &  & \bfseries 1.70 &  \\
\trimgrayrow
$\;$ + StoRM~\cite{lemercier2023storm} & \color{red} -0.01 & \hspace{-2pt}{\tiny $\pm$ 0.09} & \color{red} -0.05 & \hspace{-2pt}{\tiny $\pm$ 0.06} & \color{red} -0.09 & \hspace{-2pt}{\tiny $\pm$ 0.11} & \color{red} -0.10 & \hspace{-2pt}{\tiny $\pm$ 0.19} \\
$\;$ + Diffiner~\cite{sawata2023diffiner} & \bfseries +0.00 & \hspace{-2pt}{\tiny $\pm$ 0.24} & \color{red} -0.11 & \hspace{-2pt}{\tiny $\pm$ 0.11} & \color{red} -0.26 & \hspace{-2pt}{\tiny $\pm$ 0.18} & \color{red} -0.43 & \hspace{-2pt}{\tiny $\pm$ 0.23} \\
$\;$ + SIPS (ours) & \color{red} -0.01 & \hspace{-2pt}{\tiny $\pm$ 0.03} & \color{red} -0.02 & \hspace{-2pt}{\tiny $\pm$ 0.03} & \color{red} -0.04 & \hspace{-2pt}{\tiny $\pm$ 0.04} & \color{red} -0.05 & \hspace{-2pt}{\tiny $\pm$ 0.06} \\
\midrule\noalign{\vskip-\belowrulesep}\rowcolor{gray!15}\rule{0pt}{2.0ex}%
NCSN++~\cite{lemercier2023storm} & 1.22 &  & 1.52 &  & 2.03 &  & 2.53 &  \\
\trimgrayrow
$\;$ + StoRM~\cite{lemercier2023storm} & \color{green!50!black} \bfseries +0.06 & \hspace{-2pt}{\tiny $\pm$ 0.05} & \color{green!50!black} \bfseries +0.10 & \hspace{-2pt}{\tiny $\pm$ 0.08} & \color{green!50!black} \bfseries +0.11 & \hspace{-2pt}{\tiny $\pm$ 0.12} & \color{green!50!black} \bfseries +0.07 & \hspace{-2pt}{\tiny $\pm$ 0.12} \\
$\;$ + Diffiner~\cite{sawata2023diffiner} & \color{red} -0.08 & \hspace{-2pt}{\tiny $\pm$ 0.10} & \color{red} -0.28 & \hspace{-2pt}{\tiny $\pm$ 0.21} & \color{red} -0.65 & \hspace{-2pt}{\tiny $\pm$ 0.33} & \color{red} -1.07 & \hspace{-2pt}{\tiny $\pm$ 0.40} \\
$\;$ + SIPS (ours) & \color{red} -0.02 & \hspace{-2pt}{\tiny $\pm$ 0.04} & \color{red} -0.07 & \hspace{-2pt}{\tiny $\pm$ 0.08} & \color{red} -0.12 & \hspace{-2pt}{\tiny $\pm$ 0.13} & \color{red} -0.14 & \hspace{-2pt}{\tiny $\pm$ 0.13} \\
\midrule\noalign{\vskip-\belowrulesep}\rowcolor{gray!15}\rule{0pt}{2.0ex}%
SEMamba~\cite{chao2024investigation} & \bfseries 1.44 &  & \bfseries 1.91 &  & \bfseries 2.50 &  & \bfseries 2.95 &  \\
\trimgrayrow
$\;$ + StoRM~\cite{lemercier2023storm} & \color{red} -0.09 & \hspace{-2pt}{\tiny $\pm$ 0.08} & \color{red} -0.16 & \hspace{-2pt}{\tiny $\pm$ 0.12} & \color{red} -0.23 & \hspace{-2pt}{\tiny $\pm$ 0.13} & \color{red} -0.26 & \hspace{-2pt}{\tiny $\pm$ 0.13} \\
$\;$ + Diffiner~\cite{sawata2023diffiner} & \color{red} -0.30 & \hspace{-2pt}{\tiny $\pm$ 0.23} & \color{red} -0.67 & \hspace{-2pt}{\tiny $\pm$ 0.30} & \color{red} -0.98 & \hspace{-2pt}{\tiny $\pm$ 0.30} & \color{red} -1.11 & \hspace{-2pt}{\tiny $\pm$ 0.25} \\
$\;$ + SIPS (ours) & \color{red} -0.06 & \hspace{-2pt}{\tiny $\pm$ 0.05} & \color{red} -0.13 & \hspace{-2pt}{\tiny $\pm$ 0.07} & \color{red} -0.18 & \hspace{-2pt}{\tiny $\pm$ 0.08} & \color{red} -0.17 & \hspace{-2pt}{\tiny $\pm$ 0.07} \\
\bottomrule
\end{tabular}
\vskip 2em
\begin{tabular}{lS[table-format=+2.2,detect-weight=true,retain-explicit-plus=true]@{\hspace{\tabcolsep}}lS[table-format=+2.2,detect-weight=true,retain-explicit-plus=true]@{\hspace{\tabcolsep}}lS[table-format=+2.2,detect-weight=true,retain-explicit-plus=true]@{\hspace{\tabcolsep}}lS[table-format=+2.2,detect-weight=true,retain-explicit-plus=true]@{\hspace{\tabcolsep}}l}
\toprule
& \multicolumn{8}{c}{\textbf{SI-SDR [dB]} $\uparrow$} \\
\cmidrule(lr){2-9}
& \multicolumn{2}{c}{[-2.5, 2.5] dB} & \multicolumn{2}{c}{(2.5, 7.5] dB} & \multicolumn{2}{c}{(7.5, 12.5] dB} & \multicolumn{2}{c}{(12.5, 17.5] dB} \\ \midrule
Clean & \multicolumn{1}{c}{-} &  & \multicolumn{1}{c}{-} &  & \multicolumn{1}{c}{-} &  & \multicolumn{1}{c}{-} &  \\
Noisy & -1.99 &  & 3.02 &  & 7.96 &  & 12.91 &  \\
\midrule\noalign{\vskip-\belowrulesep}\rowcolor{gray!15}\rule{0pt}{2.0ex}%
Conv-TasNet~\cite{luo2019conv} & \bfseries -1.74 &  & \bfseries 3.18 &  & \bfseries 6.10 &  & 8.58 &  \\
\trimgrayrow
$\;$ + StoRM~\cite{lemercier2023storm} & \color{red} -0.67 & \hspace{-2pt}{\tiny $\pm$ 0.81} & \color{red} -0.70 & \hspace{-2pt}{\tiny $\pm$ 0.65} & \color{red} -0.71 & \hspace{-2pt}{\tiny $\pm$ 0.78} & \color{red} -0.76 & \hspace{-2pt}{\tiny $\pm$ 1.13} \\
$\;$ + Diffiner~\cite{sawata2023diffiner} & \color{red} -6.78 & \hspace{-2pt}{\tiny $\pm$ 5.81} & \color{red} -5.94 & \hspace{-2pt}{\tiny $\pm$ 2.44} & \color{red} -6.07 & \hspace{-2pt}{\tiny $\pm$ 1.96} & \color{red} -6.23 & \hspace{-2pt}{\tiny $\pm$ 1.51} \\
$\;$ + SIPS (ours) & \color{red} -0.21 & \hspace{-2pt}{\tiny $\pm$ 0.29} & \color{red} -0.13 & \hspace{-2pt}{\tiny $\pm$ 0.20} & \color{red} -0.04 & \hspace{-2pt}{\tiny $\pm$ 0.28} & \color{green!50!black} \bfseries +0.03 & \hspace{-2pt}{\tiny $\pm$ 0.25} \\
\midrule\noalign{\vskip-\belowrulesep}\rowcolor{gray!15}\rule{0pt}{2.0ex}%
NCSN++~\cite{lemercier2023storm} & \bfseries 6.38 &  & \bfseries 11.65 &  & \bfseries 16.00 &  & \bfseries 19.33 &  \\
\trimgrayrow
$\;$ + StoRM~\cite{lemercier2023storm} & \color{red} -0.57 & \hspace{-2pt}{\tiny $\pm$ 0.48} & \color{red} -0.63 & \hspace{-2pt}{\tiny $\pm$ 0.50} & \color{red} -0.82 & \hspace{-2pt}{\tiny $\pm$ 0.64} & \color{red} -0.98 & \hspace{-2pt}{\tiny $\pm$ 0.60} \\
$\;$ + Diffiner~\cite{sawata2023diffiner} & \color{red} -8.05 & \hspace{-2pt}{\tiny $\pm$ 3.32} & \color{red} -11.23 & \hspace{-2pt}{\tiny $\pm$ 3.28} & \color{red} -15.26 & \hspace{-2pt}{\tiny $\pm$ 3.68} & \color{red} -19.18 & \hspace{-2pt}{\tiny $\pm$ 4.52} \\
$\;$ + SIPS (ours) & \color{red} -0.60 & \hspace{-2pt}{\tiny $\pm$ 1.24} & \color{red} -0.94 & \hspace{-2pt}{\tiny $\pm$ 1.40} & \color{red} -1.13 & \hspace{-2pt}{\tiny $\pm$ 1.56} & \color{red} -1.19 & \hspace{-2pt}{\tiny $\pm$ 1.52} \\
\midrule\noalign{\vskip-\belowrulesep}\rowcolor{gray!15}\rule{0pt}{2.0ex}%
SEMamba~\cite{chao2024investigation} & \bfseries 5.01 &  & \bfseries 9.80 &  & \bfseries 13.87 &  & \bfseries 17.12 &  \\
\trimgrayrow
$\;$ + StoRM~\cite{lemercier2023storm} & \color{red} -0.37 & \hspace{-2pt}{\tiny $\pm$ 0.24} & \color{red} -0.37 & \hspace{-2pt}{\tiny $\pm$ 0.26} & \color{red} -0.44 & \hspace{-2pt}{\tiny $\pm$ 0.31} & \color{red} -0.56 & \hspace{-2pt}{\tiny $\pm$ 0.41} \\
$\;$ + Diffiner~\cite{sawata2023diffiner} & \color{red} -5.40 & \hspace{-2pt}{\tiny $\pm$ 2.40} & \color{red} -6.22 & \hspace{-2pt}{\tiny $\pm$ 1.82} & \color{red} -7.01 & \hspace{-2pt}{\tiny $\pm$ 1.85} & \color{red} -7.35 & \hspace{-2pt}{\tiny $\pm$ 2.11} \\
$\;$ + SIPS (ours) & \color{red} -0.02 & \hspace{-2pt}{\tiny $\pm$ 0.18} & \color{red} -0.06 & \hspace{-2pt}{\tiny $\pm$ 0.16} & \color{red} -0.12 & \hspace{-2pt}{\tiny $\pm$ 0.22} & \color{red} -0.12 & \hspace{-2pt}{\tiny $\pm$ 0.32} \\
\bottomrule
\end{tabular}
\endgroup
\vspace{-1em}
\end{table*}

\begin{table*}[t]\centering
\caption{\textbf{Results per input SNR (DNSMOS [P.808] and NISQA)}. Results on the EARS-WHAM (v2) test set for non-intrusive quality metrics, segmented into the four input-SNR bins specified in the column headers. Clean, noisy, and predictor entries show absolute means. Refiner entries show the mean change from their predictor, with the population standard deviation of paired per-utterance changes in smaller text. Changes are computed from unrounded scores and rounded only for display. Green changes indicate improvement and red changes indicate degradation. Best absolute means within each predictor group and SNR bin, including the predictor, are bold. Predictors have a gray background.}
\label{tab:earswham_per_snr_2}
\begingroup
\setlength{\tabcolsep}{3.5pt}
\begin{tabular}{lS[table-format=+1.2,detect-weight=true,retain-explicit-plus=true]@{\hspace{\tabcolsep}}lS[table-format=+1.2,detect-weight=true,retain-explicit-plus=true]@{\hspace{\tabcolsep}}lS[table-format=+1.2,detect-weight=true,retain-explicit-plus=true]@{\hspace{\tabcolsep}}lS[table-format=+1.2,detect-weight=true,retain-explicit-plus=true]@{\hspace{\tabcolsep}}l}
\toprule
& \multicolumn{8}{c}{\textbf{DNSMOS [P.808]} $\uparrow$} \\
\cmidrule(lr){2-9}
& \multicolumn{2}{c}{[-2.5, 2.5] dB} & \multicolumn{2}{c}{(2.5, 7.5] dB} & \multicolumn{2}{c}{(7.5, 12.5] dB} & \multicolumn{2}{c}{(12.5, 17.5] dB} \\ \midrule
Clean & 3.87 &  & 3.89 &  & 3.90 &  & 3.89 &  \\
Noisy & 2.39 &  & 2.64 &  & 2.85 &  & 3.05 &  \\
\midrule\noalign{\vskip-\belowrulesep}\rowcolor{gray!15}\rule{0pt}{2.0ex}%
Conv-TasNet~\cite{luo2019conv} & 2.57 &  & 2.77 &  & 3.04 &  & 3.25 &  \\
\trimgrayrow
$\;$ + StoRM~\cite{lemercier2023storm} & \color{green!50!black} \bfseries +0.58 & \hspace{-2pt}{\tiny $\pm$ 0.24} & \color{green!50!black} \bfseries +0.58 & \hspace{-2pt}{\tiny $\pm$ 0.22} & \color{green!50!black} \bfseries +0.54 & \hspace{-2pt}{\tiny $\pm$ 0.19} & \color{green!50!black} \bfseries +0.44 & \hspace{-2pt}{\tiny $\pm$ 0.18} \\
$\;$ + Diffiner~\cite{sawata2023diffiner} & \color{green!50!black} +0.01 & \hspace{-2pt}{\tiny $\pm$ 0.24} & \color{green!50!black} +0.13 & \hspace{-2pt}{\tiny $\pm$ 0.24} & \color{green!50!black} +0.13 & \hspace{-2pt}{\tiny $\pm$ 0.23} & \color{green!50!black} +0.09 & \hspace{-2pt}{\tiny $\pm$ 0.19} \\
$\;$ + SIPS (ours) & \color{green!50!black} +0.15 & \hspace{-2pt}{\tiny $\pm$ 0.09} & \color{green!50!black} +0.19 & \hspace{-2pt}{\tiny $\pm$ 0.11} & \color{green!50!black} +0.22 & \hspace{-2pt}{\tiny $\pm$ 0.10} & \color{green!50!black} +0.21 & \hspace{-2pt}{\tiny $\pm$ 0.10} \\
\midrule\noalign{\vskip-\belowrulesep}\rowcolor{gray!15}\rule{0pt}{2.0ex}%
NCSN++~\cite{lemercier2023storm} & 3.44 &  & 3.68 &  & 3.85 &  & \bfseries 3.93 &  \\
\trimgrayrow
$\;$ + StoRM~\cite{lemercier2023storm} & \color{green!50!black} \bfseries +0.26 & \hspace{-2pt}{\tiny $\pm$ 0.13} & \color{green!50!black} \bfseries +0.15 & \hspace{-2pt}{\tiny $\pm$ 0.12} & \color{green!50!black} \bfseries +0.04 & \hspace{-2pt}{\tiny $\pm$ 0.10} & \color{red} -0.01 & \hspace{-2pt}{\tiny $\pm$ 0.09}\\
$\;$ + Diffiner~\cite{sawata2023diffiner} & \color{red} -0.11 & \hspace{-2pt}{\tiny $\pm$ 0.23} & \color{red} -0.17 & \hspace{-2pt}{\tiny $\pm$ 0.23} & \color{red} -0.25 & \hspace{-2pt}{\tiny $\pm$ 0.23} & \color{red} -0.38 & \hspace{-2pt}{\tiny $\pm$ 0.23} \\
$\;$ + SIPS (ours) & \color{red} -0.07 & \hspace{-2pt}{\tiny $\pm$ 0.11} & \color{red} -0.05 & \hspace{-2pt}{\tiny $\pm$ 0.10} & \color{red} -0.02 & \hspace{-2pt}{\tiny $\pm$ 0.07} & \color{red} -0.02 & \hspace{-2pt}{\tiny $\pm$ 0.06} \\
\midrule\noalign{\vskip-\belowrulesep}\rowcolor{gray!15}\rule{0pt}{2.0ex}%
SEMamba~\cite{chao2024investigation} & 3.43 &  & 3.70 &  & 3.85 &  & 3.89 &  \\
\trimgrayrow
$\;$ + StoRM~\cite{lemercier2023storm} & \color{green!50!black} +0.03 & \hspace{-2pt}{\tiny $\pm$ 0.12} &+0.00 & \hspace{-2pt}{\tiny $\pm$ 0.08} & \color{red} -0.02 & \hspace{-2pt}{\tiny $\pm$ 0.06} & \color{red} -0.03 & \hspace{-2pt}{\tiny $\pm$ 0.05} \\
$\;$ + Diffiner~\cite{sawata2023diffiner} & \color{red} -0.29 & \hspace{-2pt}{\tiny $\pm$ 0.24} & \color{red} -0.31 & \hspace{-2pt}{\tiny $\pm$ 0.20} & \color{red} -0.27 & \hspace{-2pt}{\tiny $\pm$ 0.15} & \color{red} -0.26 & \hspace{-2pt}{\tiny $\pm$ 0.15} \\
$\;$ + SIPS (ours) & \color{green!50!black} \bfseries +0.04 & \hspace{-2pt}{\tiny $\pm$ 0.11} & \color{green!50!black} \bfseries +0.02 & \hspace{-2pt}{\tiny $\pm$ 0.08} & \color{green!50!black} \bfseries +0.02 &\hspace{-2pt}{\tiny $\pm$ 0.06} & \color{green!50!black} \bfseries +0.02 & \hspace{-2pt}{\tiny $\pm$ 0.05} \\
\bottomrule
\end{tabular}
\vskip 2em
\begin{tabular}{lS[table-format=+1.2,detect-weight=true,retain-explicit-plus=true]@{\hspace{\tabcolsep}}lS[table-format=+1.2,detect-weight=true,retain-explicit-plus=true]@{\hspace{\tabcolsep}}lS[table-format=+1.2,detect-weight=true,retain-explicit-plus=true]@{\hspace{\tabcolsep}}lS[table-format=+1.2,detect-weight=true,retain-explicit-plus=true]@{\hspace{\tabcolsep}}l}
\toprule
& \multicolumn{8}{c}{\textbf{NISQA} $\uparrow$} \\
\cmidrule(lr){2-9}
& \multicolumn{2}{c}{[-2.5, 2.5] dB} & \multicolumn{2}{c}{(2.5, 7.5] dB} & \multicolumn{2}{c}{(7.5, 12.5] dB} & \multicolumn{2}{c}{(12.5, 17.5] dB} \\ \midrule
Clean & 4.05 &  & 4.07 &  & 4.16 &  & 4.07 &  \\
Noisy & 1.22 &  & 1.71 &  & 2.28 &  & 2.63 &  \\
\midrule\noalign{\vskip-\belowrulesep}\rowcolor{gray!15}\rule{0pt}{2.0ex}%
Conv-TasNet~\cite{luo2019conv} & 1.01 &  & 1.18 &  & 1.58 &  & 1.99 &  \\
\trimgrayrow
$\;$ + StoRM~\cite{lemercier2023storm} & \color{green!50!black} \bfseries +1.78 & \hspace{-2pt}{\tiny $\pm$ 0.67} & \color{green!50!black} \bfseries +2.04 & \hspace{-2pt}{\tiny $\pm$ 0.67} & \color{green!50!black} \bfseries +2.04 & \hspace{-2pt}{\tiny $\pm$ 0.59} & \color{green!50!black} \bfseries +1.73 & \hspace{-2pt}{\tiny $\pm$ 0.58} \\
$\;$ + Diffiner~\cite{sawata2023diffiner} & \color{green!50!black} +1.39 & \hspace{-2pt}{\tiny $\pm$ 0.91} & \color{green!50!black} +1.80 & \hspace{-2pt}{\tiny $\pm$ 0.85} & \color{green!50!black} +1.73 & \hspace{-2pt}{\tiny $\pm$ 0.76} & \color{green!50!black} +1.45 & \hspace{-2pt}{\tiny $\pm$ 0.75} \\
$\;$ + SIPS (ours) & \color{green!50!black} +0.40 & \hspace{-2pt}{\tiny $\pm$ 0.26} & \color{green!50!black} +0.50 & \hspace{-2pt}{\tiny $\pm$ 0.32} & \color{green!50!black} +0.64 & \hspace{-2pt}{\tiny $\pm$ 0.39} & \color{green!50!black} +0.66 & \hspace{-2pt}{\tiny $\pm$ 0.39} \\
\midrule\noalign{\vskip-\belowrulesep}\rowcolor{gray!15}\rule{0pt}{2.0ex}%
NCSN++~\cite{lemercier2023storm} & 3.04 &  & 3.76 &  & 4.24 &  & \bfseries 4.34 &  \\
\trimgrayrow
$\;$ + StoRM~\cite{lemercier2023storm} & \color{green!50!black} \bfseries +0.70 & \hspace{-2pt}{\tiny $\pm$ 0.39} & \color{green!50!black} \bfseries +0.34 & \hspace{-2pt}{\tiny $\pm$ 0.34} & \color{green!50!black} \bfseries +0.09 & \hspace{-2pt}{\tiny $\pm$ 0.25} & \color{red} -0.03 & \hspace{-2pt}{\tiny $\pm$ 0.20}\\
$\;$ + Diffiner~\cite{sawata2023diffiner} & \color{green!50!black} +0.26 & \hspace{-2pt}{\tiny $\pm$ 0.74} & \color{red} -0.32 & \hspace{-2pt}{\tiny $\pm$ 0.68} & \color{red} -0.73 & \hspace{-2pt}{\tiny $\pm$ 0.53} & \color{red} -0.99 & \hspace{-2pt}{\tiny $\pm$ 0.55} \\
$\;$ + SIPS (ours) & \color{green!50!black} +0.12 & \hspace{-2pt}{\tiny $\pm$ 0.28} & \color{green!50!black} +0.03 & \hspace{-2pt}{\tiny $\pm$ 0.21} & \color{red} -0.03 & \hspace{-2pt}{\tiny $\pm$ 0.15} & \bfseries +0.00 & \hspace{-2pt}{\tiny $\pm$ 0.10} \\
\midrule\noalign{\vskip-\belowrulesep}\rowcolor{gray!15}\rule{0pt}{2.0ex}%
SEMamba~\cite{chao2024investigation} & 2.69 &  & 3.46 &  & 4.03 &  & 4.15 &  \\
\trimgrayrow
$\;$ + StoRM~\cite{lemercier2023storm} & \color{green!50!black} +0.54 & \hspace{-2pt}{\tiny $\pm$ 0.42} &\color{green!50!black} +0.18 & \hspace{-2pt}{\tiny $\pm$ 0.30} & \color{red} -0.03 & \hspace{-2pt}{\tiny $\pm$ 0.18} & \color{red} -0.11 & \hspace{-2pt}{\tiny $\pm$ 0.14} \\
$\;$ + Diffiner~\cite{sawata2023diffiner} & \color{green!50!black} \bfseries +0.65 & \hspace{-2pt}{\tiny $\pm$ 0.82} & \color{green!50!black} +0.10 & \hspace{-2pt}{\tiny $\pm$ 0.57} & \color{red} -0.17 & \hspace{-2pt}{\tiny $\pm$ 0.33} & \color{red} -0.22 & \hspace{-2pt}{\tiny $\pm$ 0.29} \\
$\;$ + SIPS (ours) & \color{green!50!black} +0.52 & \hspace{-2pt}{\tiny $\pm$ 0.37} & \color{green!50!black} \bfseries +0.27 & \hspace{-2pt}{\tiny $\pm$ 0.28} & \color{green!50!black} \bfseries +0.13 & \hspace{-2pt}{\tiny $\pm$ 0.15} & \color{green!50!black} \bfseries +0.08 & \hspace{-2pt}{\tiny $\pm$ 0.12} \\
\bottomrule
\end{tabular}
\endgroup
\vspace{-1em}
\end{table*}

\begin{table*}[t]\centering
\caption{\textbf{Results per input SNR (UTMOS and WER [\%])}. Results on the EARS-WHAM (v2) test set for speech quality and recognition metrics, segmented into the four input-SNR bins specified in the column headers. Clean, noisy, and predictor entries show absolute means. Refiner entries show the mean change from their predictor, with the population standard deviation of paired per-utterance changes in smaller text. Changes are computed from unrounded scores and rounded only for display. Green changes indicate improvement and red changes indicate degradation. Best absolute means within each predictor group and SNR bin, including the predictor, are bold. Predictors have a gray background. WER means are corpus-level within each SNR bin; changes are in percentage points and their standard deviations use paired per-utterance WER differences.}
\label{tab:earswham_per_snr_3}
\begingroup
\setlength{\tabcolsep}{3.5pt}
\begin{tabular}{lS[table-format=+1.2,detect-weight=true,retain-explicit-plus=true]@{\hspace{\tabcolsep}}lS[table-format=+1.2,detect-weight=true,retain-explicit-plus=true]@{\hspace{\tabcolsep}}lS[table-format=+1.2,detect-weight=true,retain-explicit-plus=true]@{\hspace{\tabcolsep}}lS[table-format=+1.2,detect-weight=true,retain-explicit-plus=true]@{\hspace{\tabcolsep}}l}
\toprule
& \multicolumn{8}{c}{\textbf{UTMOS} $\uparrow$} \\
\cmidrule(lr){2-9}
& \multicolumn{2}{c}{[-2.5, 2.5] dB} & \multicolumn{2}{c}{(2.5, 7.5] dB} & \multicolumn{2}{c}{(7.5, 12.5] dB} & \multicolumn{2}{c}{(12.5, 17.5] dB} \\ \midrule
Clean & 3.68 &  & 3.67 &  & 3.71 &  & 3.65 &  \\
Noisy & 1.33 &  & 1.29 &  & 1.74 &  & 2.39 &  \\
\midrule\noalign{\vskip-\belowrulesep}\rowcolor{gray!15}\rule{0pt}{2.0ex}%
Conv-TasNet~\cite{luo2019conv} & 1.25 &  & 1.35 &  & 1.82 &  & 2.27 &  \\
\trimgrayrow
$\;$ + StoRM~\cite{lemercier2023storm} & \color{green!50!black} +0.35 & \hspace{-2pt}{\tiny $\pm$ 0.30} &\color{green!50!black} \bfseries +0.63 & \hspace{-2pt}{\tiny $\pm$ 0.39} & \color{green!50!black} \bfseries +0.67 & \hspace{-2pt}{\tiny $\pm$ 0.31} & \color{green!50!black} \bfseries +0.53 & \hspace{-2pt}{\tiny $\pm$ 0.29} \\
$\;$ + Diffiner~\cite{sawata2023diffiner} & \color{green!50!black} \bfseries +0.56 & \hspace{-2pt}{\tiny $\pm$ 0.28} & \color{green!50!black} +0.48 & \hspace{-2pt}{\tiny $\pm$ 0.27} & \color{green!50!black} +0.23 & \hspace{-2pt}{\tiny $\pm$ 0.36} & \color{red} -0.04 & \hspace{-2pt}{\tiny $\pm$ 0.37} \\
$\;$ + SIPS (ours) & \color{green!50!black} +0.03 & \hspace{-2pt}{\tiny $\pm$ 0.08} & \color{green!50!black} +0.08 & \hspace{-2pt}{\tiny $\pm$ 0.10} & \color{green!50!black} +0.10 & \hspace{-2pt}{\tiny $\pm$ 0.10} & \color{green!50!black} +0.07 & \hspace{-2pt}{\tiny $\pm$ 0.10} \\
\midrule\noalign{\vskip-\belowrulesep}\rowcolor{gray!15}\rule{0pt}{2.0ex}%
NCSN++~\cite{lemercier2023storm} & 2.06 &  & 2.59 &  & 3.10 &  & 3.30 &  \\
\trimgrayrow
$\;$ + StoRM~\cite{lemercier2023storm} & \color{green!50!black} \bfseries +0.10 & \hspace{-2pt}{\tiny $\pm$ 0.13} & \color{green!50!black} \bfseries +0.13 & \hspace{-2pt}{\tiny $\pm$ 0.12} & \color{green!50!black} \bfseries +0.12 & \hspace{-2pt}{\tiny $\pm$ 0.10} & \color{green!50!black} \bfseries +0.10 & \hspace{-2pt}{\tiny $\pm$ 0.09} \\
$\;$ + Diffiner~\cite{sawata2023diffiner} & \color{red} -0.09 & \hspace{-2pt}{\tiny $\pm$ 0.42} & \color{red} -0.42 & \hspace{-2pt}{\tiny $\pm$ 0.48} & \color{red} -0.75 & \hspace{-2pt}{\tiny $\pm$ 0.50} & \color{red} -0.93 & \hspace{-2pt}{\tiny $\pm$ 0.55} \\
$\;$ + SIPS (ours) & \color{green!50!black} +0.03 & \hspace{-2pt}{\tiny $\pm$ 0.11} & \color{green!50!black} +0.02 & \hspace{-2pt}{\tiny $\pm$ 0.11} & \color{green!50!black} +0.01 & \hspace{-2pt}{\tiny $\pm$ 0.10} & +0.00 & \hspace{-2pt}{\tiny $\pm$ 0.09} \\
\midrule\noalign{\vskip-\belowrulesep}\rowcolor{gray!15}\rule{0pt}{2.0ex}%
SEMamba~\cite{chao2024investigation} & 2.18 &  & 2.78 &  & 3.31 &  & 3.48 &  \\
\trimgrayrow
$\;$ + StoRM~\cite{lemercier2023storm} & \color{red} -0.04 & \hspace{-2pt}{\tiny $\pm$ 0.12} & \color{red} -0.03 & \hspace{-2pt}{\tiny $\pm$ 0.10} & \color{red} -0.04 & \hspace{-2pt}{\tiny $\pm$ 0.08} & \color{red} -0.04 & \hspace{-2pt}{\tiny $\pm$ 0.06} \\
$\;$ + Diffiner~\cite{sawata2023diffiner} & \color{red} -0.18 & \hspace{-2pt}{\tiny $\pm$ 0.49} & \color{red} -0.46 & \hspace{-2pt}{\tiny $\pm$ 0.42} & \color{red} -0.50 & \hspace{-2pt}{\tiny $\pm$ 0.31} & \color{red} -0.45 & \hspace{-2pt}{\tiny $\pm$ 0.26} \\
$\;$ + SIPS (ours) & \color{green!50!black} \bfseries +0.07 & \hspace{-2pt}{\tiny $\pm$ 0.09} & \color{green!50!black} \bfseries +0.07 & \hspace{-2pt}{\tiny $\pm$ 0.09} & \color{green!50!black} \bfseries +0.03 &\hspace{-2pt}{\tiny $\pm$ 0.07} & \color{green!50!black} \bfseries +0.01 & \hspace{-2pt}{\tiny $\pm$ 0.05} \\
\bottomrule
\end{tabular}
\vskip 2em
\begin{tabular}{lS[table-format=+2.2,detect-weight=true,retain-explicit-plus=true]@{\hspace{\tabcolsep}}lS[table-format=+2.2,detect-weight=true,retain-explicit-plus=true]@{\hspace{\tabcolsep}}lS[table-format=+2.2,detect-weight=true,retain-explicit-plus=true]@{\hspace{\tabcolsep}}lS[table-format=+2.2,detect-weight=true,retain-explicit-plus=true]@{\hspace{\tabcolsep}}l}
\toprule
& \multicolumn{8}{c}{\textbf{WER [\%]} $\downarrow$} \\
\cmidrule(lr){2-9}
& \multicolumn{2}{c}{[-2.5, 2.5] dB} & \multicolumn{2}{c}{(2.5, 7.5] dB} & \multicolumn{2}{c}{(7.5, 12.5] dB} & \multicolumn{2}{c}{(12.5, 17.5] dB} \\ \midrule
Clean & 9.11 &  & 9.32 &  & 8.49 &  & 8.88 &  \\
Noisy & 63.24 &  & 35.52 &  & 20.14 &  & 13.13 &  \\
\midrule\noalign{\vskip-\belowrulesep}\rowcolor{gray!15}\rule{0pt}{2.0ex}%
Conv-TasNet~\cite{luo2019conv} & \bfseries 86.56 &  & \bfseries 64.92 &  & \bfseries 39.93 &  & \bfseries24.47 &  \\
\trimgrayrow
$\;$ + StoRM~\cite{lemercier2023storm} & \color{red} +5.13 & \hspace{-2pt}{\tiny $\pm$ \tablenum[table-format=2.2]{10.37}} & \color{red} +12.63 & \hspace{-2pt}{\tiny $\pm$ \tablenum[table-format=2.2]{12.36}} & \color{red} +14.03 & \hspace{-2pt}{\tiny $\pm$ \tablenum[table-format=2.2]{15.33}} & \color{red} +10.80 & \hspace{-2pt}{\tiny $\pm$ \tablenum[table-format=2.2]{12.54}} \\
$\;$ + Diffiner~\cite{sawata2023diffiner} & \color{red} +11.96 & \hspace{-2pt}{\tiny $\pm$ \tablenum[table-format=2.2]{13.40}} & \color{red} +29.97 & \hspace{-2pt}{\tiny $\pm$ \tablenum[table-format=2.2]{18.04}}& \color{red} +46.25 & \hspace{-2pt}{\tiny $\pm$ \tablenum[table-format=2.2]{20.53}} & \color{red} +47.10& \hspace{-2pt}{\tiny $\pm$ \tablenum[table-format=2.2]{16.26}} \\
$\;$ + SIPS (ours) & \color{red} +3.53 & \hspace{-2pt}{\tiny $\pm$ \tablenum[table-format=2.2]{6.82}} & \color{red} +6.79 & \hspace{-2pt}{\tiny $\pm$ \tablenum[table-format=2.2]{7.85}} & \color{red} +6.27 & \hspace{-2pt}{\tiny $\pm$ \tablenum[table-format=2.2]{7.87}} & \color{red} +4.29 & \hspace{-2pt}{\tiny $\pm$ \tablenum[table-format=2.2]{6.60}} \\
\midrule\noalign{\vskip-\belowrulesep}\rowcolor{gray!15}\rule{0pt}{2.0ex}%
NCSN++~\cite{lemercier2023storm} & \bfseries 54.75 &  & \bfseries 30.83 &  & \bfseries 18.45 &  & \bfseries 13.06 &  \\
\trimgrayrow
$\;$ + StoRM~\cite{lemercier2023storm} & \color{red} +4.34 & \hspace{-2pt}{\tiny $\pm$ \tablenum[table-format=2.2]{10.63}} & \color{red} +3.59 & \hspace{-2pt}{\tiny $\pm$ \tablenum[table-format=2.2]{9.35}} & \color{red} +1.76 & \hspace{-2pt}{\tiny $\pm$ \tablenum[table-format=2.2]{6.98}} & \color{red} +0.67 & \hspace{-2pt}{\tiny $\pm$ \tablenum[table-format=2.2]{5.36}} \\
$\;$ + Diffiner~\cite{sawata2023diffiner} & \color{red} +29.56 & \hspace{-2pt}{\tiny $\pm$ \tablenum[table-format=2.2]{20.04}} & \color{red} +41.88 & \hspace{-2pt}{\tiny $\pm$ \tablenum[table-format=2.2]{19.55}}& \color{red} +43.17 & \hspace{-2pt}{\tiny $\pm$ \tablenum[table-format=2.2]{22.11}} & \color{red} +44.76& \hspace{-2pt}{\tiny $\pm$ \tablenum[table-format=2.2]{24.38}} \\
$\;$ + SIPS (ours) & \color{red} +2.22 & \hspace{-2pt}{\tiny $\pm$ \tablenum[table-format=2.2]{9.62}} & \color{red} +1.91 & \hspace{-2pt}{\tiny $\pm$ \tablenum[table-format=2.2]{8.02}} & \color{red} +1.82 & \hspace{-2pt}{\tiny $\pm$ \tablenum[table-format=2.2]{6.55}} & \color{red} +1.35 & \hspace{-2pt}{\tiny $\pm$ \tablenum[table-format=2.2]{6.50}} \\
\midrule\noalign{\vskip-\belowrulesep}\rowcolor{gray!15}\rule{0pt}{2.0ex}%
SEMamba~\cite{chao2024investigation} & \bfseries 53.42 &  & \bfseries 29.72 &  & \bfseries 16.66 &  & \bfseries 12.99 &  \\
\trimgrayrow
$\;$ + StoRM~\cite{lemercier2023storm} & \color{red} +2.81 & \hspace{-2pt}{\tiny $\pm$ \tablenum[table-format=2.2]{7.61}} & \color{red} +2.15 & \hspace{-2pt}{\tiny $\pm$ \tablenum[table-format=2.2]{6.28}} & \color{red} +0.85 & \hspace{-2pt}{\tiny $\pm$ \tablenum[table-format=2.2]{5.02}} & \color{red} +0.42 & \hspace{-2pt}{\tiny $\pm$ \tablenum[table-format=2.2]{2.98}} \\
$\;$ + Diffiner~\cite{sawata2023diffiner} & \color{red} +36.33 & \hspace{-2pt}{\tiny $\pm$ \tablenum[table-format=2.2]{18.76}} & \color{red} +44.55 & \hspace{-2pt}{\tiny $\pm$ \tablenum[table-format=2.2]{16.74}}& \color{red} +35.10 & \hspace{-2pt}{\tiny $\pm$ \tablenum[table-format=2.2]{18.05}} & \color{red} +28.40& \hspace{-2pt}{\tiny $\pm$ \tablenum[table-format=2.2]{17.14}} \\
$\;$ + SIPS (ours) & \color{red} +1.52 & \hspace{-2pt}{\tiny $\pm$ \tablenum[table-format=2.2]{6.93}} & \color{red} +1.61 & \hspace{-2pt}{\tiny $\pm$ \tablenum[table-format=2.2]{5.54}} & \color{red} +0.97 & \hspace{-2pt}{\tiny $\pm$ \tablenum[table-format=2.2]{3.82}} & \color{red} +0.68 & \hspace{-2pt}{\tiny $\pm$ \tablenum[table-format=2.2]{4.07}} \\
\bottomrule
\end{tabular}
\endgroup
\vspace{-1em}
\end{table*}

\clearpage

\subsection{Subjective listening test}
\label{app:ccr_listening_test}

We evaluated the perceptual preference between SEMamba and SEMamba + SIPS in the mismatched EARS-WHAM setting using a Comparison Category Rating (CCR) listening test following the ITU-T P.808 style protocol.
The SEMamba + SIPS stimuli were generated using $M=15$ sampling steps, $\kappa=0.4$, and predictor post-processing, consistent with the mismatched speech enhancement evaluation.
The CCR scale ranges from -3 to +3, where positive values indicate preference for SEMamba + SIPS, negative values indicate preference for SEMamba, and zero indicates no preference.
We used the public P.808 implementation of Naderi and Cutler~\cite{naderi2020p808} to run the test on Amazon Mechanical Turk and to perform automated setup checks, quality control, data cleansing, and result aggregation.

The study collected 534 assignments from 174 workers, with 11 comparisons per assignment including one control item.
After screening for control-question performance, setup checks, and rating consistency, 421 assignments (78.84\%) were retained.
Removing the control items yielded 4,210 accepted CCR ratings from 134 retained workers across 886 unique comparison pairs.

\paragraph{Preference by input SNR.}
Table~\ref{tab:ccr_per_snr} reports the CCR results by input-\ac{SNR} bin.
The mean CCR is positive in every bin, with \acp{CI} above zero throughout.
The largest preference for SEMamba + SIPS occurs in the lowest-\ac{SNR} bin and decreases monotonically with increasing \ac{SNR}.
Figure~\ref{fig:mean_ccr_snr} in the main text visualizes these means and their 95\% \acp{CI}.
The per-bin intervals use the normal approximation $\bar{c} \pm 1.96s/\sqrt{n}$, where $\bar{c}$ is the mean CCR, $s$ the sample standard deviation, and $n$ the number of accepted ratings in the bin.
These are rating-level intervals and do not account for repeated judgments by the same worker or of the same comparison pair.

Figure~\ref{fig:ccr_preference_by_snr} separates the ratings into preferences for SEMamba, ties, and preferences for SEMamba + SIPS, using the same percentage scale in all panels.
SEMamba + SIPS receives the largest share of ratings in every bin, although this share decreases from 50.8\% at Very Low \ac{SNR} to 37.5\% at High \ac{SNR}.
Over the same range, ties increase from 22.0\% to 32.1\%, while the share favoring SEMamba changes from 27.3\% to 30.4\%.
Thus, the smaller mean preference at higher \ac{SNR} is accompanied mainly by more ties, rather than a comparable increase in preferences for SEMamba.
The margin between ratings favoring SEMamba + SIPS and those favoring SEMamba decreases from 23.5 to 7.1 percentage points.

\begin{table}[t]
\centering
\caption{CCR listening-test summary by input-SNR bin. Preference percentages are computed over accepted ratings in each bin. 95\% \acp{CI} use a rating-level normal approximation.}
\label{tab:ccr_per_snr}
\small
\begin{tabular}{@{}lrrrrrl@{}}
\toprule
SNR range &
Ratings &
SEMamba &
+ SIPS &
Tie &
CCR &
95\% CI \\
\midrule
{[-2.5, 2.5]} dB
& 1,097
& 27.3\%
& \textbf{50.8\%}
& 22.0\%
& 0.390
& [0.318, 0.462] \\
(2.5, 7.5] dB
& 1,038
& 30.3\%
& \textbf{47.9\%}
& 21.8\%
& 0.291
& [0.218, 0.364] \\
(7.5, 12.5] dB
& 1,065
& 29.0\%
& \textbf{40.2\%}
& 30.8\%
& 0.158
& [0.091, 0.224] \\
(12.5, 17.5] dB
& 1,010
& 30.4\%
& \textbf{37.5\%}
& 32.1\%
& 0.085
& [0.018, 0.153] \\
\bottomrule
\end{tabular}
\end{table}

% Shared plot for the overall and per-SNR preference counts.
% Argument: data file; the total is calculated from its counts.
% Optional argument: label for SEMamba + SIPS.
\newcommand{\listeningPreferencePlot}[2][SEMamba + SIPS]{%
\begingroup
\pgfplotstableread{#2}\listeningPreferenceData
\pgfplotstablegetelem{0}{count}\of\listeningPreferenceData
\let\semambaCount\pgfplotsretval
\pgfplotstablegetelem{1}{count}\of\listeningPreferenceData
\let\tieCount\pgfplotsretval
\pgfplotstablegetelem{2}{count}\of\listeningPreferenceData
\let\sipsCount\pgfplotsretval
\pgfmathsetmacro{\preferenceTotal}{\semambaCount+\tieCount+\sipsCount}
\pgfmathsetmacro{\semambaShare}{\semambaCount/\preferenceTotal*100}
\pgfmathsetmacro{\tieShare}{\tieCount/\preferenceTotal*100}
\pgfmathsetmacro{\sipsShare}{\sipsCount/\preferenceTotal*100}
\begin{tikzpicture}
\begin{axis}[
    width=\linewidth,
    height=4.5cm,
    font=\small,
    ylabel={Ratings [\%]},
    ymin=0, ymax=60,
    ytick={0,20,40,60},
    xmin=0.5, xmax=3.5,
    xtick={1,2,3},
    xticklabels={SEMamba,Tie,#1},
    tick align=outside,
    xtick pos=bottom,
    ytick pos=left,
    ymajorgrids=true,
    grid style={dashed,gray!50},
    ybar,
    bar width=18pt,
    bar shift=0pt,
    nodes near coords={\pgfmathprintnumber[fixed,precision=1,zerofill]{\pgfplotspointmeta}\%},
    every node near coord/.append style={font=\small},
]
\addplot[fill=cbBlue,draw=cbBlue] coordinates {(1,\semambaShare)};
\addplot[fill=gray!40,draw=gray!60] coordinates {(2,\tieShare)};
\addplot[fill=cbOrange,draw=cbOrange] coordinates {(3,\sipsShare)};
\end{axis}
\end{tikzpicture}
\endgroup
}

\begin{figure}[t]
\centering
\input{figures/listening_test/preference_by_snr.tex}
\caption{Listener preference by input-SNR bin in the mismatched EARS-WHAM setting. Negative CCR ratings favor SEMamba, zero denotes a tie, and positive ratings favor SEMamba + SIPS. Bars show percentages of the accepted ratings in each bin.}
\label{fig:ccr_preference_by_snr}
\end{figure}

\paragraph{Aggregate preference.}
Table~\ref{tab:ccr_aggregate} summarizes the aggregate result, and Figure~\ref{fig:ccr_preference_overall} shows the preference distribution over all 4,210 accepted ratings.
Of these, 1,861 (44.2\%) favor SEMamba + SIPS, 1,230 (29.2\%) favor SEMamba, and 1,119 (26.6\%) indicate a tie.
Positive ratings therefore exceed negative ratings by 631, or 15.0 percentage points.
Among the 3,091 non-tied ratings, 60.2\% favor SEMamba + SIPS.
The aggregate mean CCR is 0.234, computed by weighting the per-bin means by their respective numbers of ratings.
The aggregate 95\% confidence interval [0.152, 0.323] is a percentile interval from 5,000 two-way cluster bootstrap replicates.
Workers and comparison pairs are resampled independently with replacement, with each rating weighted by the product of its worker and pair sampling multiplicities.
A one-sided sign-flip test for positive preference uses the unweighted mean of worker-mean CCRs and 5,000 randomizations, assigning each worker mean an independent sign of $\pm$1 with equal probability.
The p-value is $(b+1)/(5000+1)=0.00020$, where $b$ counts randomized statistics at least as large as the observed statistic.
The preference proportions complement the mean: they describe how often each system is preferred, whereas the mean also reflects the strength of the preference on the seven-point scale.

\begin{figure}[t]
\centering
\begin{minipage}{0.64\columnwidth}
    \centering
    \listeningPreferencePlot{figures/listening_test/preference_sign_counts_overall.dat}
\end{minipage}
\caption{Overall listener preference across 4,210 accepted CCR ratings in the mismatched EARS-WHAM setting. Ties are included in the denominator.}
\label{fig:ccr_preference_overall}
\end{figure}

\begin{table}[t]
\centering
\caption{Aggregate CCR result over all accepted listening-test ratings.}
\label{tab:ccr_aggregate}
\begin{tabular}{@{}lr@{}}
\toprule
Statistic & Value \\
\midrule
Mean CCR & 0.234 \\
95\% bootstrap CI & [0.152, 0.323] \\
Worker-level sign-flip test & p = 0.00020 \\
Total ratings & 4,210 \\
SEMamba preferred & 1,230 (29.2\%) \\
Tie & 1,119 (26.6\%) \\
SEMamba + SIPS preferred & 1,861 (44.2\%) \\
\bottomrule
\end{tabular}
\end{table}

\end{document}